\documentclass[prd,eqsecnum,nofootinbib,twocolumn]{revtex4-1}

\usepackage{fbb} 
\usepackage{cancel, mathtools}
\usepackage[libertine,bigdelims,vvarbb]{newtxmath}
\usepackage[cal=boondoxo]{mathalfa}
\usepackage[T1]{fontenc}
\useosf

\usepackage{bm,graphicx,mathrsfs,multirow}
\usepackage{amssymb,amsmath}
\usepackage[ugly]{units}
\usepackage[colorlinks=true,citecolor=blue,bookmarks=true]{hyperref}

\renewcommand{\Re}{\operatorname{Re} }
\renewcommand{\Im}{\operatorname{Im} }

\begin{document}

\title{Local symmetries as constraints on the motion of freely-falling extended bodies}

\author{Abraham I. Harte}

\author{David Dwyer}

\affiliation{Centre for Astrophysics and Relativity
	\\
	School of Mathematical Sciences
	\\
	Dublin City University, Glasnevin, Dublin 9, Ireland}

\begin{abstract} 
	Different extended objects can fall in different ways, depending on their internal structures. Some motions are nevertheless impossible, regardless of internal structure. This paper derives universal constraints on extended-body motion, both in Newtonian gravity and in general relativity. In both theories, we identify a weak notion of ``local symmetry'' which precludes certain force and torque combinations. Local symmetries imply that certain components of a body's quadrupole moment cannot affect its motion. They also imply that some forces can arise only in combination with appropriate torques. Many of these symmetries are shown to be determined by the algebraic structure of the tidal tensor. In general relativity, we thus relate qualitative features of extended-body motion to the Petrov type of the spacetime. Doing so shows that local symmetries are in fact ubiquitous. In general relativity, there are at least two such symmetries in all algebraically-special spacetimes. Some of these are generated by Killing vectors and some by conformal Killing-Yano tensors. However, many local symmetries do not fall into either of these classes.
\end{abstract}

\maketitle

\vskip 2pc

\section{Introduction}

One of the most fundamental results in general relativity is that freely-falling objects move on geodesics \cite{Ehlers2004, Yang2013, Geroch2018, GrallaWald}. However, geodesic motion is only an approximation. Real objects have finite mass, and are therefore affected by self-interaction \cite{PoissonRev, HarteReview, BarackPoundReview}. Real objects also have finite size, and are thus  affected by internal structure \cite{Dix79, ThorneHartle}. Regardless, geodesic motion has the property that all future trajectories are uniquely determined by initial positions and initial velocities. To the extent that objects \textit{do} move along geodesics, free-fall is therefore ``universal:'' Every object with the same initial conditions falls on the same trajectory.

Expanding in powers of an object's size, the first extended-body correction to geodesic motion involves an object's angular momentum. However, taking this into account does not spoil the universality of free-fall. Once an appropriate center-of-mass definition has been fixed, a unique trajectory follows from a given position, a given linear momentum, and a given angular momentum at some initial time. While more initial data is required than in the geodesic case, all bodies with the same initial conditions still follow the same trajectories. We may thus view this as a refinement rather than a failure of universal free-fall. Failure occurs when expanding through one higher order in an object's size.

More precisely, the universality of free-fall breaks down once quadrupole moments are taken into account. Although the trajectories of \textit{some} objects can be uniquely predicted given, e.g., their initial positions, initial linear momenta, initial angular momenta, and initial quadrupole moments, equations of motion can still differ from one object to another. Moreover, many bodies can be said only to satisfy \textit{laws} of motion rather than \textit{equations} of motion: Although trajectories are constrained by the laws of motion, they may fail to be uniquely \textit{determined} from any reasonable initial data set\footnote{One generally expects that unique predictions are always possible given a sufficiently detailed model and sufficiently detailed initial data. However, the question addressed here is whether or not the future can be uniquely predicted, to adequate precision, using only the small amount of  initial data which might be ascribed to an ``effective point particle'' (possibly with structure).}. Regardless, different bodies can fall in different ways, depending on the evolution of their quadrupole and higher-order moments. Our goal here is to constrain these differences, both in general relativity and in Newtonian gravity.  For simplicity, we focus only on quadrupolar effects.

Although extended-body effects are instantaneously small in most astrophysical systems, their effects can grow over time. Perhaps the best-known Newtonian example is the phenomenon of tidal locking \cite{HutTideReview, Tremaine, PoissonWill, Ferroglia2020}; another is the chaotic tumbling of Saturn's moon Hyperion \cite{Tremaine, Wisdom1984, WisdomChaosReview}. Indeed, tidal effects are known to produce secular changes in, e.g., orbital eccentricities, inclinations, and radii \cite{HutTideReview, Tremaine, Ogilvie2014}. In general relativity, tidal effects are often studied as something which affects the late-stage evolution of inspiraling binaries, potentially allowing gravitational wave observations to probe the composition of neutron star interiors \cite{Flanagan2008, Chatziioannou2018, Baiotti2017}.

Regardless, the primary goal here is not to model any particular astrophysical system. Instead, we identify fundamental, model-independent limitations on extended-body motion: What is possible and what is not? From this perspective, it is useful to imagine a hypothetical spacecraft which has been designed to be able to control its internal mass distribution. Changes in that distribution can then be used to modulate extended-body forces, and thus to control a spacecraft's motion. Such systems have been analyzed before, both in Newtonian gravity \cite{Murakami1981, MartinezSanchez, Beletsky, GratusTucker, HarteNewtonian} and in general relativity \cite{Wisdom, Harte2007, Avron, Bergamin, Silva2016, PoissonSwim, Vesely2019, HarteExtendedtypeD, Silva2022}. In both contexts, shape-changing spacecraft have been found to be able to produce large orbital changes simply by modulating extended-body forces over many orbits. Roughly speaking, this is accomplished by exchanging internal energy with orbital energy. It is also possible (and in fact simpler) for a spacecraft to control its rotation by similarly manipulating its internal mass distribution.

There are limitations, however. Certain course corrections cannot be produced, no matter how cleverly a spacecraft has been engineered. This can be seen most easily  in a uniform Newtonian gravitational field, where internal structure has no effect whatsoever; all objects fall identically. Relativistically, the same is true in all maximally-symmetric spacetimes \cite{Dixon70a, Dix79}. Different objects can fall differently only when there is some inhomogeneity to ``grab onto.'' This can be made more precise in general relativity by noting that for each Killing field which may exist, certain force and torque combinations are impossible, regardless of an object's internal structure \cite{Dixon70a, Dixon74, EhlersRudolph, Dix79}. Analogous results are also known in Newtonian gravity \cite{HarteNewtonian}. However, it is natural to ask if these are the \textit{only} fundamental constraints on extended-body forces and torques. 

They are not. At least in vacuum spacetimes which are of Petrov type D, certain torques are known to be impossible even when there are no Killing fields which exclude them \cite{HarteExtendedtypeD}. Such constraints can in fact be related to the presence of conformal Killing-Yano tensors, which describe a different kind of symmetry. Indeed, we show below that any conformal Killing-Yano tensor in a (not necessarily type D) vacuum spacetime precludes certain torque components. However, even this does not exhaust all fundamental restrictions on extended-body motion.  

We find that constraints due to Killing vectors and constraints due to conformal Killing-Yano tensors are both special cases of a certain type of ``local symmetry.'' Crucially, these symmetries are very common. In general relativity, we show that local symmetries exist in all algebraically-special spacetimes, and in many algebraically-general ones as well. We also find that every Newtonian gravitational field admits at least one local symmetry.

A related theme in this paper is to describe how extended-body effects depend qualitatively on the algebraic structure of the relevant tidal tensor. In general relativity, we derive local symmetries, and discuss quadrupolar forces and torques, in each of the Petrov types which can be associated with four-dimensional vacuum spacetimes. There are essentially three types of result: First, what is the space of possible torques which can arise due to extended-body effects? This is either four or six-dimensional, depending on the Petrov type (where the six-dimensional case allows any torque whatsoever). Our second type of result asks for the space of possible forces which can be varied without simultaneously varying the torque. This lies between zero and four dimensions. Our third type of result asks how many of a body's ten quadrupole components can affect its motion. The answer here lies between four and ten. In some Petrov type I spacetimes, an appropriately-engineered spacecraft could vary its ten quadrupole components in order to arbitrarily control all four force components and all six torque components. In other spacetimes, considerably less is possible. 

Although our main motivation here is to understand motion in general relativity, the Newtonian case is already rich and largely unexplored. In fact, all of the concepts which appear in general relativity are already present in Newtonian gravity. We thus begin in Sect. \ref{Sect:Newtonian} by describing Newtonian extended bodies. Newtonian tidal tensors are classified in terms of their algebraic structure, and the corresponding constraints on extended-body motion are derived. A concept of local symmetry is introduced as well. Sect. \ref{Sect:GR} performs the same analysis for extended bodies in general relativity, introducing local symmetries in that context and explaining how extended-body motion depends on the Petrov type. Appendix \ref{App:Notation} summarizes our notational conventions and provides a table of symbols.  Appendix \ref{App:PNDs} reviews some material on principal null directions and the Petrov classification. Appendix \ref{App:Tides} describes objects with tidally-induced quadrupole moments.

\section{Extended bodies in Newtonian gravity}
\label{Sect:Newtonian}

Before analyzing extended-body motion in general relativity, we first discuss motion in Newtonian gravity. This is partially because model-independent features of Newtonian extended-body motion do not appear to have been explored before, and are interesting in their own right. However, a thorough understanding of the Newtonian problem also allows us to better understand which effects are ``fundamentally'' relativistic and which are not. 

Regardless, Sect. \ref{Sect:NewtReview} reviews the Newtonian theory of extended-body motion in a form which emphasizes the role of symmetry and which easily carries over into general relativity. Sect. \ref{Sect:NewtSyms} considers the effects of symmetry on objects with arbitrary quadrupole moments. Sect. \ref{Sect:NewtAlgConstraints} then classifies different tidal tensors according to their eigenvalues and discusses how extended-body motion differs in each case. Lastly, Sect. \ref{Sect:NewtExamples} applies our formalism in order to describe motion in certain example gravitational fields.

\subsection{Generalized momentum and generalized force}
\label{Sect:NewtReview}

We begin by reviewing a perspective on Newtonian motion which was developed in \cite{HarteScalar, HarteReview, HarteNewtonian}, and which grew out of Dixon's formulation of extended-body motion in general relativity \cite{Dixon70a, Dixon74, Dix79, DixonReview}. The central object of study is the ``generalized momentum,'' which unifies a body's linear momentum and angular momentum into a single object. If an extended Newtonian body has momentum density $\rho_a$, its generalized momentum at time $t$ is defined to be
\begin{equation}
	\mathcal{P}_\xi (t) \equiv \int \rho_a (x,t) \xi^a(x) d V,
	\label{PDefNewt}
\end{equation}
where $\xi^a(x)$ is any Euclidean Killing field. At fixed $t$, the generalized momentum may be viewed as a linear map from the space of Killing fields into $\mathbb{R}$, and may therefore be interpreted as a vector in the six-dimensional space which is dual to the space of Euclidean Killing fields. Three of those six dimensions describe a body's linear momentum; the remaining three describe its angular momentum. 

Extracting linear and angular momenta from the generalized momentum requires a choice of origin which is not required for $\mathcal{P}_\xi$ itself. Letting $\gamma_t$ be such an origin at time $t$, the associated linear momentum $p_a (t;\gamma_t)$ and angular momentum $S^{ab} = S^{[ab]}(t;\gamma_t)$ are implicitly defined by
\begin{equation}
	\mathcal{P}_\xi(t) = p_a(t; \gamma_t) \xi^a(\gamma_t) + \frac{1}{2} S^{ab} ( t; \gamma_t) \nabla_a \xi_b(\gamma_t) .
	\label{PpSNewt}
\end{equation}
In this Euclidean context, $\nabla_a \nabla_b \xi_c = 0$ so $\nabla_a p_b = 0$. The angular momentum does however depend on the choice of origin, as is familiar even from elementary discussions of Newtonian mechanics. In fact, the linear and the angular momenta defined by \eqref{PDefNewt} and \eqref{PpSNewt} are essentially\footnote{The only difference is that it is more conventional to consider the angular momentum \textit{vector} $S^a \equiv \tfrac{1}{2} \epsilon^{abc} S_{bc}$ in place of the bivector $S^{ab}$. Both $S^a$ and $S^{ab}$ nevertheless encode the same information in Newtonian mechanics.} equivalent to elementary textbook definitions: Using Cartesian coordinates $x^i$,
\begin{subequations}
\begin{gather}
	p_i(t;\gamma_t) = \int \rho_i (x, t) d^3 x,
	 \\
	 S^{ij}(t;\gamma_t) = 2 \int (x-\gamma_t)^{[i} \rho^{j]} (x, t) d^3 x.
\end{gather}
\label{pSNewtInt}
\end{subequations}

The generalized momentum may be viewed as describing a body's ``bulk'' state. Mass and momentum conservation constrain the evolution of that state, and therefore the evolution of the generalized momentum: Differentiating \eqref{PDefNewt} may be shown to yield the ``generalized force'' \cite{HarteScalar, HarteReview}
\begin{equation}
	\mathcal{F}_\xi (t) \equiv \frac{ d}{dt} \mathcal{P}_\xi(t) = -\int \rho(x,t) \mathcal{L}_\xi \Phi(x,t) d^3x,
	\label{GenForceNewt}
\end{equation}
where $\rho$ denotes the body's mass density and $\Phi$ the Newtonian gravitational potential. Like the generalized momentum, the generalized force is, at fixed $t$, a six-dimensional vector in the space which is dual to the space of Euclidean Killing fields.

In the same way that the generalized momentum can be decomposed into a linear momentum and an angular momentum, the generalized force can be decomposed into an ordinary force $F_a(t;\gamma_t)$ and a torque $N^{ab} = N^{[ab]} (t;\gamma_t)$, both of which satisfy
\begin{align}
	\mathcal{F}_\xi (t)  = F_a(t;\gamma_t) \xi^a(\gamma_t) + \frac{1}{2} N^{ab} ( t;\gamma_t) \nabla_a \xi_b(\gamma_t) .
	\label{GenForceNewtFN}
\end{align}
Comparing this expression to the time derivative of \eqref{PpSNewt} recovers the laws of motion
\begin{equation}
	\frac{D}{ dt } p_a = F_a , \qquad \frac{ D }{ dt } S^{ab} = 2 p^{[a} \dot{\gamma}_t^{b]} + N^{ab}.
 \label{EOMNewtonian}
\end{equation}
The time derivatives here act on both arguments of $p_a(t; \gamma_t)$ and $S^{ab}(t; \gamma_t)$. Also note that the $p^{[a} \dot{\gamma}_t^{b]}$ term which affects the angular momentum is purely kinematic, and is therefore natural to separate from the ``dynamical'' torque $N^{ab}$. It vanishes when, e.g., $\gamma_t$ is placed at an object's center of mass. Although the force and the torque which appear here are equivalent to elementary expressions, we find it convenient to work with the more-abstract concepts of generalized momentum and generalized force. One reason for this is that doing so allows forces and torques to be considered simultaneously in a single calculation. Another advantage is that the Lie derivative in \eqref{GenForceNewt} provides an immediate connection with symmetries and conservation laws. 

Regardless, the gravitational potential $\Phi$ which appears in the generalized force is, \textit{a priori}, the sum of an external field and a self-field. The gravitational self-field may nevertheless be shown not to contribute to the generalized force \cite{HarteScalar, HarteReview}. The $\Phi$ which appears there can thus be reinterpreted as a purely-external potential. Doing so while further assuming that all lengthscales associated with the external field are large compared with the size of the body, it becomes useful to Taylor expand $\mathcal{L}_\xi \Phi$ in the generalized force \eqref{GenForceNewt}. Doing so around $\gamma_t$ results in
\begin{align}
	\mathcal{F}_\xi(t)  = - M \mathcal{L}_\xi \Phi(\gamma_t,t) - D^a (t;\gamma_t) \mathcal{L}_\xi \nabla_a \Phi (\gamma_t,t)
	\nonumber
	\\
	~ + \tfrac{1}{2} \tilde{Q}^{ab}(t;\gamma_t) \mathcal{L}_\xi \mathcal{E}_{ab}(\gamma_t,t) + \ldots,
	\label{GenForceNewt2}
\end{align}
where $M$ is the body's mass,
\begin{equation}
	D^i(t;\gamma_t) \equiv \int  (x-\gamma_t)^i \rho(x,t) d^3x 
\end{equation}
is its mass dipole moment, and 
\begin{equation}
	\tilde{Q}^{ij} (t;\gamma_t) \equiv \int  (x-\gamma_t)^i (x-\gamma_t)^j \rho(x,t) d^3 x
	\label{QfullDef}
\end{equation}
is its ``full'' (not necessarily trace-free) quadrupole moment. We have also used
\begin{equation}
	\mathcal{E}_{ab}(x,t)  \equiv - \nabla_a \nabla_b \Phi(x,t)
	\label{tideDef}
\end{equation}
to denote the Newtonian tidal tensor. This tensor is always symmetric and trace-free, where the latter property follows from the vacuum field equation $\nabla^2 \Phi = 0$. The definition \eqref{tideDef} also implies that
\begin{equation}
    \nabla_{[a} \mathcal{E}_{b]c } = 0,
    \label{BianchiNewt}
\end{equation}
which may be viewed as a Newtonian analog of the Bianchi identity.  

Our focus here is on the quadrupolar contribution to the force and torque, which is given by the second line of \eqref{GenForceNewt2}. That term can, however, be simplified by noting that since $\mathcal{E}_{ab}$ is trace-free and $\mathcal{L}_\xi g_{ab} = 0$, where $g_{ab}$ denotes the Euclidean  metric, arbitrary multiples of $g^{ab}$ can be added to $\tilde{Q}^{ab}$ without affecting $\tilde{Q}^{ab} \mathcal{L}_\xi \mathcal{E}_{ab}$. The quadrupole moment in that expression may therefore be replaced by its trace-free counterpart 
\begin{equation}
	Q^{ab} \equiv (\delta^a_c \delta^b_d - \tfrac{1}{3} g^{ab} g_{cd} ) \tilde{Q}^{cd}.
	\label{QtfDef}	
\end{equation}
From now on, we refer to $Q^{ab}$ (and not $\tilde{Q}^{ab}$) as ``the'' quadrupole moment. Like $\mathcal{E}_{ab}$, this moment is symmetric and trace-free. In terms of it, the quadrupolar contribution to the generalized force is 
\begin{equation}
	\mathcal{F}_\xi^{(q)} = \frac{1}{2} Q^{ab} \mathcal{L}_\xi \mathcal{E}_{ab}.
	\label{FNewt}
\end{equation}
Quadrupolar forces and torques therefore arise only when the tidal field fails to share the same symmetries as the background Euclidean space. Combining \eqref{GenForceNewtFN} and \eqref{FNewt}, the ordinary force and torque are given by
\begin{equation}
    F_a^{(q)} = \frac{1}{2} Q^{bc} \nabla_a \mathcal{E}_{bc}, \qquad N_{ab}^{(q)} = 2 Q^{c}{}_{[a} \mathcal{E}_{b]c} 
    \label{FNNewt}
\end{equation}
at quadrupolar order.

A body which does not eject or absorb mass has no control over the monopolar generalized force $-M\mathcal{L}_\xi \Phi$. Moreover, the dipolar force $- D^a \mathcal{L}_\xi \nabla_a \Phi$ can always be set to zero by placing $\gamma_t$ at the center of mass. The first nontrivial contribution to ``non-universal'' free-fall therefore arises at quadrupolar order, which is our focus. In astrophysical contexts, it is often assumed that all quadrupole moments are induced by the tidal field. Such cases are discussed briefly in Appendix \ref{App:Tides}, where it is shown that introducing an effective potential which depends on $\mathcal{E}^{ab} \mathcal{E}_{ab}$ can allow the quadrupolar force to be absorbed into the monopole. However, our goal here is not to model any particular system: Unless otherwise noted, we allow for arbitrary quadrupole moments below. 

\subsection{Constraints from symmetry}
\label{Sect:NewtSyms}

Intuitively, extended-body effects arise from inhomogeneities in the gravitational field. Depending on a body's internal mass distribution, different parts of it may interact with slightly different gravitational fields, resulting in different net effects. Indeed, no extended-body effects are possible in a uniform gravitational field where $\nabla_a \Phi = \mathrm{constant}$. This suggests that extended-body effects should be constrained by any symmetries which may exist. 

The simplest such constraints arise from symmetries of the potential. It is immediately clear from \eqref{PpSNewt} and \eqref{GenForceNewt} that if there exists a Killing field $\Xi^a(x)$ such that\footnote{We use $\xi^a$ here to denote a generic Killing field, but $\Xi^a$ to denote a specific Killing field which  also generates a symmetry of $\Phi$.} 
\begin{equation}
    \mathcal{L}_\Xi \Phi(x,t) = 0   
    \label{LiePhi} 
\end{equation}
throughout the body of interest, one component of the generalized momentum must be conserved:
\begin{equation}
	\mathcal{P}_\Xi = p_a  \Xi^a + \frac{1}{2} S^{ab} \nabla_a \Xi_b = \mathrm{constant}.
\end{equation}
This is in fact not restricted to the quadrupole approximation. It is exact. As a consequence, 
\begin{equation}
	  \mathcal{F}_\Xi = F_a  \Xi^a + \frac{1}{2} N^{ab} \nabla_a \Xi_b = 0.
	  \label{SymConstr}
\end{equation} 
This too is exact. It implies that when $\Phi$ shares a symmetry with the background Euclidean space, \textit{certain force and torque combinations are impossible}, regardless of a body's internal structure. Such constraints hold regardless of whether or not $\gamma_t$ lies at the center-of-mass. 

One simple example concerns the motion of an object in a spherically-symmetric gravitational field. In that case, three generalized momentum components are conserved, one for each of the three rotational symmetries. Similarly, three generalized force components vanish. In more elementary language, the angular momentum 3-vector which is associated with motion around the origin is conserved. As a consequence, non-radial forces---which affect an object's orbital angular momentum---can arise only in combination with torques which produce compensating changes in the spin angular momentum. Certain linear combinations of force and torque components therefore vanish, and these are precisely the generalized force components which are associated with the rotational symmetries.

Returning to our discussion of generic gravitational fields (which are not necessarily spherically-symmetric), symmetry in the sense of \eqref{LiePhi} is a fairly strong requirement. It is therefore interesting to ask if that requirement can be weakened while still retaining interesting physical consequences. Can constraints such as \eqref{SymConstr} continue to hold even when $\mathcal{L}_\Xi \Phi \neq 0$? Indeed they can. If there is a 1-parameter family of Killing fields $\Xi^a_{t}(x)$ such that
\begin{equation}
    \mathcal{L}_{\Xi_{t}} \mathcal{E}_{ab}(\gamma_t,t) = -\nabla_a \nabla_b \mathcal{L}_{\Xi_{t}} \Phi(\gamma_t,t) = 0,
    \label{LieE}
\end{equation}
inspection of \eqref{FNewt} shows that at least the quadrupolar contribution to the generalized force must vanish:
\begin{equation}
    \mathcal{F}^{(q)}_{\Xi_{t}} = F^{(q)}_a  \Xi_{t}^a + \frac{1}{2} N^{(q)}_{ab} \nabla^a \Xi^b_{t} = 0.
    \label{localSymConstr}
\end{equation}
The quadrupolar component of the constraint \eqref{SymConstr} therefore generalizes in three ways: First, we may consider Killing fields which are symmetries of the tidal field but not of the potential. Second, we may consider Killing fields which preserve the tidal field \textit{only at} $\gamma_t$. Third, we may consider different Killing fields at different times. Although these generalizations are straightforward, they considerably weaken our notion of symmetry while still implying that certain force and torque combinations are impossible. We describe a 1-parameter family of Killing fields $\Xi^a_t$ which satisfy \eqref{LieE} as the generators of a \textit{local symmetry}. Somewhat more precisely, these are local symmetries of the tidal field. The ``ordinary'' symmetries which satisfy \eqref{LiePhi} are special cases. We refer to local symmetries which do not preserve $\Phi$ as ``proper.''

Unlike ordinary symmetries of the potential, proper local symmetries are not necessarily associated with conservation laws. A natural candidate for a potentially-conserved conserved quantity in this context is $\mathcal{P}_{\Xi_t}$. However, the rate of change of this quantity is not quite given by the generalized force $\mathcal{F}_{\Xi_t}$, since now the Killing fields may depend on time. Instead,
\begin{equation}
    \frac{d}{dt} \mathcal{P}_{\Xi_t} = \mathcal{F}_{\Xi_t} + \mathcal{P}_{\dot{\Xi}_t}.
    \label{localsymConsNewt}
\end{equation}
The first term on the right-hand simplifies due to \eqref{localSymConstr}, but does not necessarily disappear. In some cases, both terms simplify when $\gamma_t$ is placed at an object's center of mass; one such example is given in Sect. \ref{Sect:type2Ex} below.

What is interesting here is not so much that local symmetries imply force and torque constraints; that much is obvious from \eqref{FNewt}. What is more important is that \textit{proper local symmetries are ubiquitous}. We show below that at least one (not necessarily proper) local symmetry exists in every Newtonian gravitational field, and in many cases there are more. Local symmetries therefore play an important role in constraining extended-body motion. We now identify these symmetries and their consequences in different types of tidal fields.

\subsection{Constraints from algebraic structure}
\label{Sect:NewtAlgConstraints}

Any nonzero Newtonian tidal tensor can be classified, at each point, in terms of its eigenvalues. These tensors must be real, symmetric, and trace-free, and therefore admit three real eigenvalues (counting multiplicity) which sum to zero. Either,
\begin{enumerate}
	\item[1.] $\mathcal{E}_{ab}$ has three distinct and nonzero eigenvalues.
	\item[2.] $\mathcal{E}_{ab}$ has two distinct nonzero eigenvalues and one vanishing eigenvalue.
	\item[3.] $\mathcal{E}_{ab}$ has one doubly-degenerate nonzero eigenvalue and one non-degenerate nonzero eigenvalue.
\end{enumerate}
Depending on which of these descriptions hold, we describe $\mathcal{E}_{ab}$ as being of algebraic type 1, 2, or 3. Type 3 tidal tensors can be described as ``algebraically special.'' Type 1 and type 2 tidal tensors are instead ``algebraically general.'' It is shown in Appendix \ref{App:PNDs} that the algebraically-special Newtonian tidal tensors may be viewed as approximations to Petrov type D spacetimes in general relativity. Type 1 and 2 tidal fields instead correspond to Petrov type I spacetimes, which are conventionally described as algebraically general.

Regardless, the three eigenvalues $\mathcal{E}_+$, $\mathcal{E}_-$, and $-\mathcal{E}_+ - \mathcal{E}_-$ of the tidal tensor can all be encoded in the complex ``tidal scalar''
\begin{align}
	\mathcal{E} \equiv (\mathcal{E}_+ + \mathcal{E}_- ) + i (\mathcal{E}_+ - \mathcal{E}_- ),
	\label{Ecomplex}
\end{align}
which is analogous to the Weyl scalars used in general relativity\footnote{A four-dimensional Weyl tensor is associated, in general, with five complex Weyl scalars. Without aligning the triad, a Newtonian tidal tensor would be associated with two complex scalars and one real scalar. In both cases, however, certain scalars can be made to vanish by appropriately aligning the basis vectors. In the Newtonian case, doing so leaves only $\mathcal{E}$. In the relativistic case, simplifications which arise when aligning the tetrad are discussed in Sect. \ref{Sect:RelBasis} below.}. However, $\mathcal{E}$ depends on the ordering of the eigenvalues. If $\mathcal{E}_+$ and $\mathcal{E}_-$ are swapped, $\mathcal{E} \mapsto \bar{\mathcal{E}}$; if $\mathcal{E}_+$ and $-\mathcal{E}_+ - \mathcal{E}_-$ are swapped, $\mathcal{E} \mapsto -\frac{1}{2} i[ \mathcal{E} + (2-i) \bar{\mathcal{E}} ]$; if $\mathcal{E}_-$ and $-\mathcal{E}_+ - \mathcal{E}_-$ are swapped, $\mathcal{E} \mapsto \frac{1}{2} i[ \mathcal{E} + (2+i) \bar{\mathcal{E}} ]$. The eigenvalues of a type 3 tidal tensor may nevertheless be ordered such that $\mathcal{E}$ real. For type 2 tidal tensors, the eigenvalues may be ordered such that $\mathcal{E}$ is imaginary. In the generic type 1 case, $\mathcal{E}$ must have both real and imaginary components.

One order-independent way to determine the algebraic type of the tidal tensor is to compute the dimensionless ratio
\begin{align}
    \frac{ 4 (\det \mathcal{E}_{ab} ) ^2 }{  ( \mathcal{E}_{cd} \mathcal{E}^{cd} )^3 } &= \frac{ [(\mathcal{E}  + \bar{\mathcal{E}}  )( \mathcal{E}^2 + \bar{\mathcal{E}}^2 )]^2 }{ (\mathcal{E}^2 + 4 |\mathcal{E}|^2 + \bar{\mathcal{E}}^2)^3}.
\end{align}
If this vanishes, the tidal tensor is of type 2; if it is equal to $2/27$, the tidal tensor is of type 3; in all other cases, the tidal tensor is of type 1.

Forces and torques which arise in gravitational fields with each of the three algebraic types may be understood by diagonalizing $\mathcal{E}_{ab}$. If $e^a_+$ and $e^a_-$ are real orthonormal eigenvectors associated with the eigenvalues $\mathcal{E}_+$ and $\mathcal{E}_-$, it will be useful to define the complex null vector
\begin{equation}
	m^a \equiv \frac{1}{\sqrt{2}} ( e^a_+ + i e^a_- )	.
\end{equation}
Also defining $\ell_a \equiv i \epsilon_{abc} m^b  \bar{m}^c = \epsilon_{abc} e_+^b  e_-^c$, which is an eigenvector of $\mathcal{E}_{ab}$ with eigenvalue $-\mathcal{E}_+ - \mathcal{E}_-$,  the triad $(\ell^a, m^a, \bar{m}^a)$ forms a convenient basis with inner products
\begin{equation}
	m^a m_a = m^a \ell_a = 0, \qquad m^a \bar{m}_a = \ell^a \ell_a = 1.
	\label{Triad}
\end{equation}
Using it, the tidal tensor can be written as
\begin{equation}
	\mathcal{E}_{ab} = \frac{1}{2}  ( g_{ab} - 3 \ell_a\ell_b ) \Re \mathcal{E} + \Re ( m_a m_b ) \Im \mathcal{E}.
	\label{Ebasis}
\end{equation}

The triad here is adapted to the tidal tensor, not the quadrupole moment, so the latter can look more complicated when written in an analogous form: Introducing the three ``quadrupole scalars,'' $Q_{\ell m} \equiv Q_{ab} \ell^a m^b$, $Q_{m m} \equiv Q_{ab} m^a m^b$, and $Q_{\ell\ell} \equiv Q_{ab} \ell^a \ell^b$,
\begin{align}
    Q_{ab} = \frac{1}{2} Q_{\ell\ell} \big( 3\ell_a \ell_b -  g_{ab}\big) + 2 \Re \big[ \big( Q_{mm} \bar{m}_{(a}
    \nonumber
    \\
    ~  + 2 Q_{\ell m } \ell_{(a} \big) \bar{m}_{b)} \big]  .
    \label{Qbasis}
\end{align}
While $Q_{\ell\ell}$ is real, both $Q_{\ell m}$ and $Q_{mm}$ can be complex. Together, these scalars encode all five real components of $Q_{ab}$.

Eqs. \eqref{Ebasis} and \eqref{Qbasis} can now be substituted into \eqref{FNewt} in order to show that the quadrupolar generalized force is
\begin{align}
	\mathcal{F}_\xi^{(q)}  = ( \Im Q_{mm} )  ( \Im \mathcal{E} ) i \bar{m}^a \mathcal{L}_\xi m_a - \Re [ Q_{\ell m} (3 \bar{m}^a \Re \mathcal{E}
	\nonumber
	\\
	~  + m^a \Im \mathcal{E} ) ] \mathcal{L}_\xi \ell_a  - \frac{1}{4} \Re [ ( 3 Q_{\ell\ell} 
 + 2 i \Re Q_{mm}) \mathcal{L}_\xi \mathcal{E} ]  .
	\label{FNewtGen}
\end{align}
This holds for all tidal tensors and for all quadrupole moments. Using it and \eqref{GenForceNewtFN} shows that
\begin{align}
	F_a^{(q)} =  ( \Im Q_{mm} ) ( \Im \mathcal{E} ) i \bar{m}^b \nabla_a m_b - \Re [ Q_{\ell m} (3 \bar{m}^b \Re \mathcal{E}
	\nonumber
	\\
	~  + m^b \Im \mathcal{E} ) ] \nabla_a \ell_b  - \frac{1}{4} \Re [ ( 3 Q_{\ell\ell} + 2 i \Re Q_{mm}) \nabla_a \mathcal{E} ] ,
	\label{FNewtGen2}
\end{align}
and
\begin{align}
	N^{ab}_{(q)} = 2 \Re [Q_{\ell m}   \ell^{[a} ( 3 \bar{m}^{b]} \Re \mathcal{E} + m^{b]} \Im \mathcal{E} ) ] 
	\nonumber
	\\
	~ + 2 (\Im Q_{mm}) ( \Im \mathcal{E})  i \bar{m}^{[a} m^{b]}.
	\label{NNewtGen}
\end{align}
Although the quadrupole components $Q_{\ell \ell}$ and $\Re Q_{mm}$ can (at least sometimes) affect the force, these expressions show that they can never affect the torque. By contrast, both the force and the torque can depend on $Q_{\ell m}$ and on $\Im Q_{mm}$.

This shared dependence on $Q_{\ell m}$ and on $\Im Q_{mm}$ can be used to write the force partially in terms of the torque. From \eqref{NNewtGen}, first note that
\begin{subequations}
%\begin{equation}
\begin{gather}
	\Re [ Q_{\ell m } ( 3 \bar{m}^a \Re \mathcal{E} + m^a \Im \mathcal{E} )  ] = -N^{ab}_{(q)} \ell_b,
	\\
	(\Im Q_{mm})\Im \mathcal{E}  = -i N^{ab}_{(q)} m_a \bar{m}_b.
\end{gather}
%\end{equation}
\end{subequations}
Substituting these expressions into \eqref{FNewtGen} then results in
\begin{align}
	F_a^{(q)} =  ( \ell^d \nabla_a \ell^c + \bar{m}^b m^c \bar{m}^d \nabla_a m_b ) N_{cd}^{(q)} + \Re [\mathcal{Q} \nabla_a \mathcal{E} ],
	\label{FlinNNewt}
\end{align}
where 
\begin{equation}
	\mathcal{Q} \equiv - \frac{1}{4} ( 3 Q_{\ell\ell}  + 2 i \Re Q_{mm}) 
    \label{QcalDef}
\end{equation}
is a complex quadrupole component which does not affect the torque. The quadrupolar force is therefore an affine function of the quadrupolar torque. Furthermore, the space of forces which can be varied independently of the torque is spanned by the real and the imaginary components of $\nabla_a \mathcal{E}$. These forces are all that can be produced if, e.g., the torque vanishes.

One interesting implication of this is that if the torque vanishes, and if $\gamma_t$ is chosen such that $D^a = 0$, the \textit{total} force, up to quadrupolar order, may be viewed as a purely-\textit{monopolar} force in the effective potential
\begin{equation}
	\Phi_\mathrm{eff} = \Phi - \Re [( \mathcal{Q} /M)\mathcal{E}] .
        \label{PhiEffGen}
\end{equation}
Even in this restricted regime, a shape-changing spacecraft can exert considerable control over its motion simply by modulating $\mathcal{Q}$ at appropriate points in its orbit \cite{HarteNewtonian}. It may also be noted that this $\Phi_\mathrm{eff}$ is ``physically equivalent'' (but not equal) to the effective potential \eqref{phiEff} when the quadrupole moment is tidally induced.

\subsubsection{Type 3 tidal tensors}
\label{Sect:Newt3}

We have now determined the quadrupolar contributions to the generalized force \eqref{FNewtGen}, the torque \eqref{NNewtGen}, and the ordinary force \eqref{FlinNNewt}. These expressions hold for any extended body in any gravitational field, but can now be specialized to discuss extended-body motion in each of the three types of tidal field discussed above. Type 3 tidal tensors are the simplest, so we begin with them. 

Every type 3 tidal tensor admits a degenerate eigenvalue, and the orthonormal eigenvectors $e^a_\pm$ may be chosen to span the associated eigenspace. Then $\mathcal{E}_+ = \mathcal{E}_-$ and $\mathcal{E} = 2 \mathcal{E}_+$. The eigenvector $\ell^a$ is associated with the non-degenerate  eigenvalue $-2 \mathcal{E}_+$, and the tidal tensor \eqref{Ebasis} reduces to
\begin{equation}
    \mathcal{E}_{ab} = ( g_{ab} - 3 \ell_{a} \ell_{b} ) \mathcal{E}_+ .
    \label{Etype3}
\end{equation}
It follows from \eqref{FlinNNewt} that at fixed torque, the quadrupolar force in a type 3 field can be modulated only in the direction parallel to $\nabla_a \mathcal{E}_+$.  

Assuming that the tidal tensor remains type 3 in a neighborhood of the relevant point, \eqref{FNewtGen} reduces to
\begin{align}
         \mathcal{F}_\xi^{(q)} = - \frac{3}{2} Q_{\ell\ell} \mathcal{L}_\xi \mathcal{E}_+  -6 \mathcal{E}_+ \Re ( Q_{\ell m} \bar{m}^a) \mathcal{L}_\xi \ell_a  .
     \label{FNewt3}
\end{align}
The motion is therefore unaffected by $Q_{mm}$; at least two of the five (real) quadrupole components are irrelevant in type 3 tidal fields. It can also be observed that the torque \eqref{NNewtGen} reduces to
\begin{equation}
    N_{(q)}^{ab} = - 12 \mathcal{E}_+ \Re \big( Q_{\ell m} \bar{m}^{[a}  \big) \ell^{b]},
    \label{TorqueNewt3}
\end{equation}
which is controlled only by $Q_{\ell m}$. 

As $Q_{\ell m}$ encodes only two real control parameters, it is not possible for an extended body to use its quadrupole moment in order to control all three torque components: For any such moment,
\begin{equation}
    N^{ab}_{(q)} m_{[a} \bar{m}_{b]} = 0.
    \label{NconstrNewt}
\end{equation}
In terms of a \textit{vector} torque $N_c$ which satisfies $N^{ab} = \epsilon^{abc} N_c$, this is equivalent to
\begin{equation}
    N^a_{(q)} \ell_a = 0.
    \label{TorqueConstNewt}
\end{equation}
Regardless, quadrupole moments cannot affect a body's torque within the degenerate eigenplane of the tidal tensor. They do however affect the other two torque components. 

One way to understand this torque constraint, and also the fact that forces depend in part on torques, is via local symmetries. Type 3 tidal fields admit at least three local symmetries with the properties discussed in Sect. \ref{Sect:NewtSyms} above. To find them, first use \eqref{GenForceNewtFN}, \eqref{FlinNNewt}, and \eqref{NconstrNewt} to note that
\begin{align}
	\mathcal{F}^{(q)}_\xi = \frac{1}{2} \big[ ( 2\xi^a \ell^c \nabla_a \ell^b + \nabla^b \xi^c ) N_{bc}^{(q)}  - 3 Q_{\ell\ell} \mathcal{L}_\xi \mathcal{E}_+ \big].
\end{align}
Eq. \eqref{localSymConstr} implies that we would like to find a 1-parameter family of Killing fields $\Xi^a_t(x)$ such that $\mathcal{F}_{\Xi_t}(t) = 0$ for all possible quadrupole moments. Varying $Q_{\ell\ell}$ while noting that that quadrupole component cannot affect the torque, one necessary condition is clearly
\begin{equation}
	\Xi_t^a (\gamma_t) \nabla_a \mathcal{E}_+(\gamma_t) = 0.
	\label{type3Constr1}
\end{equation}
Moreover, since the torque can be varied throughout the two-dimensional space which is not excluded by \eqref{NconstrNewt}, the other necessary condition is that
\begin{equation}
	\nabla^a \Xi^b_t (\gamma_t) = 2 \Xi^c_t (\gamma_t) \ell^{[a} \nabla_c \ell^{b]} + i \lambda_t m^{[a} \bar{m}^{b]},
	\label{type3Constr2}
\end{equation}
where $\lambda_t$ is real but otherwise arbitrary. Each Killing field which satisfies these constraints preserves the tidal tensor at $\gamma_t$, and therefore generates a local symmetry. Choosing $\Xi^a_t (\gamma_t)=0$ while varying $\lambda_t$ produces a family of pure rotations about $\gamma_t$; these imply the torque constraint \eqref{NconstrNewt}. Setting $\lambda_t =0$ while varying $\Xi^a_t$ throughout the space which is consistent with \eqref{type3Constr1} generates at least two more local symmetries; these imply that there are at least two force components which cannot arise without accompanying torques. 

Our discussion thus far has assumed only that $\mathcal{E}_{ab}$ is given by \eqref{Etype3}. However, somewhat more can be said by recalling that a tidal tensor must arise as two derivatives of a scalar field. That implies the ``Bianchi identity'' \eqref{BianchiNewt}, which may be used to show that for type 3 tidal fields,
\begin{subequations}
\label{Etype3Bianchi}
\begin{align}
    \nabla_a \mathcal{E}_+ &= - \frac{3}{2} \mathcal{E}_+ (\nabla \cdot  \ell) \ell_a, 
    \\
    \nabla_a \ell_b &=  (\nabla \cdot  \ell) m_{(a} \bar{m}_{b)}. 
\end{align}
\end{subequations}
The non-degenerate eigenvector is therefore geodesic, shear-free, and twist-free, which is reminiscent of the Goldberg-Sachs theorem for repeated principal null directions in a four-dimensional spacetime. Regardless, applying these expressions to \eqref{FlinNNewt} shows that 
\begin{equation}
    F^{(q)}_a =  \frac{1}{4} (\nabla \cdot  \ell )  [ 9 Q_{\ell\ell} \mathcal{E}_+ \ell_a  + 2 N_{ab}^{(q)} \ell^b ] .
\end{equation}
The force component which is independent of the torque must therefore be parallel to the non-degenerate eigenvector $\ell^a$. Although force components which are orthogonal to $\ell^a$ can be controlled as well (at least when $\nabla \cdot \ell \neq 0$), they can arise only with accompanying torques.

\subsubsection{Type 2 tidal tensors}

By definition, one of the eigenvalues of a type 2 tidal tensor must vanish. Identifying $\ell^a$ with the unit eigenvector which corresponds to that eigenvalue, $\mathcal{E}_+ = - \mathcal{E}_-$ and $\mathcal{E} = 2 i \mathcal{E}_+$. Eq. \eqref{Ebasis} then reduces to
\begin{equation}
    \mathcal{E}_{ab} = 2 \mathcal{E}_+ \Re (m_a m_b) . 
    \label{Etype2}
\end{equation}
Inspection of \eqref{FNewtGen} shows that in a type 2 field, $Q_{\ell\ell}$ cannot affect the motion; up to four (out of five) real quadrupole components matter in these cases. It also follows from \eqref{NNewtGen} that the quadrupolar torque can be controlled arbitrarily in type 2 fields. However, the force can be varied in only one direction without also varying the torque.

These are consequences of the fact that there are at least two local symmetries in each type 2 tidal field. To identify them, note from \eqref{FlinNNewt} that
\begin{align}
	\mathcal{F}_\xi^{(q)} = \frac{1}{2} [\xi^a (2 \ell^d \nabla_a \ell^c + \bar{m}^b m^c \bar{m}^d \nabla_a m_b) + \nabla^c \xi^d ] N_{cd}^{(q)}
	\nonumber
	\\
	~ + \Re Q_{mm} \mathcal{L}_\xi \mathcal{E}_+.
\end{align}
As $\Re Q_{mm}$ cannot affect the torque, ensuring that this vanishes for all possible quadrupole moments implies that the local symmetries are given  by all Killing fields which satisfy
\begin{subequations}
	\label{localSymType2}
\begin{gather}
	\Xi^a_t (\gamma_t) \nabla_a \mathcal{E}_+ (\gamma_t) = 0,
	\\
	\nabla^a \Xi^b_t (\gamma_t) = 2 \Xi^c_t (\gamma_t) ( \ell^{[a} \nabla_c \ell^{b]} +  m^{[a} \bar{m}^{b]} m^d \nabla_c \bar{m}_d).
\end{gather}
\end{subequations}
This space is at least two dimensional.

\subsubsection{Type 1 tidal tensors}

Type 1 tidal tensors admit three distinct and nonzero eigenvalues. In these cases, the quadrupolar torque can be varied arbitrarily. It also follows from \eqref{FlinNNewt} that at fixed torque, forces can be varied through the space spanned by $\nabla_a \mathcal{E}_+$ and by $\nabla_a \mathcal{E}_-$. That space is at most two dimensional, so quadrupolar effects can never be used to fully control all force and torque components. At least one force component cannot be varied without an accompanying torque. This is a consequence of the fact that type 1 tidal fields admit at least one local symmetry. Indeed, \textit{all} Newtonian tidal tensors admit at least one local symmetry. This symmetry can be found by determining all Killing fields which satisfy
\begin{subequations}
\begin{gather}
	\Xi^a_t (\gamma_t) \nabla_a \mathcal{E}_+ (\gamma_t) = \Xi^a_t (\gamma_t) \nabla_a \mathcal{E}_- (\gamma_t) = 0,
	\\
	\nabla^a \Xi^b_t (\gamma_t) = 2 \Xi^c_t (\gamma_t) ( \ell^{[a} \nabla_c \ell^{b]} +  m^{[a} \bar{m}^{b]} m^d \nabla_c \bar{m}_d).
\end{gather}
\end{subequations}

The impossibility of completely controlling all forces and torques could have been anticipated by a counting argument: The five components of the quadrupole moment are not sufficient to independently control all six generalized force components. Nevertheless, there are cases in which all force and torque components \textit{can} be independently controlled using octupole and higher-order moments. The situation is different in general relativity, where there are ten quadrupole components and also ten generalized force components. We shall see in Sect. \ref{Sect:RelAlgConstraints} below that complete control of all relativistic forces and torques \textit{is} possible in the quadrupole approximation, at least in some spacetimes. This suggests that there may be a sense in which, at quadrupolar order, some ``essentially Newtonian'' force or torque components can be controlled only relativistically, via a body's current (rather than mass) quadrupole.

\subsubsection{Summarizing the Newtonian constraints}
\label{Sect:NewtSumm}

We may now summarize our Newtonian results by describing how much control would be available to a spacecraft which has been engineered to arbitrarily control its quadrupole moment. First, we have found that in a type $n$ tidal field, there are at least $n$ local symmetries (with $n=1,2,3$). In algebraically general tidal fields,  which are of types 1 or 2, appropriate spacecraft have complete control over the quadrupolar torques which are exerted upon them. In the algebraically-special type 3 case, torque vectors can instead be controlled only within the 2-plane which is orthogonal to the non-degenerate eigenvector of $\mathcal{E}_{ab}$. How such a spacecraft can control the quadrupolar forces which act upon it is more complicated. However, one general statement is that without changing the torque, suitable spacecraft can arbitrarily control forces only throughout the space which is spanned by the gradients of the eigenvalues $\mathcal{E}_{\pm}$. For tidal tensors of types 2 and 3, this space is at most one-dimensional. For tidal tensors of type 1, it is at most two-dimensional. These and related results are collected in Table \ref{Table:NewtDims}.

\begin{table}
\setlength{\tabcolsep}{8pt}
  \begin{tabular}{c | c |  c | c | c}
%  \hline
   Alg. type & $\big\{\Xi^a_t \big\}$ & $\big\{ Q_{ab} \big\}$  &
   $\big\{ N_{ab}^{(q)} \big\}$  &  $\big\{ F_a^{(q)} \big| N_{bc}^{(q)} \big\}$  \\[.7ex]
    \hline\hline
    1	& 1--3 & 3--$5^*$ & 3*  & 0--2  \\	 
    2	& 2--3 & 3--4 	  & 3* & 0--1 \\
    3  &  3--4 & 2--3 	  & 2     &   0--1  \\[.2ex]
 %   \hline
	\end{tabular}
	\caption{Forces, torques, and local symmetries in Newtonian tidal fields with different algebraic types. The second column displays the number of local symmetries $\Xi^a_t$. The third column specifies the number of real quadrupole components which can affect the force or torque. The fourth column specifies the dimension of the space of possible quadrupolar torques. The rightmost column displays the dimension of the space of quadrupolar forces which can be varied at fixed torque. Starred numbers are used to indicate that there is no constraint. All ranges which appear here depend on the dimension of the space which is spanned $\nabla_a \mathcal{E}_+$ and $\nabla_a \mathcal{E}_-$.
	} 
	\label{Table:NewtDims}
\end{table}

\subsection{Examples of Newtonian tidal fields}
\label{Sect:NewtExamples}

In order to illustrate our results, we now discuss some simple examples of Newtonian tidal fields, including their local symmetries.

\subsubsection{Type 3 examples}

The prototypical example of a type 3 tidal tensor is generated by the spherically-symmetric potential $\Phi = - k/r$, where $k$ is a constant and $r$ is a radial coordinate. Then,
\begin{equation}
    \mathcal{E}_{ab} = \frac{ k }{ r^3 } ( 3 \hat{r}_a \hat{r}_b - g_{ab}),
\end{equation}
where $\hat{r}_a \equiv \nabla_a r$ denotes the radial unit vector and $g_{ab}$ is again the Euclidean metric. The vector $\hat{r}_a$ is a non-degenerate eigenvector of $\mathcal{E}_{ab}$, so we may identify it with $\ell_a$. Doing so, the tidal scalar \eqref{Ecomplex} reduces to $\mathcal{E} = - 2k/r^3$. At fixed torque, only the radial force can thus be controlled using extended-body effects. Additionally, the torque constraint \eqref{TorqueConstNewt} implies that there can be no quadrupolar torque along the radial direction. These constraints are intuitively clear given the conservation of angular momentum. What is perhaps less clear is that even with these conservation laws, an object which controls its quadrupole moment can still exert considerable control over its orbit \cite{HarteNewtonian}. In this case, the three local symmetries determined by \eqref{type3Constr1} and \eqref{type3Constr2} are in fact the ordinary rotational symmetries of $\Phi$. They are not ``proper'' local symmetries. 

Another type 3 example is provided by $\Phi = k (r^2 - 3z^2)$, where $z$ is a Cartesian coordinate and $k$ is again a constant. This describes a constant tidal field, and is essentially the $r \to \infty$ limit of the spherically-symmetric example above. Although quadrupolar forces vanish in this case, torque vectors can be arbitrarily controlled in all directions which are orthogonal to the $z$ axis. Also, since $\nabla_a \mathcal{E} =0$, there are four local symmetries rather than three---three translations and a rotation around the $z$ axis. Only the rotation is however a symmetry of $\Phi$. Each translation $\partial_i$ is a proper local symmetry, and the corresponding generalized momentum varies according to 
\begin{equation}
	\frac{d}{d t} \mathcal{P}_{\partial_i} = \frac{d p_i}{dt}  = - M \partial_i \Phi - D^j \partial_i \partial_j \Phi,
\end{equation}
at least through quadrupolar order. The linear momentum $p_a$ is therefore unaffected by a body's quadrupole moment; its behavior is (at least instantaneously) ``universal.''

\subsubsection{Type 2 examples}
\label{Sect:type2Ex}

The simplest example of a non-constant type 2 tidal tensor is generated by the cylindrically-symmetric potential $\Phi = k \ln r$, where $k$ is another constant and $r$ is now the distance away from the axis of symmetry. Let $\ell^a$ be a unit eigenvector of $\mathcal{E}_{ab}$ with eigenvalue 0, which must be parallel to the symmetry axis. Also defining $\hat{r}_a \equiv \nabla_a r$,
\begin{equation}
    \mathcal{E}_{ab} = \frac{ k  }{ r^2 } ( 2 \hat{r}_a \hat{r}_b + \ell_a \ell_b - \delta_{ab}  )
\end{equation}
and $\mathcal{E} = 4i k/r^2$. The two local symmetries which can be found by solving \eqref{localSymType2} are in fact the ordinary translational and azimuthal symmetries of $\Phi$. The translational symmetry along $\ell^a$ clearly precludes any force in that direction. The azimuthal symmetry instead requires that any azimuthal force be accompanied by a torque along the symmetry axis; one cannot exist without the other. Moreover, because $\nabla_a \mathcal{E}$ is radial, only the radial force can be controlled independently of the torque. 

More interesting type 2 examples can be found by superposing the potentials from multiple long, parallel cylinders. Although doing so breaks the azimuthal symmetry of $\Phi$, a proper local symmetry takes its place. The geometric significance of this symmetry is, however, difficult to visualize.

We therefore consider a simpler example instead: Letting $y$ and $z$ denote Cartesian coordinates, and letting $k$ and $a$ be constants, suppose that
\begin{equation}
	\Phi = k e^{z/ a} \sin (y/a) .
        \label{PhiWave}
\end{equation}
The corresponding tidal tensor is then
\begin{align}
    \mathcal{E}_{ab} = \frac{k}{a^2 } e^{z/a} [ (\nabla_a y \nabla_b y - \nabla_a z \nabla_b z ) \sin (y/a) 
    \nonumber
    \\
    ~ - 2 \nabla_{(a}y \nabla_{b)}z \cos (y/a) ],
    \label{EStandWave}
\end{align}
which clearly admits $\ell^a = \epsilon^{abc} \nabla_b y \nabla_c z$ as an eigenvector with vanishing eigenvalue. That eigenvector generates an ordinary translational symmetry of $\Phi$, which implies that it is not possible to produce a quadrupolar force orthogonal to the $yz$ plane. The other two eigenvectors of $\mathcal{E}_{ab}$ may be arranged such that $\mathcal{E} = 2 i k e^{z/a} /a^2$, which implies that the force can be controlled independently of the torque only in the $z$ direction. Quadrupolar forces in the $y$ direction can be controlled as well, but only at the cost of accompanying torques. 

This last statement is a consequence of the proper local symmetry 
\begin{align}
    \Xi^i_{t}(x)  = \epsilon^{ijk} (x - \gamma_t )_j \ell_k  + 2 a \nabla^i y ,
    \label{localsymEx}
\end{align}
which may be found by applying \eqref{localSymType2} to \eqref{EStandWave}. It may be verified that although $\mathcal{L}_{\Xi_{t} } \mathcal{E}_{ab}(x)$ vanishes when $x=\gamma_t$, it does not vanish more generally. Geometrically, $\Xi^a_t$ corresponds to a rotation in the $yz$ plane, with origin $\gamma_t$, together with a translation in the $y$ direction. Alternatively, it may be interpreted as a pure rotation around a point which is displaced from $\gamma_t$ by a distance $2a$ in the $z$ direction. Regardless, the combination of translational and rotational components here is what links forces to torques. 

While proper local symmetries are not necessarily associated with conservation laws, we can again see how close they can get. Using the local symmetry \eqref{localsymEx} together with \eqref{GenForceNewt2}, \eqref{localSymConstr}, and \eqref{localsymConsNewt},
\begin{equation}
	\frac{d}{dt} \mathcal{P}_{\Xi_t}(t) = -M \mathcal{L}_{\Xi_t} \Phi - D^a \mathcal{L}_{\Xi_t} \nabla_a \Phi - \epsilon_{abc} p^a \dot{\gamma}_t^b \ell^c
\end{equation}
through quadrupolar order. If $\gamma_t$ is chosen to lie at the body's center of mass, the last two terms here vanish, leaving only the monopolar contribution $-M \mathcal{L}_{\Xi_t} \Phi$. While $\mathcal{P}_{\Xi_t}$ can change, it does so only in the same way as for a monopolar particle. In this sense, its behavior is universal.

\section{Extended-body effects in general relativity}
\label{Sect:GR}

We now move on to discussing extended-body constraints in general relativity. Following essentially the same steps as in the Newtonian theory, we begin in Sect. \ref{Sect:MotionReview} by reviewing the generalized momentum and the generalized force in a relativistic context. Sect. \ref{Sect:RelSyms} applies these concepts to determine how symmetries constrain extended-body motion. It focuses on local symmetries in general relativity and shows that some of these are generated by conformal Killing-Yano tensors. Sect. \ref{Sect:RelAlgConstraints} then analyzes quadrupolar forces and torques in vacuum spacetimes, deriving local symmetries and their physical consequences for each of the possible Petrov types. Lastly, Sect. \ref{Sect:RelExamples} uses pp-wave and Kasner spacetimes as examples with which to illustrate our results.

\subsection{Generalized momentum and generalized force}
\label{Sect:MotionReview}

As in Newtonian theory, the bulk state of an extended body in general relativity can be described in terms of a generalized momentum $\mathcal{P}_\xi(s)$ \cite{HarteSyms, HarteScalar, HarteGrav, HarteReview}. At least for a test body with stress-energy tensor $T^{ab}$, it is useful to define this as
\begin{equation}
	\mathcal{P}_\xi(s) \equiv \int_{\Sigma_s} T^{a}{}_{b}(x) \xi^b(x) dS_a
	\label{PDefRel}
\end{equation}
at ``time'' $s$, where the hypersurfaces $\Sigma_s$ are chosen to foliate the body's worldtube. Since there may not be any  Killing fields here, the vector fields $\xi^a$ must be chosen more broadly than in the Newtonian setting: They are  ``generalized Killing fields,'' or ``GKFs.'' A complete definition for the GKFs may be found in \cite{HarteSyms, HarteReview}, but for our purposes, it suffices to note that they require for their specification the aforementioned hypersurfaces $\Sigma_s$, as well as a reference worldline which we parameterize by $\gamma_s$. Both of these structures can, e.g., be fixed using center-of-mass conditions \cite{EhlersRudolph, Dix79, CM1, CM2}. In that case, $\gamma_s$ would be regarded as a point on the body's center-of-mass worldline. Regardless, any GKF is uniquely determined by its value and that of its first derivative anywhere on the reference worldline---both of which can be chosen arbitrarily so long as $\nabla_{(a} \xi_{b)} (\gamma_s) = 0$. The space of possible choices for $\xi^a(\gamma_s)$ and $\nabla_{[a} \xi_{b]}(\gamma_s)$ is ten dimensional, so the space of GKFs is also ten dimensional. Noting that $\mathcal{P}_\xi$ is linear in the GKFs, the generalized momentum may be viewed as an $s$-dependent vector in the ten-dimensional space which is dual to the space of generalized Killing fields.

These ten dimensions encode the four components of a body's linear momentum $p_a$ and the six components of its angular momentum $S^{ab} = S^{[ab]}$, both of which are tensors on the reference worldline. As in the Newtonian setting, the linear and the angular momenta can be defined implicitly by using \eqref{PpSNewt} to relate them to $\mathcal{P}_\xi$. For extended test bodies, doing so results in the same momenta as those found by Dixon \cite{Dixon70a, Dixon74, Dix79, DixonReview}. When self-fields are significant, those fields finitely renormalize the generalized momentum \eqref{PDefRel}, and therefore $p_a$ and $S^{ab}$ as well \cite{HarteGrav, HarteReview}.

Whether or not self-interaction is significant, we may again introduce the generalized force $\mathcal{F}_\xi \equiv d \mathcal{P}_\xi / ds$ in order to describe changes in a body's generalized momentum. Both an ordinary force $F_a$ and a torque $N^{ab} = N^{[ab]}$ can be extracted from $\mathcal{F}_\xi$ using \eqref{GenForceNewt}. However, the force and the torque are not simply the rates of change of $p_a$ and $S^{ab}$. To see this, first note that all GKFs satisfy Killing's equation at least through first order on the reference worldline \cite{HarteSyms, HarteReview}:
\begin{equation}
	\mathcal{L}_\xi g_{ab} (\gamma_s) = \nabla_c \mathcal{L}_\xi g_{ab} (\gamma_s) = 0.
	\label{Killing}
\end{equation}
Applying this and \eqref{GenForceNewt} while differentiating \eqref{PpSNewt}
results in the Mathisson-Papapetrou-Dixon equations
\begin{subequations}
\label{MP}
	\begin{align}
	\frac{D }{ds} p_a &= - \frac{1}{2} R_{abcd} \dot{\gamma}_s^b S^{cd} + F_a,
	\\
	\frac{ D }{ds} S^{ab} &= 2 p^{[a} \dot{\gamma}_s^{b]} + N^{ab}.
\end{align}
\end{subequations}
The force and the torque, or equivalently $\mathcal{F}_\xi$, encode only \textit{dynamical} contributions to the evolution. The terms $- \frac{1}{2} R_{abcd} \dot{\gamma}_s^b S^{cd}$ and $2 p^{[a} \dot{\gamma}_s^{b]}$ which appear in \eqref{MP} are instead kinematical. They are related to the fact that the approximate Poincar\'{e} symmetry which is encoded in \eqref{Killing} mixes translations, rotations, and boosts from one moment in time to the next. This  mixing of approximate symmetries physically manifests as a mixing of linear and angular momenta over time. However, our focus here is only on understanding dynamical contributions which can differ from one extended body to another, all of which are encoded in $F_a$ and in $N^{ab}$.

Assuming that $\nabla_a T^{ab} = 0$, differentiation of \eqref{PDefRel} shows that at least for test bodies, 
\begin{equation}
	\mathcal{F}_\xi = \frac{1}{2} \int_{\Sigma_s} T^{ab} \mathcal{L}_\xi g_{ab} w^c dS_c,
	\label{FGenRel}
\end{equation}
where $w^c$ is a time evolution vector field for the foliation $\Sigma_s$. The generalized force component which is associated with a particular GKF $\xi^a$ therefore measures the degree by which that GKF fails to be a genuine Killing field. However, this ``measurement'' is weighted by $T^{ab}$, which can vary from one extended body to another. Different weightings allows different objects to experience different forces and different torques in the same spacetime. 

It is inconvenient to analyze these differences using an integral expression for $\mathcal{F}_\xi$. We instead assume that all bodies we consider are sufficiently small that multipolar expansions can be employed. Since $\mathcal{L}_\xi g_{ab}$ vanishes through first order around $\gamma_s$, the first nontrivial contribution in such an expansion arises at second---i.e., quadrupolar---order. A calculation shows that in fact \cite{Dixon70a, Dixon74, HarteTrenorm, HarteReview}
\begin{equation}
	\mathcal{F}_\xi^{(q)} = - \frac{1}{6} \tilde{J}^{abcd} \mathcal{L}_\xi R_{abcd},
	\label{QuadrupoleForce}
\end{equation} 
where $\tilde{J}^{abcd}$ denotes the body's full (not necessarily trace-free) quadrupole moment. This moment has all of the same algebraic properties as a Riemann tensor. For a test body, it is the quadrupole moment derived by Dixon; see Eq. (9.12) of \cite{Dixon74} or Eq. (2.8) of \cite{HarteExtendedtypeD}. If self-interaction is significant, the relevant $\tilde{J}^{abcd}$ is finitely renormalized with respect to Dixon's definition \cite{HarteGrav, HarteReview}. Also, the $R_{abcd}$ which appears in the generalized force must then be understood as the Riemann tensor which is associated with a certain effective metric. Although the details are not important here, we assume that the metric which appears in all of our equations is this effective one. It reduces to the physical metric when considering test bodies, but more generally includes both ``external'' and ``self-field'' contributions. In a Newtonian limit, it incorporates only the external gravitational field.

Regardless, all of our discussion is confined to spacetimes which satisfy the vacuum Einstein equation, perhaps with a cosmological constant\footnote{The cosmological constant has no direct influence on extended-body effects, so there is no downside to including it.} $\Lambda$. The Ricci tensor is therefore
\begin{equation}
    R_{ab} = \Lambda g_{ab},
    \label{vacuum}
\end{equation}
and the Weyl tensor is related to the Riemann tensor by
\begin{align}
	C_{abcd} = R_{abcd} + \frac{1}{6} \Lambda g_{a[d} g_{c]b}.
	\label{WeylfromRiemann}
\end{align}
Substituting this into \eqref{QuadrupoleForce} while using \eqref{Killing} shows that $\mathcal{L}_\xi R_{abcd}(\gamma_t) = \mathcal{L}_\xi C_{abcd}(\gamma_t)$ for every GKF $\xi^a$. Similarly, $\tilde{J}^{abcd}$ may be replaced in \eqref{QuadrupoleForce} by its trace-free counterpart $J^{abcd} \equiv (\tilde{J}^{abcd})_{\mathrm{TF}}$ without affecting $\mathcal{F}_\xi^{(q)}$ \cite{HarteTrenorm, Bini2014, HarteExtendedtypeD}. The quadrupolar generalized force in any vacuum spacetime may therefore be written as
\begin{equation}
	\mathcal{F}^{(q)}_\xi= - \frac{1}{6} J^{abcd} \mathcal{L}_\xi C_{abcd} .
	\label{QuadrupoleForce2}
\end{equation} 
Using \eqref{GenForceNewt}, the corresponding force and torque are
\begin{subequations}
	\begin{align}
		F^{(q)}_a &= - \frac{1}{6} J^{bcde} \nabla_a C_{bcde} ,
		\\
		N_{ab}^{(q)} &= - \frac{4}{3} J_{[a}{}^{cde} C_{b]cde} . 
	\end{align}
\end{subequations}
Much of our discussion below is focused on these expressions. 

In astrophysical contexts, it is often assumed that $J^{abcd}$ vanishes for isolated and non-spinning bodies, but that nonzero quadrupole moments can be induced either by external tidal fields or by rotation. The former case is briefly discussed in Appendix \ref{App:Tides}, where the simplest models are shown to result in very simple forces and torques. Although the mass 
\begin{equation}
    M \equiv \sqrt{- p_a p^a}
    \label{mass}
\end{equation}
can vary in these (and other) models, a certain \textit{effective} mass is conserved at least for some bodies with tidally-induced quadrupole moments; see \eqref{Meff}. That is in turn analogous to the existence of the Newtonian effective potential \eqref{phiEff}. Nevertheless, our main goal here is to understand model-independent features of the extended-body problem. We therefore make no assumptions below regarding the specific form of the quadrupole moment.  

One model-independent feature which can already be deduced is that in a vacuum spacetime, many objects with differing internal structures can experience identical forces and torques. To see this, first note that $J^{abcd}$ has all of the same algebraic properties as a Weyl tensor, and  therefore has ten independent components. This contrasts with the twenty independent components of $\tilde{J}^{abcd}$. Einstein's equation thus implies that at least ten components of the full quadrupole moment cannot affect an object's motion\footnote{As noted in Sect. \ref{Sect:NewtReview} above, the Newtonian analog of this statement is that the vacuum field equation implies that one of the six components of $\tilde{Q}^{ab}$ cannot affect the motion.}. Depending on the Petrov type of the spacetime, we shall see below that even more quadrupole components can fail to affect the motion. As our focus is only on vacuum spacetimes, we now refer only to $J^{abcd}$ (and not to $\tilde{J}^{abcd}$) as ``the'' quadrupole moment in relativistic contexts.

\subsection{Constraints from symmetry}
\label{Sect:RelSyms}

If there exists a Killing field $\Xi^a$, it is clear from \eqref{FGenRel} that regardless of an object's internal structure, $\mathcal{P}_\Xi$ is conserved and $\mathcal{F}_\Xi = 0$. Killing fields therefore place universal constraints on extended-body effects. In fact, these constraints hold not only for the full generalized force, but also for each term in its multipole expansion. This much has been known since at least the 1970s \cite{Dixon70a, Dixon74,EhlersRudolph, Dix79}. What is new here is the concept of a local symmetry.

\subsubsection{Local symmetries}

A Killing field in a curved spacetime is analogous, in Newtonian gravity, to a vector field which preserves not only the Euclidean metric, but also the potential $\Phi$. However, we found in Sect. \ref{Sect:NewtSyms} above that it was useful to generalize this by considering symmetries of the tidal tensor $\mathcal{E}_{ab}$ which are not necessarily symmetries of $\Phi$. We also found it useful to allow the tidal tensor to perhaps be preserved at only a single point.

Applying these ideas in a relativistic context, we now define the generator of a local symmetry to be a 1-parameter family of generalized Killing fields $\Xi^a_s$ which locally preserve the curvature:
\begin{equation}
	\mathcal{L}_{\Xi_s} R_{abcd} (\gamma_s ) = 0.
	\label{localsymRel}
\end{equation}
If such a family exists, it follows from \eqref{QuadrupoleForce} that
\begin{equation}
	\mathcal{F}_{\Xi_s}^{(q)} = F_a^{(q)} \Xi^a_s + \frac{1}{2} N_{ab}^{(q)} \nabla^a \Xi^b_s = 0
	\label{forceConstrRel}
\end{equation}
at time $s$. Conversely, a family of GKFs generates a local symmetry whenever $\mathcal{F}_{\Xi_s}^{(q)}=0$ for all possible quadrupole moments. 

As noted in Sect. \ref{Sect:MotionReview} above, GKFs always satisfy Killing's equation through first order along the reference worldline. Eq. \eqref{localsymRel} asks if it is possible to extend this through one higher order, in which case \eqref{Killing} is supplemented by
\begin{equation}
	\nabla_a \nabla_b \mathcal{L}_{\Xi_s} g_{cd} (\gamma_s) = 0.
	\label{localsymRel2}
\end{equation} 
This is not possible in general. However, cases where it \textit{is} possible are not uncommon and are physically interesting.

It is clear that every Killing field which may exist generates a local symmetry, and each of these is associated with a conservation law. A ``proper'' local symmetry, which is a local symmetry which is \textit{not} generated by a Killing field, might fail to be associated with any conservation law. 

\subsubsection{Conformal Killing-Yano tensors as local symmetries}
\label{App:CKY}

We now show that some proper local symmetries are generated by conformal Killing-Yano (CKY) tensors. By definition, a CKY tensor $f_{ab} = f_{[ab]}(x)$ must satisfy
\begin{equation}
	\nabla_{(a} f_{b)c} = g_{ab} f_{c} - f_{(a} g_{b)c} , 	
	\label{CKY}
\end{equation}
where $f_a \equiv \frac{1}{3} \nabla^b f_{ba}$. A CKY tensor for which $f_a =0$ is called a Killing-Yano tensor. Regardless, the vacuum Einstein equation implies that $\nabla_{(a} f_{b)} = 0$ \cite{jezierski2006}. Using this as well as \eqref{CKY} and the Ricci identity, every CKY tensor may be shown to satisfy
\begin{align}
	\nabla_c \nabla_d f_{ab} = - \nabla_a \nabla_b f_{cd} + 2 R_{ca[d}{}^e f_{b]e} + \nabla_a ( 2 g_{bc} f_d 
	\nonumber
	\\
	~ - g_{cd} f_b) + \nabla_c ( 2 g_{ad} f_b - g_{ab} f_d) . 
\end{align}
Antisymmetrizing over the index pairs $ab$ and $cd$, it then follows that
\begin{equation}
    R_{ab[c}{}^{e} f_{d]e}  = - R_{cd[a}{}^{e} f_{b]e}.
    \label{RIdentity}
\end{equation}
This can be viewed as an integrability condition for the existence of a CKY tensor. Introducing the 1-parameter family of GKFs which are determined by
\begin{equation}
	\Xi^a_s (\gamma_s) = 0, \qquad \nabla^a \Xi^b_s(\gamma_s) = f^{ab}(\gamma_s),
	\label{CKYGKF}
\end{equation}
it implies that $\mathcal{L}_{\Xi_s} R_{abcd}(\gamma_s) = 0$. Every CKY tensor therefore generates a local symmetry. Since $\Xi^a_s (\gamma_s) = 0$, these symmetries have no translational components. They may be viewed as generating a family of curvature-preserving Lorentz transformations on the reference worldline.

It follows from \eqref{forceConstrRel} and \eqref{CKYGKF} that the physical consequence of such a symmetry is that one component of the quadrupolar torque must vanish. In fact, since the Hodge dual $f^*_{ab} \equiv \frac{1}{2} \epsilon_{ab}{}^{cd} f_{cd}$ of any CKY tensor is also a CKY tensor \cite{jezierski2006}, \textit{two} real torque components must vanish
\begin{equation}
	N^{ab}_{(q)} f_{ab} = N^{ab}_{(q)} f_{ab}^* = 0.
	\label{CKYconstr}
\end{equation}
These constraints have previously been derived in Petrov type D spacetimes \cite{HarteExtendedtypeD}, although it was not clear  then whether or not the connection with CKY tensors was anything more than coincidence. Our derivation here shows that it was not a coincidence, that these constraints are not restricted only to type D spacetimes, and also that they can be related to curvature-preserving vector fields.

CKY tensors have seen a number of other applications in the literature, perhaps most prominently in the Kerr spacetime. There, the square of a Killing-Yano tensor can be used to construct a quadratic conserved quantity for geodesics: the Carter constant. Nontrivial Killing-Yano tensors nevertheless exist in spacetimes which are not Kerr, and Carter-like constants may be constructed in those cases as well. Killing-Yano tensors can also be used in order to construct conserved quantities---or at least conserved currents---for various field equations \cite{Carter1977, Carter1979, Kastor2004, LarsSpinReview, Grant2020EM, Grant2020, Jezierski2002, jezierski2006}. However, there are reasons to expect that exact generalizations of the Carter constant do not exist for generic compact objects \cite{Grant2015}.

In the context of this paper, it is interesting to ask if the torque constraint \eqref{CKYconstr} can nevertheless be used to construct a quantity which is at least approximately conserved. In the pole-dipole approximation where quadrupole and higher-order moments are neglected, the existence of conserved quantities which are either linear or quadratic in the momenta have been investigated \cite{Ruediger1, Ruediger2, CompereKT1} at least in combination with the Tulczyjew spin supplementary (or center-of-mass) condition
\begin{equation}
	p_a S^{ab} = 0.
	\label{SSC}
\end{equation}
In that context, an approximate conservation law was found which directly generalized the Carter constant. Additionally, the quantity
\begin{equation}
	\mathcal{P}_{\Xi_s} = \frac{1}{2} S^{ab} f_{ab}
	\label{PConsGen}
\end{equation}
was found to be conserved when $f_{ab}^*$ is Killing-Yano and when it satisfies certain other conditions as well. These extra conditions are not satisfied in Kerr \cite{Ruediger1, Santos2020}, although their failure there merely reduces $\mathcal{P}_{\Xi_s}$ to a quantity which is \textit{approximately} conserved within the pole-dipole approximation: In powers of the spin, $d\mathcal{P}_{\Xi_s}/ds = \mathcal{O}(S^2)$. If quadrupole moments are included, but are assumed to be spin-induced and with the deformability which is expected for black holes, this quantity is instead conserved up to terms of order $S^3$ \cite{CompereKT2}. However, $\mathcal{P}_{\Xi_s}$ is not so well preserved for objects with any other deformabilities.

Our comment on this is that the situation may differ with different spin supplementary conditions. Relaxing the Tulczyjew condition \eqref{SSC} while also allowing for an arbitrary angular momentum and an arbitrary (not necessarily spin-induced) quadrupole moment, the torque constraint \eqref{CKYconstr} implies that for an arbitrary CKY tensor $f_{ab}$,
\begin{align}
	\frac{d}{ds} \mathcal{P}_{\Xi_s} &= \mathcal{P}_{\dot{\Xi}_s}
	= \dot{\gamma}_s^b (p^a f_{ab} + \tfrac{1}{2} S^{ac} \nabla_b f_{ac} ),
\end{align} 
at least if the octupole and higher-order moments are neglected. It follows that $\mathcal{P}_{\Xi_s}$ is conserved whenever $\dot{\gamma}_s^b$ is orthogonal to $p^a f_{ab} + \tfrac{1}{2} S^{ac} \nabla_b f_{ac}$. It may be possible to construct spin supplementary conditions in which this is guaranteed to occur. If so, we would have a new conserved quantity for objects with \textit{arbitrary} quadrupole moments. 

Another point to note is that although $\mathcal{P}_{\Xi_s}$ is the most obvious guess for a conserved quantity associated with the local symmetry $\Xi_s^a$, it is not necessarily optimal. In the Schwarzschild spacetime, one of the torque constraints which is associated with a conformal Killing-Yano tensor can in fact be derived from ordinary Killing symmetries \cite{HarteExtendedtypeD}. That implies that there \textit{is} an associated conservation law. However, the quantity which is conserved in that case does not coincide with $\mathcal{P}_{\Xi_s}$, except at one moment in time. We find a similar result when discussing motion in plane wave spacetimes in Sect. \ref{Sect:ppWave} below.

\subsection{Constraints from algebraic structure}
\label{Sect:RelAlgConstraints}

The majority of local symmetries we consider are not related either to Killing vector fields or to conformal Killing-Yano tensors. We now  use the algebraic structure of the Weyl tensor to identify these symmetries and to determine their physical consequences. We consider general vacuum spacetimes and determine how extended-body motion depends on the Petrov type of the spacetime in which it moves. No spin supplementary condition is assumed, and there might not be any Killing vectors or CKY tensors. 
 
 \subsubsection{A convenient basis}
 \label{Sect:RelBasis}

Simple expressions for forces and torques require decompositions which are adapted to the spacetime geometry, and not, e.g., to an object's rest frame. Mathematically, this corresponds to expressing forces and torques in terms of a basis which is adapted to the principal null directions (PNDs) of the spacetime's Weyl tensor. As reviewed in Appendix \ref{App:PNDs}, Weyl tensors can be classified according to their Petrov type, which is determined by the numbers of PNDs with different multiplicities; see the second column in Table \ref{Table:Petrov}. Although it is possible for the Petrov type to vary from point to point, we assume below that it does not.

In order to introduce an appropriate basis, first let $(\ell^a, n^a, m^a, \bar{m}^a)$ be a null tetrad where $\ell^a$ and $n^a$ are real, $\bar{m}^a$ is the complex conjugate of $m^a$, and the only non-vanishing scalar products are 
\begin{equation}
	m^a \bar{m}_a = - \ell^a n_a = 1.
	\label{tetradProd}
\end{equation}
Many such tetrads exist. However, the space of possibilities can be reduced by aligning $\ell^a$ and $n^a$ with PNDs of the Weyl tensor. More precisely, we choose $\ell^a$ and $n^a$ to be parallel to ``the'' PNDs with the largest and the second-largest multiplicities, respectively\footnote{Even if rescalings of $\ell^a$ and $n^a$ are ignored, this prescription is not necessarily unique. In the type N case, $\ell^a$ is aligned with the sole PND while $n^a$ is essentially free. In type I spacetimes where there are four PNDs with equal multiplicity, twelve alignments are possible. Two possible alignments are possible in type D and in type II spacetimes.}; see the third column of Table \ref{Table:Petrov}.

\begin{table}
\setlength{\tabcolsep}{8pt}
  \begin{tabular}{c | c | c | c }
   Petrov type & PNDs & Multiplicities & Vanishing $\Psi_I$\\[.4ex] 
    \hline\hline
    I 	& 1, 1, 1, 1 & 1, 1 & $\Psi_0, \Psi_4$	\\	
    II	& 2, 1, 1 & 2, 1 & $\Psi_0, \Psi_1, \Psi_4$ \\
    D	& 2, 2 & 2, 2 & $\Psi_0, \Psi_1, \Psi_3, \Psi_4$	\\
    III	& 3, 1 & 3, 1 & $\Psi_0, \Psi_1, \Psi_2, \Psi_4$  \\	
    N	& 4 & 4, - & $\Psi_0, \Psi_1, \Psi_2, \Psi_3$
	\end{tabular}
	\caption{The Petrov classification and the Weyl scalars which vanish with an appropriately-aligned tetrad. The second column summarizes the multiplicities of all principal null directions which are associated with a given Weyl tensor. The third column gives the respective multiplicities of the PNDs which are tangent to $\ell^a$ and $n^a$ when these vectors are chosen as described in the text.  The fourth column lists all Weyl scalars which necessarily vanish with this alignment.} 
		\label{Table:Petrov}
\end{table}

Given an arbitrary null tetrad, the Weyl tensor can be described in terms of the five Weyl scalars $\Psi_0, \ldots, \Psi_4$ which are defined by \eqref{PsiDef} in Appendix \ref{App:PNDs}. However, some of these scalars vanish when $\ell^a$ and $n^a$ are aligned as described in the previous paragraph. In type D spacetimes, only $\Psi_2$ can be nonzero; in type III spacetimes, it is only $\Psi_3$ which fails to vanish; in type N spacetimes, it is only $\Psi_4$; in type II spacetimes, both $\Psi_2$ and $\Psi_3$ are nonzero; in type I spacetimes, $\Psi_1$, $\Psi_2$, and $\Psi_3$ can all fail to vanish. These simplifications justify our choices for $\ell^a$ and $n^a$, and  are summarized in the final column of Table \ref{Table:Petrov}.

Further simplifications can sometimes be performed by employing the type III tetrad transformations which are described in Appendix \ref{App:PNDs}. These rescale $\ell^a$ and $n^a$ while rotating $m^a$ and $\bar{m}^a$. Referring to \eqref{bivectorIII} and \eqref{PsiDef}, a type III transformation which is generated by the complex scalar $c$ transforms the Weyl scalars via
\begin{equation}
		\Psi_I \mapsto c^{2-I} \Psi_I, \qquad I=0,\ldots,4.
		\label{PsiXform}
\end{equation}
In the Petrov type D case, where aligning $\ell^a$ and $n^a$ with the two PNDs leaves only $\Psi_2$ nonzero, type III tetrad transformations have no effect. In the Petrov type III case, a type III tetrad transformation can be used to set $\Psi_3$ equal to any nonzero constant. Similarly, $\Psi_4$ can be made constant in any Petrov type N spacetime. In the Petrov type I case, a type III tetrad transformation can be used to ensure that $\Psi_1 = \Psi_3$. Although the simplifications afforded by type III transformations are occasionally useful, we employ them below only in type I spacetimes. 

However a particular tetrad has been fixed, it is convenient to define from it the complex bivectors 
\begin{equation}
\begin{gathered}
	X^{ab} \equiv 2 \ell^{[a} m^{b]} , \qquad Y^{ab} \equiv 2n^{[a} \bar{m}^{b]} ,
	\\
	Z^{ab} \equiv 2 ( \ell^{[a} n^{b]} - m^{[a} \bar{m}^{b]} ) .
\end{gathered}
\label{2formBasis}
\end{equation}
These and and their complex conjugates form a basis for all bivectors. The basis elements $X^{ab}$, $Y^{ab}$, and $Z^{ab}$ are self-dual, meaning that, e.g., $X^{*ab} = i X^{ab}$. Their complex conjugates are  instead anti self-dual. It can also be shown, using \eqref{tetradProd}, that the non-vanishing antisymmetrized products between members of this basis are
\begin{equation}
	X^{[a}{}_{c} Y^{b]c} = \tfrac{1}{2} Z^{ab} , \quad 
	X^{[a}{}_{c} Z^{b]c} = -X^{ab}, \quad Y^{[a}{}_{c} Z^{b]c} = Y^{ab}.
\label{bivectorProd1}
\end{equation}
If both pairs of indices are contracted, the only non-vanishing inner products are
\begin{equation}
	Z_{ab} Z^{ab} = 2 X_{ab} Y^{ab} = -4.
	\label{bivectProd2}
\end{equation}
The main motivation for introducing this bivector basis is that it allows the curvature and the quadrupole moment to be written down and manipulated without having to perform coordinate computations; see, e.g., \eqref{Weyl}. These bivectors also provide a convenient basis for the torque which is experienced by an extended body. 

\subsubsection{Quadrupolar forces and torques in general vacuum spacetimes}
\label{Sect:Forces}

The vector and bivector bases which have just been described can now be used to compute forces and torques. We allow for general vacuum spacetimes which are not conformally flat, and also for arbitrary quadrupole moments (but no octupole or higher moments). As observed in \cite{HarteExtendedtypeD}, the trace-free quadrupole moment $J_{abcd}$ has the same algebraic properties as a Weyl tensor, and may therefore be described in terms of five complex scalars $J_0, \ldots, J_4$ which are analogous to the five Weyl scalars. Comparing with \eqref{PsiDef} and \eqref{Weyl}, any quadrupole moment can thus be written as
\begin{align}
	J_{abcd} = 2 \Re \big[ J_0 Y_{ab} Y_{cd} + J_1 (Y_{ab} Z_{cd} + Z_{ab} Y_{cd}) 
	\nonumber
	\\
	~ + J_2 (Z_{ab} Z_{cd} - X_{ab} Y_{cd} - Y_{ab} X_{cd} )
	\nonumber
	\\
	~ -  J_3 ( X_{ab} Z_{cd} + Z_{ab} X_{cd}) + J_4 X_{ab} X_{cd} \big],
	\label{quad}
\end{align}
where
\begin{subequations}
\label{Ji}
\begin{gather}
	J_0 \equiv \tfrac{1}{4} J_{abcd} X^{ab} X^{cd}, \quad J_1 \equiv \tfrac{1}{8} J_{abcd} X^{ab} Z^{cd},
	\\
	J_2 \equiv -\tfrac{1}{4} J_{abcd} X^{ab} Y^{cd} = \tfrac{1}{16} J_{abcd} Z^{ab} Z^{cd},
	\\
	J_3 \equiv -\tfrac{1}{8} J_{abcd} Y^{ab} Z^{cd}, \quad J_4 = \tfrac{1}{4} J_{abcd} Y^{ab} Y^{cd}.
\end{gather}
\end{subequations}

At quadrupolar order, the generalized force may now be computed by substituting these equations and the Weyl expansion \eqref{Weyl} into \eqref{QuadrupoleForce2}. As $\ell^a$ has already been assumed to have been aligned with one of the PNDs, $\Psi_0$ will always vanish. Contributions from the other Weyl scalars are
\begin{widetext}
\begin{align}
	\mathcal{F}^{(q)}_\xi =  \frac{4}{3} \Re \big\{ \big[ \Psi_1 ( J_4 X^{ab} - 3 J_2 Y^{ab} ) \mathcal{L}_\xi Z_{ab} - 2 J_3 X^{ab} \mathcal{L}_\xi (\Psi_1 Y_{ab}) \big] + 3 \big[ \Psi_2 ( J_1 Y^{ab} - J_3 X^{ab} ) \mathcal{L}_\xi Z_{ab} - 2 J_2 \mathcal{L}_\xi \Psi_2 \big]
	\nonumber
	\\
	~ + \big[ \Psi_3 ( 3 J_2 X^{ab} - J_0 Y^{ab} ) \mathcal{L}_\xi Z_{ab} - 2  J_1 Y^{ab} \mathcal{L}_\xi (\Psi_3 X_{ab} ) \big] + \big[  \Psi_4 ( J_0 Y^{ab} + J_1 Z^{ab}) \mathcal{L}_\xi X_{ab}  - J_0 \mathcal{L}_\xi \Psi_4 \big]\big\}.
	\label{ForceFull}
\end{align}
\end{widetext}
The terms here which involve $\Psi_2$ and $\Psi_4$ were already obtained in \cite{HarteExtendedtypeD}; the others are new. Much of the remainder of this paper analyses the implications of this expression. One consequence which is already apparent is that each Weyl scalar can couple only to certain quadrupole scalars, as summarized in Table \ref{Table:Jcouple}.

\begin{table}
\setlength{\tabcolsep}{8pt}
  \begin{tabular}{c |  c }
   Weyl scalar & Quadrupole scalars \\[.3ex]
    \hline\hline
    $\Psi_1$ 	& $J_2$, $J_3$, $J_4$   \\	
    $\Psi_2$	 	& $J_1$, $J_2$, $J_3$ \\	
    $\Psi_3$	 	& $J_0$, $J_1$, $J_2$ \\
    $\Psi_4$		& $J_0$, $J_1$
	\end{tabular}
	\caption{Summary of which quadrupole scalars can couple to which Weyl scalars in the generalized force \eqref{ForceFull}. The scalar $\Psi_0$ has been omitted as it always vanishes with our tetrad choice.}
	\label{Table:Jcouple}
\end{table}

\subsubsection{Type N spacetimes}
\label{Sect:TypeN}

The simplest nontrivial Weyl tensors are of Petrov type N. In that case, aligning $\ell^a$ with the sole PND results in only $\Psi_4$ being nonzero. Eq. \eqref{ForceFull} then reduces to
\begin{align}
	\mathcal{F}^{(q)}_\xi = \frac{4}{3} \Re \left[ \Psi_4 ( J_0 Y^{ab} + J_1 Z^{ab}) \mathcal{L}_\xi X_{ab}  - J_0 \mathcal{L}_\xi \Psi_4 \right].
	\label{genForceN}
\end{align}
Equivalently, \eqref{GenForceNewt} and \eqref{bivectorProd1} can be used to convert this into the force 
\begin{equation}
	F^{(q)}_a = \frac{4}{3} \Re \big[ \Psi_4 ( J_0 Y^{bc} + J_1 Z^{bc}) \nabla_a X_{bc}  - J_0 \nabla_a \Psi_4 \big],
	\label{forceN}
\end{equation}
and the torque
\begin{equation}
	N^{(q)}_{ab} = \frac{8}{3} \Re \big[ \Psi_4 ( 2 J_1 X_{ab} - J_0 Z_{ab}) \big]. 
	\label{torqueN}
\end{equation}
At most $J_0$ and $J_1$ can thus affect the motion, meaning that there are only four real quadrupole components which must be considered in type N spacetimes (rather than the ten components which might affect motion in a general vacuum spacetime).

It also follows from \eqref{bivectProd2} and \eqref{torqueN} that regardless of the quadrupole moment,
\begin{equation}
	N_{ab}^{(q)} X^{ab} = 0. 
        \label{torqueConstrN}
\end{equation} 
The real and the imaginary components of this constraint imply that there is a two-dimensional space of real torques which cannot be produced by any quadrupole moment in a type N spacetime. Torques can, however, be controlled throughout the four-dimensional space which is spanned by real combinations of $X_{ab}$, $Z_{ab}$, and their complex conjugates. Moreover, given any torque within this  space, the quadrupole scalars $J_0$ and $J_1$ which produce it are uniquely determined. This can be used to show that the force is a linear function of the torque:
\begin{align}
	F_a^{(q)} = \frac{3}{4} \Re \big[  (Z^{bc} Y^{df} - Y^{bc} Z^{df}) \nabla_a X_{df} 
	 \nonumber
	 \\
	 ~ - Z^{bc} \nabla_a \ln \Psi_4  \big] N_{bc}^{(q)}.
	 \label{ForceTorqueN}
\end{align}
It is therefore impossible to vary the force without simultaneously varying the torque. 

Both \eqref{torqueConstrN} and \eqref{ForceTorqueN} are consequences of local symmetries. One way to identify these symmetries is to note that the generalized force can be written as
\begin{align}
	\mathcal{F}^{(q)}_{\xi} =  \frac{1}{4} \big\{ 3 \xi^a \Re  \big[ ( Z^{bc} Y^{df} - Y^{bc} Z^{df} )  \nabla_a X_{df} 
	\nonumber
	\\
	 ~ - Z^{bc} \nabla_a \ln \Psi_4 \big] + 2 \nabla^b \xi^c \big\} N_{bc}^{(q)},
\end{align}
where $\xi^a$ is any GKF. Recalling that local symmetries are generated by families of GKFs $\Xi^a_s$ whose associated generalized forces vanish for all possible quadrupole moments, it follows that
\begin{align}
	\nabla^a \Xi^b_s = \frac{3}{2} \Xi^f_s \Re \big[ ( Y^{ab} Z^{cd} - Z^{ab} Y^{cd} )  \nabla_f X_{cd} 
	\nonumber
	\\
	~ + Z^{ab} \nabla_f \ln \Psi_4 \big]  + \bar{\lambda}_s X^{ab} + \lambda_s \bar{X}^{ab}
	\label{localSymN}
\end{align}
at $\gamma_s$, where $\lambda_s$ is any family of complex scalars and $\Xi^a_s(\gamma_s)$ is arbitrary. Setting $\Xi^a_s(\gamma_s) = 0$ while varying $\lambda_s$ recovers the two real symmetries which generate the torque constraint \eqref{torqueConstrN}. Setting $\lambda_s = 0$ while varying $\Xi^a_s(\gamma_s)$ results in four more local symmetries; these imply that forces and torques must be linked via \eqref{ForceTorqueN}. In total, there are six local symmetries in each type N spacetime.

\subsubsection{Type III spacetimes}
\label{Sec:TypeIII}

In Petrov type III spacetimes, we align the tetrad such that $\Psi_3$ is the only non-vanishing Weyl scalar. The generalized force \eqref{ForceFull} then reduces to
\begin{align}
	\mathcal{F}^{(q)}_\xi =\frac{4}{3} \Re  
 \big[ \Psi_3 ( 3 J_2 X^{ab} - J_0 Y^{ab} ) \mathcal{L}_\xi Z_{ab} 
 \nonumber
 \\
 ~ - 2  J_1 Y^{ab} \mathcal{L}_\xi (\Psi_3 X_{ab} ) \big],
\end{align}
which depends only on $J_0$, $J_1$, and $J_2$. Six real quadrupole components can therefore affect motion in type III spacetimes. Converting the generalized force into an ordinary force and a torque, 
\begin{align}
	F^{(q)}_a =\frac{4}{3} \Re  
 \big[ \Psi_3 ( 3 J_2 X^{bc} - J_0 Y^{bc} ) \nabla_a Z_{bc} 
 \nonumber
 \\
 ~ - 2  J_1 Y^{bc} \nabla_a (\Psi_3 X_{bc} ) \big] ,
\end{align}
and
\begin{equation}
	N_{ab}^{(q)} = \frac{16}{3} \Re \big[ \Psi_3 ( J_1 Z_{ab} - J_0 Y_{ab}-3 J_2 X_{ab}) \big].
\end{equation}
Any torque whatsoever can thus be generated by an appropriately-structured object.

 In fact, a given torque uniquely determines $J_0$, $J_1$ and $J_2$. That may be used to show that the force is again a linear function of the torque:
\begin{align}
	F_a^{(q)} = \frac{1}{4} \Re \big[ ( X^{bc} Y^{df} - Y^{bc} X^{df}) \nabla_a Z_{bc} + Z^{df}  
	\nonumber
	\\
	~ \times (Y^{bc} \nabla_a X_{bc}- 2 \nabla_a \ln \Psi_3 ) \big] N_{df}^{(q)}.
	\label{ForceTorqueIII}
\end{align}
It is therefore impossible to control the force independently of the torque in type III spacetimes. 

Local symmetries may be found in type III spacetimes by first writing the generalized force which is associated with a generic GKF $\xi^a$ as
\begin{align}
	\mathcal{F}_{\xi}^{(q)} = \frac{1}{4} \Re \big\{ \xi^a \big[ ( X^{bc} Y^{df} - Y^{bc} X^{df}) \nabla_a Z_{bc} + Z^{df}  
	\nonumber
	\\
	~ \times (Y^{bc} \nabla_a X_{bc}- 2 \nabla_a \ln \Psi_3 ) \big] + \tfrac{1}{2} \nabla^d \xi^f \big\} N_{df}^{(q)}.
\end{align}
This vanishes for all quadrupole moments when we choose a 1-parameter family of GKFs which satisfy
\begin{align}
	\nabla^a \Xi^b_s = 2 \Xi^f_s \big[ (X^{ab} Y^{cd} - Y^{ab} X^{cd}) \nabla_f Z_{cd} + Z^{ab} 
	\nonumber
	\\
	~ \times ( 2 \nabla_f \ln \Psi_3 - Y^{cd} \nabla_f X_{cd}) \big]
\end{align}
at $\gamma_s$, where $\Xi^a_s$ is arbitrary. It follows that there are four local symmetries in type III spacetimes. Physically, these  imply that the force and the torque must be linked by \eqref{ForceTorqueIII}.

\subsubsection{Type D spacetimes}
\label{Sec:TypeD}

In Petrov type D spacetimes, we align the tetrad such that $\Psi_2$ is the only non-vanishing Weyl scalar. The generalized force then reduces to
\begin{align}
	\mathcal{F}^{(q)}_\xi = 4 \Re  \big[ \Psi_2 (  J_1 Y^{ab} - J_3 X^{ab} ) \mathcal{L}_\xi Z_{ab}  - 2 J_2 \mathcal{L}_\xi \Psi_2 \big],
\end{align}
which depends only on $J_1$, $J_2$, and $J_3$. Up to\footnote{In the type N and type III cases discussed above, we were able to say exactly how many quadrupole components contribute to the motion. In the type D case, the answer varies depending on the properties of $\nabla_a \Psi_2$.} six real quadrupole components therefore contribute to the motion in type D spacetimes.

This statement can be refined by first using \eqref{GenForceNewtFN} to extract the force 
\begin{equation}
	F^{(q)}_a = 4 \Re  \big[ \Psi_2 (  J_1 Y^{bc} - J_3 X^{bc} ) \nabla_a Z_{bc}  - 2 J_2 \nabla_a \Psi_2 \big],
\end{equation}
and the torque
\begin{equation}
	N_{ab}^{(q)} = 16 \Re \big[ \Psi_2 ( J_3 X_{ab} + J_1 Y_{ab}) \big].
\end{equation}	
The torque therefore depends on $J_1$ and $J_3$, but not $J_2$. It is also apparent that regardless of the quadrupole moment,
\begin{equation}
	N_{ab}^{(q)} Z^{ab} = 0.
	\label{NconstrTypeD}
\end{equation}
Quadrupolar torques can therefore be varied only within the four-dimensional space which is spanned by real combinations of $X_{ab}$, $Y_{ab}$, and their complex conjugates. 

Unlike in type N or type III spacetimes, the force in a type D spacetime is not necessarily a linear function of the torque. Instead,
\begin{align}
	F_a^{(q)} = \frac{1}{4} \Re \big[ ( X^{bc} Y^{df} - Y^{bc} X^{df}) \nabla_a Z_{bc} \big] N_{df}^{(q)}
	\nonumber
	\\
	~ - 8 \Re \big[ J_2 \nabla_a \Psi_2\big]. 
	\label{ForceTorqueD}
\end{align}
Unless $\Psi_2$ is constant, this implies that the force can be varied independently of the torque. In particular, the space of forces which can be controlled at fixed torque is spanned by the real and the imaginary components of $\nabla_a \Psi_2$.

If $\Psi_2$ is constant, the force is instead linear in the torque and $J_2$ disappears from the laws of motion. Such an example (necessarily with a nonzero cosmological constant $\Lambda$) is provided by the Nariai or anti-Nariai spacetimes\footnote{This is described as a Bertotti-Robinson spacetime in \cite{Barnes2015} but as a Nariai or anti-Nariai spacetime (depending on the sign of $\Lambda$) in Sect. 18.6 of \cite{GriffithsExact}.} \cite{Barnes2015} with line elements
\begin{equation}
	ds^2 = - 2 du dv + \Lambda v^2 du^2 + \frac{ dy^2 + dz^2 }{ [1 + \tfrac{1}{4} \Lambda(y^2+z^2) ]^2}.
\end{equation}
In fact, $F_a^{(q)} = 0$ in these spacetimes; extended-body effects can  influence only the torque, at least at quadrupolar order.

In more complicated type D spacetimes where $\Psi_2$ is not constant, the space of forces which can be produced at fixed torque is either one- or two-dimensional. In Kerr spacetimes with nonzero angular momentum, forces may be varied throughout a two-dimensional space without also varying the torque. In the Schwarzschild limit where the angular momentum goes to zero, forces can instead be varied in only one direction without also varying the torque \cite{HarteExtendedtypeD}.

Local symmetries may be derived in an arbitrary type D spacetimes by first writing the generalized force as
\begin{align}
	\mathcal{F}^{(q)}_\xi = \frac{1}{4} \Re \big\{ \xi^a \big[ ( X^{bc} Y^{df} - Y^{bc} X^{df}) \nabla_a Z_{bc} \big] 
	\nonumber
	\\
	~ + 2 \nabla^d \xi^f \big\} N_{df}^{(q)} - 8 \Re [J_2 \mathcal{L}_\xi \Psi_2].
\end{align}
This vanishes for all quadrupole moments when we construct a 1-parameter family of GKFs which satisfy
\begin{align}
	\nabla^a \Xi^b_s = \frac{1}{2} \Xi^f_s \Re \big[ ( X^{ab} Y^{cd} - Y^{ab} X^{cd} ) \nabla_f Z_{cd} \big]
	\nonumber
	\\
	~ + \bar{\lambda}_s Z^{ab} + \lambda_s \bar{Z}^{ab}
\end{align}
at $\gamma_s$. Here, $\lambda_s$ is arbitrary and $\Xi^a_s$ must be orthogonal to $\nabla_a \Psi_2$ at $\gamma_s$. Setting $\Xi^a_s(\gamma_s) =0$ while varying $\lambda_s$ recovers the two real symmetries which imply the torque constraint \eqref{NconstrTypeD}. These are related to the fact that the real and the imaginary components of $Z_{ab}$ are proportional to conformal Killing-Yano tensors. If we instead set $\lambda_s = 0$ and vary $\Xi^a_s(\gamma_s)$, our prescription results in two to four additional symmetries. These imply that the force components which are orthogonal to $\nabla_a \Psi_2$ and its complex conjugate must be linked to torques via \eqref{ForceTorqueD}. In total, there are between four and six local symmetries in type D spacetimes. There are, e.g., four such symmetries in Kerr (with nonzero angular momentum), five in Schwarzschild, and six in the Nariai and anti-Nariai spacetimes.

\subsubsection{Type II spacetimes}

In Petrov type II spacetimes, we choose the tetrad such that only $\Psi_2$ and $\Psi_3$ are nonzero. The generalized force \eqref{ForceFull} then reduces to
\begin{align}
	\mathcal{F}^{(q)}_\xi &= \frac{4}{3} \Re \big\{ 4 J_1 \mathcal{L}_\xi \Psi_3 -6 J_2 \mathcal{L}_\xi \Psi_2  - 2  \Psi_3 J_1 Y^{ab} \mathcal{L}_\xi X_{ab}  
	\nonumber
	\\
	& ~ + [3(\Psi_3 J_2 - \Psi_2 J_3) X^{ab}	
	+ (3 \Psi_2 J_1 - \Psi_3 J_0) Y^{ab} 
 ] \mathcal{L}_\xi Z_{ab} \big\},
\end{align}
the ordinary force is given by
\begin{align}
	F_a^{(q)} &= \frac{4}{3} \Re\big\{ 4 J_1 \nabla_a \Psi_3 - 6 J_2 \nabla_a \Psi_2 - 2 \Psi_3 J_1 Y^{bc} \nabla_a X_{bc}
	\nonumber
	\\
	&~ + [ 3 ( \Psi_3 J_2 - \Psi_2 J_3) X^{bc} + ( 3 \Psi_2 J_1 - \Psi_3 J_0) Y^{bc} ] \nabla_a Z_{bc} \big\},
\end{align}
and the torque by
\begin{align}
	N_{ab}^{(q)} = \frac{16}{3} \Re\big[ \Psi_3 J_1 Z_{ab} + ( 3 \Psi_2 J_1 - \Psi_3 J_0) Y_{ab} 
	\nonumber
	\\
	~ + 3 ( \Psi_2 J_3 - \Psi_3 J_2 ) X_{ab} \big].
\end{align}
These expressions depend on all quadrupole scalars except for $J_4$, so up to eight quadrupole components can affect motion in type II spacetimes. 

It also follows that any torque whatsoever can be produced by appropriately varying the quadrupole moment. However, unlike in the type N, type III, and type D cases discussed above, a given torque does not uniquely determine the relevant quadrupole scalars. Instead, fixing $N_{ab}^{(q)}$ fixes only $J_0$, $J_1$, and $\Psi_2 J_3 - \Psi_3 J_2$. Using this, the force can nevertheless be shown to be affine function of the torque:
\begin{align}
	F_a^{(q)} = \frac{1}{4} \Re\big[  Y^{bc} Z^{df} \nabla_a X_{bc} - 2 Z^{df} \nabla_a \ln \Psi_3 
	 \nonumber
	\\
	~ (X^{bc} Y^{df} - Y^{bc} X^{df} ) \nabla_a Z_{bc}  \big] N_{df}^{(q)} 
	\nonumber
	\\
	~ - 8 \Re [ J_2 \nabla_a \Psi_2] .
	\label{ForceTorqueII}
\end{align}
By varying $J_2$ and $J_3$ at fixed $\Psi_2 J_3 - \Psi_3 J_2$, the quadrupolar force can thus be controlled, at fixed torque, throughout the space which is spanned by the real and the imaginary components of $\nabla_a \Psi_2$. That space has at most two dimensions. If $\Psi_2$ is constant, which occurs only in certain Kundt spacetimes \cite{Barnes2015}, the force cannot be controlled independently of the torque and only six quadrupole components affect the motion. If $\nabla_a \Psi_2$ is nonzero and linearly independent of its complex conjugate, there are eight quadrupole components which affect the motion.

All local symmetries may be found in type II spacetimes by first writing the generalized force as 
\begin{align}
	\mathcal{F}_{\xi} = \frac{1}{4} \Re \big\{ \xi^a \big[ Y^{bc} Z^{df} \nabla_a X_{bc} - 2 Z^{df} \nabla_a \ln \Psi_3 + (X^{bc} Y^{df}
	 \nonumber
	\\
	~  - Y^{bc} X^{df} ) \nabla_a Z_{bc} \big]  + 2 \nabla^d \xi^f \big\} N_{df}^{(q)} - 8 \Re [ J_2 \mathcal{L}_\xi \Psi_2 ] . 
\end{align}
Ensuring that this vanishes for all possible quadrupole moments, local symmetries are seen to be generated by the 1-parameter family of GKFs whose gradients satisfy 
\begin{align}
	\nabla^a \Xi^b_s = \frac{1}{2} \Xi_s^f \Re \big[2 Z^{ab} \nabla_f \ln \Psi_3  -  Y^{cd} Z^{ab} \nabla_f X_{cd} 
	 \nonumber
	\\
	~  + (X^{ab}  Y^{cd}  -  Y^{ab} X^{cd} ) \nabla_f Z_{cd} \big]
\end{align}
at $\gamma_s$, where $\Xi^a_s(\gamma_s)$ is constrained only to be orthogonal to $\nabla_a \Psi_2$. This implies that there are  between two and four local symmetries. Their presence requires forces which are orthogonal to $\nabla_a \Psi_2$ and its complex conjugate to be linked to torques via \eqref{ForceTorqueII}.

\subsubsection{Type I spacetimes}
\label{Sect:TypeI}

In type I spacetimes, only $\Psi_0$ and $\Psi_4$ can necessarily be made to vanish by choosing an appropriate tetrad. However, as noted in Sect. \ref{Sect:RelBasis} above, a type III tetrad transformation can always be used to ensure that $\Psi_1 = \Psi_3$. Applying such a transformation for simplicity, the generalized force \eqref{ForceFull} reduces to
\begin{align}
	\mathcal{F}^{(q)}_\xi = \frac{8}{3} \Re \big\{ 2 ( J_1 +J_3) \mathcal{L}_\xi \Psi_1  + \Psi_1 (J_3 - J_1) Y^{ab} \mathcal{L}_\xi X_{ab}    
	\nonumber
	\\
	~  - 3 J_2 \mathcal{L}_\xi \Psi_2 + \tfrac{1}{2} \big[  \big( \Psi_1 (3 J_2 + J_4) - 3 \Psi_2 J_3\big) X^{ab}   
		\nonumber
	\\
	~ + \big( 3 \Psi_2 J_1 - \Psi_1  (3J_2 + J_0) \big) Y^{ab} \big] \mathcal{L}_\xi Z_{ab}   \big\}.
\end{align}
All five quadrupole scalars $J_0, \ldots , J_4$ appear here, so all of a body's quadrupole moment can affect its motion in at least some type I spacetimes. 

Using \eqref{GenForceNewtFN} to extract the torque from the generalized force,
\begin{align}
	N^{(q)}_{ab} = \frac{16}{3} \Re \big\{ [ 3 \Psi_2 J_3 - \Psi_1 ( 3 J_2 + J_4) ] X_{ab} + [ 3 \Psi_2 J_1
	\nonumber
	\\
	~ - \Psi_1 ( 3 J_2 + J_0 ) ] Y_{ab} + \Psi_1 ( J_1-J_3) Z_{ab} \big\}.
	\label{NtypeI}
\end{align}
All torques are therefore possible in type I spacetimes. However, although all five quadrupole scalars appear here, fixing the torque fixes only the three combinations
\begin{equation}
\begin{gathered}
	J_1-J_3, \qquad J_0 - J_4,
	\\
	3 \Psi_2 ( J_1 + J_3 ) - \Psi_1 ( 6 J_2 + J_0 + J_4)
\end{gathered}
\label{JtypeI}
\end{equation}
of quadrupole scalars. This observation allows the force to again be written as an affine function of the torque:
\begin{align}
	F_a^{(q)} = \frac{1}{4} \Re \big[ Y^{bc} Z^{df} \nabla_a X_{bc} + (X^{bc} Y^{df} - Y^{bc} X^{df} ) \nabla_a Z_{bc}  \big]
	\nonumber
	\\
	~ \times  N_{df}^{(q)} + \frac{8}{3} \Re\big[ 2 ( J_1 + J_3 ) \nabla_a \Psi_1 - 3 J_2 \nabla_a \Psi_2 \big].
	\label{ForceTorqueI}
\end{align}
Noting that $J_1+J_3$ and $J_2$ can be varied arbitrarily without affecting the quadrupole components \eqref{JtypeI}, forces can thus be varied, at fixed torque, throughout the space which is spanned by the real and imaginary components of $\nabla_a \Psi_1$ and $\nabla_a \Psi_2$. If these gradients are all linearly independent, that space is four-dimensional. All ten force and torque components can then be controlled independently. At the opposite extreme, $\Psi_1$ and $\Psi_2$ may both be constant \cite{Barnes2015}, in which case the force is entirely determined by the torque. 

Unlike in the algebraically-special spacetimes discussed above, there might not be any local symmetries in type I spacetimes. Any local symmetries which do exist nevertheless satisfy
\begin{align}
	\nabla^a \Xi^b_s = \frac{1}{2} \Xi^f_s \Re \big[ ( X^{ab} Y^{cd} -Y^{ab} X^{cd} ) \nabla_f Z_{cd} 
	\nonumber
	\\
	~ - Z^{ab} Y^{cd} \nabla_f X_{cd} \big]
	\label{localSymI}
\end{align}
at $\gamma_s$, where $\Xi^a_s$ can be varied arbitrarily at $\gamma_s$ as long as it is orthogonal to the real and the imaginary components of both $\nabla_a \Psi_1$ and $\nabla_a \Psi_2$. This results in between zero and four local symmetries, depending on which particular type I spacetime is considered.

\subsubsection{Summarizing the relativistic constraints}
\label{Sect:RelSumm}

We have now derived quadrupolar forces and torques in vacuum spacetimes and discussed qualitative differences which depend on the Petrov type of the relevant spacetime. As summarized in Table \ref{Table:RelSumm}, our focus has been on three characteristics of extended-body motion: the number of torque components which can be affected by internal structure, the number of force components which can be controlled independently of the torque, and the number of quadrupole components which affect the motion. These characteristics have also been related to the presence of local symmetries. Roughly speaking, there are fewer local symmetries in spacetimes which are ``more'' algebraically general, and in those cases, extended bodies can exert more control over their motion.

\begin{table}
\setlength{\tabcolsep}{8pt}
  \begin{tabular}{c | c | c | c | c}
   Petrov type & $\big\{ \Xi^a_s \big\}$ &  $\big\{ J_{abcd} \big\}$  & $\big\{ N_{ab}^{(q)} \big\}$  & $ \big\{ F_a^{(q)} \big| N_{bc}^{(q)} \big\}$ \\[.7ex]
    \hline\hline
    I 	&	0--4		&	6-10* 	& 	6*  &  $\leq 4$*   \\	
    II 	&	2--4		& 	6-8 	&  	6*	&  $\leq 2$		 \\	
    D	&   4--6		&    4-6 	&   4	&  $\leq 2$ 	\\
    III	&	4 		&	6		&  	6*	&  0  		\\
    N	&	6 		&	4 		&	4	&  0 		
	\end{tabular}
	\caption{Qualitative features of quadrupolar forces and torques in spacetimes with different Petrov types. Column 2 lists the number of local symmetries. The upper bounds there also provide an upper bound for the number of Killing vectors which can exist. Column 3 lists the number of real quadrupole components which affect the motion.  The number of controllable torque components is provided in column 4. Column 5 lists the number of force components which can be controlled at fixed torque. Stars are used to indicate that a given number is unconstrained.} 
	
	\label{Table:RelSumm}
\end{table}

Any torque whatsoever can be produced in type I, type II, and type III spacetimes, but not in type N or type D spacetimes. In the latter cases, a two-dimensional space of torques is inaccessible, regardless of the quadrupole moment. It is interesting in this context to recall that the relativistic torque is qualitatively different from its Newtonian countepart. The relativistic torque includes three components which are physically similar to the Newtonian torque, but it also involves three components which are fundamentally non-Newtonian. These additional components may be viewed as controlling the misalignment between the 4-velocity and the 4-momentum: the ``hidden momentum'' \cite{Harte2007, Bobbing, HarteExtendedtypeD}. Constraints on the relativistic torque therefore affect an object's ability not only to control its spin, but also to directly control its velocity. One example of this is given in Sect. \ref{Sect:Kasner} below, where an object in a Kasner spacetime is shown to be able to move itself arbitrarily simply by controlling its torque. More generally, since at least four torque components can be controlled in every nontrivial vacuum spacetime, the torque can always be used to control at least some of an object's velocity. 

Another of our results is that the quadrupolar force can always be written as an affine function of the quadrupolar torque: Eqs. \eqref{ForceTorqueN}, \eqref{ForceTorqueIII}, \eqref{ForceTorqueD}, \eqref{ForceTorqueII}, and \eqref{ForceTorqueI} all have the form
\begin{equation}
	F_a^{(q)} = \chi_{a}{}^{bc} N_{bc}^{(q)} + \Re \sum_{I} \alpha_I J_{4-I} \nabla_a \Psi_I,
	\label{ForceTorqueGen}
\end{equation}
where the $\alpha_I$ are coefficients and $\chi_{a}{}^{bc}$ depends only on the geometry (but not on an object's internal structure). The first term here describes that portion of the force which is universally tied to the torque. The second term provides all portions of the force which can be varied independently of the torque, and is also the only force which remains when $N_{ab}^{(q)} = 0$. Interestingly, this latter term is simply a linear combination of gradients of the Weyl scalars. It does not depend on, e.g., any gradients of the tetrad. Recalling \eqref{FlinNNewt}, a similar result holds also for quadrupolar forces in Newtonian gravity.

One generic feature of extended-body motion is that the mass \eqref{mass} is not necessarily constant at quadrupolar and higher orders. Our result \eqref{ForceTorqueGen} nevertheless implies that at least for torque-free, spin-free bodies with constant quadrupole scalars, there exists an ``effective mass'' which is conserved. As we have discussed already, it is always possible to arrange for the torque to vanish. Adopting the Tulczyjew spin supplementary condition \eqref{SSC}, doing so implies that if $S^{ab}$ is initially zero, it remains so. Then $p_a = M \dot{\gamma}_s^a$ \cite{EhlersRudolph}, and it follows from \eqref{MP} that
\begin{equation}
    \frac{DM}{ds} = - \Re \sum_I \alpha_I J_{4-I} \dot{\gamma}^a_s \nabla_a \Psi_I. 
\end{equation}
If all of the \textit{relevant} quadrupole scalars (i.e., the ones for which $\alpha_I J_{4-I} \neq 0$) are constant, the effective mass
\begin{equation}
    M_{\mathrm{eff}} \equiv M + \Re \sum_I \alpha_I J_{4-I} \Psi_I 
    \label{MeffGen}
\end{equation}
is therefore conserved. A special case of this was used in \cite{HarteExtendedtypeD} in order to understand how extended-body effects can be used to alter orbits in the Schwarzschild spacetime. Appendix \ref{App:Tides} identifies a somewhat different effective mass \eqref{Meff} which is conserved when quadrupole moments are tidally induced with constant deformabilities. Using different assumptions on the nature of the quadrupole moment, certain other effective masses can be found as well \cite{Dixon70a, Dix79}. 

\subsection{Example spacetimes}
\label{Sect:RelExamples}

We now apply the general results derived above to two specific examples: extended-body motion in pp-wave spacetimes and extended-body motion in Kasner spacetimes. Kasner spacetimes are type I while pp-waves are type N, so the Weyl tensors in these examples lie at the two extremes of algebraic speciality.

\subsubsection{Motion in pp-wave spacetimes}
\label{Sect:ppWave}

A pp-wave spacetime describes a plane-fronted gravitational wave with parallel rays \cite{ExactSolns, GriffithsExact, EhlersKundt}. Besides their interpretation as idealized gravitational waves, some pp-waves also arise as ultrarelativistic limits of other (not necessarily radiative) spacetimes. This can occur both via ``global'' boosts \cite{AichelburgSexl, Podolsk1998}, or via the Penrose limit, which locally describes the geometry near arbitrary null geodesics as effective plane waves \cite{PenroseLimit, Blau}.

Regardless of interpretation, any pp-wave spacetime can be described by the line element
\begin{equation}
	ds^2 = - 2du dv + H(u,y,z) du^2 + dy^2 + dz^2,
\end{equation}
where $H(u,x,y)$ is a dimensionless ``waveform,'' $y$ and $z$ are transverse coordinates, and $u$ is a null ``phase'' coordinate. Imposing the $\Lambda=0$ vacuum Einstein equation shows that
\begin{equation}
	(\partial_y^2 + \partial_z^2 ) H(u,y,z) = 0,
\end{equation}
so the waveform here must be harmonic\footnote{That the nonlinearity of Einstein's equation disappears in this class of spacetimes is due to the fact that pp-waves are Kerr-Schild transformations of Minkowski spacetime \cite{Xanthopoulos1978, HarteVines}.} on each $u = \mathrm{constant}$ hypersurface. Harmonic functions in two real dimensions can be related to complex analytic functions of one variable, so the waveform of an arbitrary vacuum pp-wave can be written as
\begin{equation}
	H(u,y,z) = \Re \, \mathcal{H}(u, \zeta(y,z) ),
\end{equation}
where $\mathcal{H}$ is complex and analytic in the complexified transverse coordinate $\zeta \equiv (y + i z)/\sqrt{2}$. A pp-wave is said to be linearly polarized when $\arg \mathcal{H} = \mathrm{constant}$, and other properties of $\mathcal{H}$ can be used to classify pp-waves as described in \cite{EhlersKundt}; see also Table 24.2 of \cite{ExactSolns}.

The most important category in this classification are the vacuum ``plane waves,'' which satisfy 
\begin{equation}
	\partial_\zeta^3 \mathcal{H} = 0.	
\end{equation}
In any plane wave spacetime, there exist coordinates in which 
\begin{equation}
	\mathcal{H}(u,\zeta) = \mathcal{h}(u) \zeta^2,
	\label{planeWave}
\end{equation}
where the complex function $\mathcal{h}$ determines the curvature as a function of phase; see $\Psi_4$ in \eqref{WeylPP} below. Einstein's equation does not constrain $\mathcal{h}$. 

All vacuum pp-waves, whether plane waves or not, are type N wherever they are not flat. The lone principal null direction is parallel to 
\begin{equation}
	\ell_a = -\nabla_a u,
\end{equation}
which physically describes the direction along which the gravitational wave propagates. Its integral curves are the ``rays'' of that wave. A calculation shows that
\begin{equation}
	\nabla_a \ell_b =0,
\end{equation} 
so these rays are geodesic, non-expanding, shear-free, and twist-free. It also follows that $\ell^a$ is Killing. For some pp-waves, $\ell^a$ is the only Killing field. In special cases, there can be up to five more\footnote{If the vacuum restriction is relaxed, there can be up to seven Killing vectors in total \cite{SippelSyms}.} \cite{EhlersKundt}. All plane waves admit at least five Killing fields in total, although some admit six. In this latter case, all of the local symmetries described by \eqref{localSymN} are ordinary Killing fields. In other pp-wave spacetimes, some local symmetries are Killing while some are not. Regardless, $\mathcal{P}_\ell = p_a \ell^a$ is conserved for any extended body moving in any pp-wave spacetime. Furthermore,
\begin{equation}
    \mathcal{F}_\ell = F_a \ell^a = 0.
    \label{FEll}
\end{equation}
This constraint holds not only for the full force, but also for its quadrupolar contribution. 

In any pp-wave spacetime, it is convenient to use $\ell_a$ as one element of the null tetrad $(\ell_a, n_a, m_a, \bar{m}_a)$, where $n_a \equiv - ( \nabla_a v +\tfrac{1}{2} H \ell_a)$ and $m_a \equiv \nabla_a \zeta$. Employing this to construct the $X_{ab}$ defined by \eqref{2formBasis}, a calculation shows that
\begin{equation}
	\nabla_c X_{ab} = 0.
	\label{DX}
\end{equation}
The real and imaginary components of $X_{ab}$ are therefore Killing-Yano tensors. As described in Sect. \ref{App:CKY} above, each such tensor generates a local symmetry. For generic pp-waves, these symmetries are truly local and are not necessarily associated with any conservation law. However, there are special pp-wave spacetimes where the local symmetries associated with Killing-Yano tensors are related to ordinary Killing symmetries. In those cases, there \textit{are} genuine conservation laws which can be associated with Killing-Yano tensors (regardless of, e.g., spin supplementary conditions). 

To see this, consider the special case of a plane wave spacetime. The waveform $\mathcal{H}$ is then given by \eqref{planeWave}. Letting $\lambda(u)$ be any possibly-complex solution to the differential equation $\chi''(u) = \frac{1}{2} \mathcal{h}(u) \bar{\chi}(u)$, and letting $\zeta_{\tau}$ and $u_{\tau}$ be any families of constants,  the vector fields
\begin{align}
	\Xi^a_{\tau}(x) = \Re \big\{ [ \zeta_{\tau} \chi'(u_{\tau}) - \zeta(x) \chi'(u(x)) ] \ell^a(x) 
	\nonumber
	\\
	~ - \chi(u(x)) m^a(x) \big\}
\end{align}
are Killing. At fixed $\tau$, varying over all possible $\chi(u)$ results in four real Killing fields with this form. In a flat limit, two of these Killing fields describe translations transverse to the rays of the gravitational wave. The remaining two describe mixed boosts together with rotations, and are a consequence of the fact that moving transverse to a plane wave appears only to rotate it. For all four of these Killing fields,
\begin{equation}
	\nabla^a \Xi^b_\tau = \Re [ \chi'(u) X^{ab} ].
\end{equation}
If $u_{\tau} = u(\gamma_\tau)$ and $\zeta_{\tau} = \zeta(\gamma_\tau)$ now denote an object's phase and transverse coordinates at some time $\tau$, and if $\chi(u_{\tau}) = 0$, it follows that $\Xi^a_{\tau} (\gamma_{\tau} ) = 0$. Also choosing $\chi'(u_{\tau})$ to be equal to $1$ or to $-i$ reproduces the generalized Killing fields determined by \eqref{CKYGKF}, when $s=\tau$, where the CKY tensor $f_{ab}$ which appears there is understood to be either the real or the imaginary component of $X_{ab}$. Since $\Xi^a_\tau$ is genuinely Killing, the generalized momentum component
\begin{align}
	\mathcal{P}_{\Xi_{\tau}} = \Re \big\{ \big[ \zeta_{\tau} \chi'(u_{\tau}) - \zeta_s \chi'(u_s) \big] \mathcal{P}_\ell - \chi(u_s) p_a m^a 
	\nonumber
	\\
	~ + \tfrac{1}{2} \lambda'(u_s) S_{ab} X^{ab} \big\} 
	\label{PconsPW}
\end{align}
is conserved for any fixed $\tau$; it is independent of $s$. Given the aforementioned initial conditions for $\chi(u)$, the first line here necessarily vanishes when $s= \tau$, implying that $\mathcal{P}_{\Xi_\tau} = \frac{1}{2} \Re [ \chi'(u_\tau) X^{ab} ] S_{ab}(\tau)$. However, the first line in \eqref{PconsPW} must be retained when $s \neq \tau$. It is interesting to note that in this case, where the symmetries associated with Killing-Yano tensors can definitively be associated with conservation laws, the quantities which are conserved do not coincide with the real and the imaginary components of $S_{ab} X^{ab}$, except at one moment in time. The situation here is similar to the one in the Schwarzschild spacetime, where the local symmetry associated with one (but not both) of the conformal Killing-Yano tensors can be derived from ordinary Killing symmetries \cite{HarteExtendedtypeD}. In that case as well, the associated conservation law is not trivial.

Returning to the case of a general pp-wave spacetime, use of \eqref{PsiDef} shows that with the above tetrad, the only non-vanishing Weyl scalar is
\begin{equation}
	\Psi_4 = \frac{1}{4} (\partial_z^2 - \partial_y^2 + 2 i \partial_y \partial_z ) H = - \frac{1}{4} \partial_\zeta^2 \mathcal{H}.
	\label{WeylPP}
\end{equation}
Substituting this and \eqref{DX} into the generic type N force \eqref{forceN} shows that the quadrupolar force which acts on an arbitrary extended body is
\begin{align}
	F^{(q)}_a &= - \frac{4}{3} \Re [J_0 \nabla_a \Psi_4 ] ,
        \nonumber
        \\
        & = \frac{1}{3} \Re \big[ J_0 ( m_a \partial_\zeta  -  \ell_a \partial_u  ) \partial_\zeta^2 \mathcal{H} \big].
	\label{FPPwave}
\end{align}
This is clearly consistent with the Killing constraint \eqref{FEll}. It may also be seen that although the torque \eqref{torqueN} depends on the quadrupole components $J_0$ and $J_1$, the force here depends only on $J_0$. This suggests that there are two real control parameters with which to control the force (at least if the torque is allowed to vary as well). However, these two parameters do not necessarily have independent effects. If a pp-wave is linearly polarized, for example, the quadrupolar force can be varied only in the one direction parallel to $\nabla_a |\Psi_4|$. The force can also be varied in only one direction in a plane wave spacetime---whether it is linearly polarized or not. In fact, the force is always proportional to $\ell_a$ plane wave spacetimes; it is longitudinal.

\subsubsection{Motion in Kasner spacetimes} 
\label{Sect:Kasner}

The vacuum Kasner spacetimes may be viewed as describing homogeneous but anisotropic (and empty) universes \cite{GriffithsExact, ExactSolns, HarveyKasner}. They have the line elements
\begin{equation}
	ds^2 = - dt^2 + \sum_{i=1}^3 t^{2u_i} (dx^i)^2 ,
\end{equation}
where $u_1$, $u_2$, and $u_3$ are constants\footnote{These are more commonly denoted by $p_1$, $p_2$, and $p_3$. We use a different notation in order to avoid confusion with the momentum.}. Applying the $\Lambda=0$ vacuum Einstein equation results in
\begin{equation}
	\sum_{i=1}^3 u_i = \sum_{i=1}^3 u_i^2 = 1, 
\end{equation}
which imples that the space of vacuum Kasner spacetimes may be viewed, in $\mathbb{R}^3$, as the intersection of a unit sphere with a plane. This leaves a 1-parameter family of solutions. 

Except in special cases which we do not consider, the Kasner spacetimes are of Petrov type I. Being spatially homogeneous, they admit the three translational Killing fields $\partial_i$. Since the Weyl scalars in an appropriately-adapted tetrad can at most depend on $t$, there is a three-dimensional space of local symmetries described by \eqref{localSymI}. However, these three symmetries are simply the three translational Killing fields. There are no proper local symmetries.

As in all type I spacetimes, the torque which acts on an extended body in a Kasner spacetime is unconstrained. However, the force can be varied only along $t_a \equiv -\nabla_a t$ without simultaneously varying the torque. Homogeneity implies that the three spatial components 
\begin{equation}
	\mathcal{P}_{\partial_i} = p_i + u_i t^{2u_i-1} S^{ti} 
	\label{PconsKasner}
\end{equation}
of the generalized momentum are conserved, where no sum over $i$ is implied. Similarly, the force and the torque are related via $\mathcal{F}_{\partial_i} = F_i + u_i t^{2u_i-1} N^{ti} = 0$, where again, no sum is implied.

One interesting feature of motion in Kasner spacetimes is that despite their spatial homogeneity, it is still possible for an extended body to exert essentially arbitrary control over its trajectory. To see this in a special case, first fix a centroid using the spin supplementary condition
\begin{equation}
	S_{ab} t^b = 0,
	\label{SSCKasner}
\end{equation}
which demands that $\gamma_s$ be chosen such that the mass dipole moment vanishes in the frame which is associated with the background homogeneity. Given \eqref{PconsKasner}, this spin supplementary condition implies that the three momentum components $p_i$ must be conserved. However, the velocity is not necessarily proportional to the momentum and is not necessarily conserved: Differentiating \eqref{SSCKasner} while using \eqref{MP} instead shows that
\begin{equation}
	(- p \cdot t) \dot{\gamma}^a_s = (-\dot{\gamma}_s \cdot t) p^a - N^{a}{}_{b} t^b -S^{ab} \dot{\gamma}^c_s \nabla_c t_b.
	\label{momVelKasner}
\end{equation}
Suppose for simplicity that $p_i = 0$. Choosing the body's quadrupole moment such that $N_{(q)}^{ij} = 0$, it is then possible to arrange for the angular momentum to vanish for all time, at least through quadrupolar order. Doing so, the spatial velocity becomes proportional to $N^{it}_{(q)}$, which can be controlled arbitrarily. An extended body with vanishing spatial momentum and vanishing angular momentum may therefore translate itself arbitrarily, simply by controlling its quadrupole moment. This takes advantage of the fact that although Kasner spacetimes are spatially homogeneous, they are not boost-invariant. A similar phenomenon has been discussed before in flat Friedmann-Robertson-Walker spacetimes \cite{Harte2007}, which are both homogeneous and isotropic (though not vacuum). Torques may be used to control translations in other spacetimes as well, but then changes in the momentum can complicate the interpretation.

\section{Conclusion}

We have derived universal constraints on the gravitational forces and torques which can be produced by an object's quadrupole moment, both in Newtonian gravity and in general relativity. Depending on the algebraic structure of the relevant tidal tensor, certain quadrupole moments can be irrelevant, certain torques can be impossible, and only certain forces can be produced without an accompanying torque. These results are summarized at the ends of Sects. \ref{Sect:NewtAlgConstraints} and \ref{Sect:RelAlgConstraints}, and particularly by Tables \ref{Table:NewtDims} and \ref{Table:RelSumm}. We have also found that the quadrupolar force can be viewed as an affine function of the quadrupolar torque; see \eqref{FlinNNewt} and \eqref{ForceTorqueGen}. These results are independent of any spin supplementary conditions.  

Fundamentally, our results are explained by the existence of ``local symmetries.'' In the Newtonian case, local symmetries  correspond to Euclidean Killing fields which locally preserve the tidal tensor. In general relativity, local symmetries are generalized Killing fields which locally preserve the Riemann tensor. Regardless, each local symmetry precludes certain force and torque combinations. This generalizes the well-known fact that Killing fields constrain motion in general relativity. In fact, no further generalization is possible: \textit{Every} universal constraint on extended-body motion is associated with a local symmetry, at least at quadrupolar order. Any generalized force which is not forbidden by local symmetries may be experienced by a suitably-structured object.

The local symmetries we have introduced are an essentially geometric concept, and may thus be of interest not only in the theory of motion. Roughly speaking, the generalized Killing fields introduced in \cite{HarteSyms} provide a sense in which full Poincar\'{e} symmetry can exist around a given worldline in a curved spacetime. This results in Killing's equation being satisfied at least through first order on the reference worldline. Additionally, certain geometric structures (though not the metric) are preserved even away from that worldline. However, it is natural to ask if Killing's equation can be made to hold through one higher order, at least at one point along the reference worldline. When this occurs, we have a local symmetry. Perhaps surprisingly, examples are common and have the physically-interesting consequences described above. We have shown explicitly how to construct all  local symmetries, both in general relativity and in Newtonian gravity. Their number depends, in part, on the algebraic structure of the tidal tensor, and is summarized in the second columns of Tables \ref{Table:NewtDims} and \ref{Table:RelSumm}. All ordinary Killing fields generate a local symmetry in a curved spacetime, and we have shown that conformal Killing-Yano tensors do so as well. Many local symmetries are, however, unrelated either to Killing vectors or to conformal Killing-Yano tensors.

\appendix

\section{Notation}
\label{App:Notation}

We use the same sign conventions as Wald \cite{Wald}, so, e.g., the Riemann tensor satisfies $R_{abc}{}^{d} \omega_d = 2 \nabla_{[a} \nabla_{b]} \omega_c$ for any covector $\omega_a$. The letters $a,b,\ldots$ are used to denote abstract indices in both three and four dimensions, $i,j, \ldots$ are used to denote three-dimensional coordinate components,  $\alpha, \beta, \ldots$ are used to denote four-dimensional coordinate components, and $I,J, \ldots$ are used for numerical indices which are not associated with any coordinates.  Hodge duals are indicated by $*$ and overbars are used to denote complex conjugates. 

Symbols which are commonly used in the text are summarized in Table \ref{Table:Notation}. Generalized forces, as well as ordinary forces and torques, are often supplemented with a ``$(q)$'' superscript to refer only to quadrupolar contributions. We use the abbreviations ``GKF'' (generalized Killing field), ``PND'' (principal null direction), and ``CKY'' (conformal Killing-Yano). Three classification schemes are also used: The algebraic structure of Newtonian tidal tensors is summarized in \ref{Sect:NewtAlgConstraints}, the analogous Petrov classification for four-dimensional Weyl tensors is summarized in Appendix \ref{App:PNDs}, and that Appendix also specifies the three types of tetrad transformations which may be performed in four spacetime dimensions.

\begin{table*}
  \centering
  \setlength{\tabcolsep}{9pt}

  \begin{tabular}{p{8em} p{0.4\textwidth} p{12em}}
    \hline\hline
    Symbol & Description & Reference \\[.3ex]
    \hline
    $x$	&	Generic point	&	-
    \\
    $x^i$   &   Spatial (usually Cartesian) coordinates &   -
    \\
    $\gamma_t$, $\gamma_s$	&	Reference point for object's location at time $t$ (or $s$) & 	-
    \\
    $g_{ab}$		&	Metric	&	-
    \\
    $\xi^a$		&	Generalized or ordinary Killing vector	& \eqref{Killing}
    \\
    $\Xi_t^a$, $\Xi_s^a$		&	Local symmetry generator	& \eqref{LieE}, \eqref{localsymRel}, \eqref{localsymRel2}
    \\
    $M$ &   Mass    &   \eqref{mass}
    \\
    $\mathcal{P}_\xi$	& Generalized momentum associated with $\xi^a$	& \eqref{PDefNewt}, \eqref{PpSNewt}, \eqref{PDefRel}
    \\
    $\mathcal{F}_\xi$	& Generalized force associated with $\xi^a$	& \eqref{GenForceNewt}, \eqref{EOMNewtonian},  \eqref{MP}, \eqref{FGenRel}
    \\
    $p_a$, $S^{ab}$	&	Linear and angular momenta & \eqref{PpSNewt}, \eqref{pSNewtInt}
    \\
    $F_a$, $N^{ab}$	&	Force and torque  & \eqref{GenForceNewtFN}, \eqref{EOMNewtonian}, \eqref{MP}
    \\[.2 ex] 
    \hline
    $\ell^a$, $m^a$, $\bar{m}^a$	&  Newtonian semi-null triad & \eqref{Triad}
    \\
    $\Phi$, $\mathcal{E}_{ab}$	& Newtonian potential and tidal tensor & \eqref{tideDef}, \eqref{Ebasis}
    \\
    $\mathcal{E}_\pm$, $\mathcal{E}$ & Real eigenvalues of $\mathcal{E}_{ab}$ and complex tidal scalar & \eqref{Ecomplex}, \eqref{Ebasis}
    \\
    $\tilde{Q}_{ab}$, $Q_{ab}$ 	& Full and trace-free Newtonian quadrupole moments & \eqref{QfullDef}, \eqref{QtfDef}, \eqref{Qbasis}
    \\
    $Q_{\ell\ell}$, $Q_{\ell m}$, $Q_{mm}$	&	Quadrupole scalars & \eqref{Qbasis} 
    \\[.2ex]
    \hline
       $\ell^a$, $n^a$, $m^a$, $\bar{m}^a$ & Relativistic null tetrad &	\eqref{tetradProd} 
    \\
    $X^{ab}$, $Y^{ab}$, $Z^{ab}$ & Bivector basis elements & \eqref{2formBasis}
    \\
    $\Lambda$	&	Cosmological constant	& 	\eqref{vacuum}
    \\
    $R_{abcd}$, $C_{abcd}$ &	 Riemann and Weyl tensors & \eqref{WeylfromRiemann}, \eqref{Weyl}
    \\
    $\Psi_0, \ldots, \Psi_4$ & Weyl scalars & \eqref{PsiDef}, \eqref{Weyl}
    \\
    $\tilde{J}_{abcd}$, $J_{abcd}$ 	& Full and trace-free relativistic quadrupole moments & \eqref{quad}
    \\
    $J_0, \ldots, J_4$	&	Quadrupole scalars & \eqref{quad}, \eqref{Ji}
   	\\
    $f_{ab}$		& Conformal Killing-Yano tensor	& \eqref{CKY}
    \\[.2ex]
    \hline\hline
  \end{tabular}
  
  \caption{\label{tab:symbols} Table of symbols. The first group of symbols are used in both Newtonian and relativistic contexts. The second group lists Newtonian symbols while the third lists relativistic ones.}
  \label{Table:Notation}
\end{table*}

\section{Tetrad transformations, principal null directions, and the Petrov classification} 
\label{App:PNDs}

It is convenient in a four-dimensional spacetime to introduce a complex null tetrad $(\ell^a, n^a, m^a, \bar{m}^a)$, and using \eqref{2formBasis}, any such tetrad can be used to construct the bivector basis $(X^{ab}, Y^{ab}, Z^{ab}, \bar{X}^{ab}, \bar{Y}^{ab}, \bar{Z}^{ab})$. However, different tetrads are possible, and different choices result in different bivectors. This appendix reviews some facts regarding the admissible tetrad transformations, as well as their application to the construction of principal null directions and to the Petrov classification. We also comment on relations between the algebraic classifications of relativistic and Newtonian tidal tensors.

It is explained in, e.g., Sect. 1.8(g) of \cite{Chandra} that all  null tetrads which are normalized according to \eqref{tetradProd} can be generated from a single example using three types of transformation. Type I transformations preserve $\ell^a$, and in terms of an arbitrary complex scalar $a$, these are given by
\begin{subequations}
\begin{gather}
	\ell^a \mapsto \ell^a, \qquad m^a \mapsto m^a + a \ell^a,
	\\
	n^a \mapsto n^a + \bar{a} m^a + a \bar{m}^a + |a|^2 \ell^a .	
\end{gather}	
\end{subequations}
Type II transformations instead preserve $n^a$, and in terms of an arbitrary complex scalar $b$, they are given by
\begin{subequations}
\label{tetradII}
\begin{gather}
	n^a \mapsto n^a, \qquad m^a \mapsto m^a + b n^a,
	\\
	\ell^a \mapsto \ell^a + \bar{b} m^a + b \bar{m}^a + |b|^2 n^a .	
\end{gather}	
\end{subequations}
Lastly, the type III transformations
\begin{equation}
	\ell^a \mapsto |c| \ell^a, \quad n^a \mapsto \frac{1}{|c|} n^a, \quad m^a \mapsto \frac{c}{|c|} m^a
	\label{tetradIII}
\end{equation}
preserve the directions (though not the scales) of both $\ell^a$ and $n^a$, and can be applied for any nonzero complex scalar $c$. All three types of tetrad transformation affect the bivector basis defined by \eqref{2formBasis}. Type I transformations do so via
\begin{subequations}
\begin{gather}
	X^{ab} \mapsto X^{ab}, \quad Y^{ab} \mapsto Y^{ab} -\bar{a} Z^{ab} - \bar{a}^2 X^{ab} ,
	\\
	Z^{ab} \mapsto Z^{ab} + 2 \bar{a} X^{ab},
\end{gather}
\label{bivectorI}
\end{subequations}
type II transformations via
\begin{subequations}
\begin{gather}
	X^{ab} \mapsto X^{ab} + b Z^{ab} - b^2 Y^{ab}, \quad Y^{ab} \mapsto Y^{ab} ,
	\\
	Z^{ab} \mapsto Z^{ab} - 2 b Y^{ab},
\end{gather}
\label{bivectorII}
\end{subequations}
and type III transformations via
\begin{gather}
	X^{ab} \mapsto c X^{ab}, \quad Y^{ab} \mapsto c^{-1} Y^{ab} , \quad
	Z^{ab} \mapsto Z^{ab} .
	\label{bivectorIII}
\end{gather}

The bivector basis may be used to decompose a Weyl tensor $C_{abcd}$ into the five Weyl scalars
\cite{ExactSolns, Chandra} 
\begin{subequations}
\label{PsiDef}
\begin{gather}
	\Psi_0 \equiv \tfrac{1}{4} C_{abcd} X^{ab} X^{cd}, \quad \Psi_1 \equiv \tfrac{1}{8} C_{abcd} X^{ab} Z^{cd},
	\\
	\Psi_2 \equiv \tfrac{1}{16} C_{abcd} Z^{ab} Z^{cd} = -\tfrac{1}{4} C_{abcd} X^{ab} Y^{cd} ,
	\\
	\Psi_3 \equiv -\tfrac{1}{8} C_{abcd} Y^{ab} Z^{cd}, \quad \Psi_4 \equiv \tfrac{1}{4} C_{abcd} Y^{ab} Y^{cd},
\end{gather}
\end{subequations}
which are in general complex. Working in the opposite direction, one can instead write the Weyl tensor in terms of the Weyl scalars and the given bivectors:
\begin{align}
	C_{abcd} = 2 \Re \big[ \Psi_0 Y_{ab} Y_{cd} + \Psi_1 (Y_{ab} Z_{cd} + Z_{ab} Y_{cd}) 
	\nonumber
	\\
	~ + \Psi_2 (Z_{ab} Z_{cd} - X_{ab} Y_{cd} - Y_{ab} X_{cd} )
	\nonumber
	\\
	~ -  \Psi_3 ( X_{ab} Z_{cd} + Z_{ab} X_{cd}) + \Psi_4 X_{ab} X_{cd} \big].
	\label{Weyl}
\end{align}
Regardless, it follows from \eqref{bivectorI}, \eqref{bivectorII}, and \eqref{bivectorIII} that type I tetrad transformations preserve $\Psi_0$, type II transformations preserve $\Psi_4$, and type III transformations preserve $\Psi_2$.

The algebraic structure of a Weyl tensor largely depends on its principal null directions (PNDs). Recall that each PND may be defined as parallel to a nonzero real null vector field $k^a$ which satisfies \cite{Wald, Hall, ExactSolns}
\begin{equation}
	k_{[a} C_{b]cd[e}k_{f]} k^c k^d = 0.
	\label{kPND}
\end{equation}
Noting that $\ell^a$ is tangent to a PND iff $\Psi_0 =0$, PNDs may be generated by using type II tetrad transformations to rotate $\ell^a$ until the zeroth Weyl scalar vanishes. If $n^a$ is not already aligned with a PND, which would occur only when $\Psi_4 = 0$, this method can in fact be used to identify \textit{all} PNDs. Applying it, \eqref{tetradII}, \eqref{bivectorII}, and \eqref{PsiDef} show that any $b$ satisfying
\begin{equation}
	\Psi_0 + 4 b \Psi_1 + 6 b^2 \Psi_2 + 4 b^3 \Psi_3 + b^4 \Psi_4 = 0
	\label{bPND}
\end{equation}
is associated with a PND which is tangent to
\begin{equation}
	k^a = \ell^a + \bar{b} m^a + b \bar{m}^a + |b|^2 n^a .	
\end{equation}
Assuming that $\Psi_4 \neq 0$, \eqref{bPND} is a quartic polynomial in $b$. This implies that there are at most four distinct PNDs. The multiplicity of each PND is defined to be equal to the algebraic multiplicity of the relevant root. Equivalently, multiplicities can be determined by checking whether or not \eqref{kPND} can be strengthened according to the Bel criteria \cite{Hall, GriffithsExact} which are listed in the second column of Table \ref{Table:Bel}. The final column of that table describes how multiple Weyl scalars must vanish when $\ell^a$ is aligned with a degenerate PND; we take advantage of this in Sect. \ref{Sect:RelAlgConstraints} in order to eliminate as many Weyl scalars as possible.

The Petrov type of the Weyl tensor is determined by the multiplicities of its PNDs. Assuming that $C_{abcd} \neq 0$, there are five possibilities \cite{Chandra, Hall, GriffithsExact, ExactSolns}, described as Petrov types I, II, D, III, and N:
\begin{enumerate}
	\item[I.] Four multiplicity-1 PNDs.
	\item [II.] One multiplicity-2 and two multiplicity-1 PNDs.
	\item [D.] Two multiplicity-2 PNDs.
	\item [III.] One multiplicity-3 and one multiplicity-1 PND.
	\item [N.] One multiplicity-4 PND.
\end{enumerate}
These cases are summarized in the second column of Table \ref{Table:Petrov} on page \pageref{Table:Petrov}. ``Generic'' (or ``algebraically general'') spacetimes are of Petrov type I; all other possibilities are referred to as ``algebraically special.''

\begin{table}
\setlength{\tabcolsep}{8pt}
  \begin{tabular}{c | c | c }
%  \hline\hline
   Multiplicity & Weyl constraint & Vanishing $\Psi_I$ \\[.3 ex]
    \hline\hline
    1 	&	$\ell_{[a} C_{b]cd[e} \ell_{f]} \ell^c \ell^d  =0$ & $\Psi_0 $ \\	
    2	& 	$\ell_{[a} C_{b]cde}  \ell^c \ell^d  =0$ & $\Psi_0, \Psi_1$\\
    3	&	$C_{bcde} \ell^c \ell^d = 0$	& $\Psi_0, \Psi_1, \Psi_2$	\\
    4	&	$C_{bcde} \ell^c = 0$	& $\Psi_0, \Psi_1, \Psi_2, \Psi_3$	
    %\\[.2ex]
%    \hline\hline
	\end{tabular}
	\caption{Summary of Bel criteria and vanishing Weyl scalars when $\ell^a$ is aligned with PNDs of differing multiplicities.} 
	\label{Table:Bel}
\end{table}

The Kerr family of spacetimes are all type D, and the two PNDs which appear there are associated with shear-free families of ingoing and outgoing null geodesics. In the Schwarzschild case, these geodesics are purely radial and twist-free; more generally, they are twisted by the rotation of the black hole. Type N solutions include, e.g., gravitational plane waves, where the lone PND is parallel to the rays of the gravitational wave. Although many solutions are known with Petrov types II and III \cite{GriffithsExact, ExactSolns}, most of their interpretations are physically obscure. Nevertheless, there is a sense in which all Petrov types appear generically when expanding the Weyl tensor at large distances in an asymptotically-flat spacetime: The peeling property states that as one approaches future null infinity along an outgoing null geodesic with increasing affine parameter $r$, 
\begin{equation}
	C_{abcd} = \frac{ N_{abcd} }{ r } + \frac{ III_{abcd} }{ r^2 } + \frac{ II_{abcd} }{ r^3 } + \frac{ I_{abcd} }{ r^4 } + \mathcal{O}(r^{-5}),
\end{equation}
where $N_{abcd}$ is of Petrov type N, $III_{abcd}$ is of Petrov type III, $II_{abcd}$ is either of Petrov type II or Petrov type D, and $I_{abcd}$ is of Petrov type I; see Sect. 11.1 of \cite{Wald}. 

The Petrov classification may be related to the classification of Newtonian tidal tensors which is presented in Sect. \ref{Sect:NewtAlgConstraints} above. First recall that a Newtonian gravitational potential $\Phi$ can be associated with the approximate line element \cite{Wald, PoissonWill}
\begin{equation}
	ds^2 = -(1 + 2 \Phi ) dt^2 + (1 - 2 \Phi) (dx^2 +dy^2 + dz^2).
\end{equation}
If $\nabla^2 \Phi = 0$, the corresponding Weyl tensor is
\begin{equation}
	C_{abcd} = 2 [ \mathcal{E}_{d[a} ( \eta_{b]c} + 2 t_{b]} t_c ) -  \mathcal{E}_{c[a} ( \eta_{b]d} + 2 t_{b]} t_d )]
\end{equation}
through first order in $\Phi$, where $t_a \equiv -\nabla_a t$ and $\mathcal{E}_{ab} \equiv -\nabla_a \nabla_b \Phi$ again denotes the Newtonian tidal tensor. If that tidal tensor is of type 3, meaning that it admits a doubly-degenerate eigenvalue, and if $\ell^a$ corresponds to the unit spacelike eigenvector of $\mathcal{E}_{ab}$ which is associated with the non-degenerate eigenvalue, the two null vectors $k_\pm^a \equiv t^a \pm \ell^a$ both satisfy
\begin{equation}
	k^a_\pm k^c_\pm C_{abc}{}^{[d} k^{e]}_\pm = 0.
\end{equation}
Referring to Table \ref{Table:Bel}, this is the Bel criterion for a multiplicity-2 PND. Type 3 Newtonian tidal fields are therefore associated with approximate Petrov type D spacetimes. Newtonian tidal tensors with types 1 and 2 instead correspond to approximate Petrov type I spacetimes.

\section{Tidally-induced quadrupole moments}
\label{App:Tides}

Although this paper is concerned primarily with model-independent constraints on extended-body motion, the formalism can easily be  applied to specific models. This Appendix describes a simple class of models which describe what happens when an object's quadrupole moment is quasi-statically induced by the applied tidal field. We begin with the Newtonian case and then discuss its relativistic counterpart.

\subsection{Newtonian motion}
\label{Sect:NewtTidal}

Introducing a tidal deformability parameter $\kappa$, which is proportional to an object's Love number, one of the simplest nontrivial models for a Newtonian extended body supposes that
\begin{equation}
    Q_{ab} = \kappa \mathcal{E}_{ab}.
    \label{QLove}
\end{equation}
This can describe the approximate structure of a self-gravitating, near-equilibrium fluid which is in a slowly-varying tidal field. Regardless, substitution into the generalized force \eqref{FNewt} shows that
\begin{equation}
    \mathcal{F}_\xi^{(q)} = \frac{1}{4} \kappa \mathcal{L}_\xi (\mathcal{E}^{ab} \mathcal{E}_{ab} ) . 
\end{equation}
Since the Lie derivative here is acting on a scalar, the quadrupolar torque vanishes. The quadrupolar force is instead proportional to the gradient of $\mathcal{E}^{ab} \mathcal{E}_{ab}$, so these bodies act as though they were monopolar particles moving in the effective potential
\begin{equation}
    \Phi_{\mathrm{eff}} = \Phi - \frac{ \kappa }{ 4 M } \mathcal{E}^{ab} \mathcal{E}_{ab}. 
    \label{phiEff}
\end{equation}
In a spherically-symmetric gravitational field where $\Phi$ falls off like $1/r$, extended-body effects thus contribute a $1/r^6$ correction when all quadrupole moments are tidally induced.

Regardless of $\Phi$, the effective potential \eqref{phiEff} is closely related---but not identical to---the more general effective potential \eqref{PhiEffGen}. Using \eqref{QcalDef} and \eqref{QLove}, the quadrupolar term in that latter potential reduces to
\begin{equation}
    - \frac{\kappa(t) }{ 2 M } \mathcal{E}^{ab} ( \gamma_t, t) \mathcal{E}_{ab} (x,t)
    \label{phiEffDiff}
\end{equation}
when the quadrupole moment is tidally induced. This appears to differ by a factor of two from the quadrupolar term 
\begin{equation}
    -\frac{ \kappa(t) }{ 4 M } \mathcal{E}^{ab}(x,t) \mathcal{E}_{ab}(x,t),
\end{equation}
which appears in \eqref{phiEff}. Nevertheless, the \textit{gradients} of both potentials agree when evaluated at $x= \gamma_t$. This means that they are physically equivalent. In the context of tidally-induced quadrupole moments, it is awkward for the two tidal tensors in \eqref{phiEffDiff} to have different arguments. But in the more general context from which that potential arises, it makes sense to assume that the quadrupole component $\mathcal{Q}$ depends only on time.

The quadrupole moment \eqref{QLove} is, in any case, highly idealized, even for the astrophysically-relevant case of a self-gravitating fluid. Somewhat more realistically, internal dissipation can result in the quadrupole moment depending not only on the current value of $\mathcal{E}_{ab}$, but also on its past history; see, e.g., Sect. 2.5 of \cite{PoissonWill}. Allowing for this makes it possible to obtain a nonzero torque. Indeed, refinements of this sort are necessary to explain tidal locking and other astrophysical phenomena \cite{HutTideReview, PoissonWill, Ogilvie2014, Tremaine}.

\subsection{Relativistic motion}
\label{Sect:Tidal}

It is also possible to construct relativistic models for bodies with tidally-induced quadrupole moments. Unlike in the Newtonian case, however, we can easily introduce two deformability (or ``Love-type'') parameters here: Suppose that 
\begin{equation}
	J_{abcd} = \kappa_+ C_{abcd} + \kappa_- C^*_{abcd},
	\label{JLove}
\end{equation}
where $\kappa_\pm$ are two deformability parameters which describe the body's even and odd parity responses. Recalling \eqref{Killing}, substitution into the generalized force \eqref{QuadrupoleForce2} shows that
\begin{equation}
	\mathcal{F}^{(q)}_\xi = - \frac{1}{12} [ \kappa_+ \mathcal{L}_\xi ( C^{abcd} C_{abcd} ) + \kappa_- \mathcal{L}_\xi ( C^{abcd} C^*_{abcd} ) ].
	\label{FgenLove}
\end{equation}
Since this involves only Lie derivatives of scalars, bodies whose quadrupole moments are described by \eqref{JLove} cannot experience  any torque. Forces are however determined by gradients of the curvature scalars $C^{abcd} C_{abcd}$ and $C^{abcd} C^*_{abcd}$. This is similar to the Newtonian case described above.

In the Schwarzschild spacetime, $C^{abcd} C^*_{abcd}$ vanishes while $C^{abcd} C_{abcd}$ does not, implying that only $\kappa_+$ can affect the motion. By contrast, both $\kappa_+$ and $\kappa_-$ contribute to the force in Kerr spacetimes with nonzero angular momentum. In some curved backgrounds, $C^{abcd} C_{abcd}$ and $C^{abcd} C_{abcd}^*$ both vanish, implying that there is no force or torque at all. This occurs in, e.g., all ``vanishing-scalar-invariant'' spacetimes, which are known to be in the Kundt class \cite{Pravda2002}. All pp-waves are special cases. 

Regardless of the spacetime, at least some effects of deformability can be related to changing masses. To see this, suppose that $\kappa_\pm$ are both constant and that the centroid has been chosen by enforcing the Tulczyjew spin supplementary condition \eqref{SSC}. Since $N_{ab}^{(q)} =0$ here, we may focus on non-spinning objects for which $S^{ab} = 0$. The momentum-velocity relation derived in \cite{EhlersRudolph} then reduces to the trivial $p^a = M \dot{\gamma}^a_s$. However, \eqref{MP} implies that although the mass $M$ is not necessarily constant, the ``effective mass''
\begin{equation}
	M_\mathrm{eff} = M - \frac{1}{12}  ( \kappa_+ C^{abcd} C_{abcd} + \kappa_- C^{abcd} C^*_{abcd} ) 
    \label{Meff}
\end{equation}
is. The constancy of $M_\mathrm{eff}$ for relativistic deformable bodies is closely analogous to the appearance of the effective potential \eqref{phiEff} for deformable Newtonian bodies. Some related senses in which conserved Newtonian energies are interpreted relativistically as conserved masses have been discussed in \cite{HarteExtendedtypeD}. We also note that this effective mass differs from \eqref{MeffGen}. The latter assumes that the relevant quadrupole scalars are constant along the object's worldline, which is not typically the case for an object with a tidally-induced quadrupole moment.

Although our model \eqref{JLove} for tidally-induced quadrupole moments is simple, it differs from what is commonly considered in the literature. There, one first introduces a unit timelike vector $u^a$ which is interpreted as describing the body's instantaneous rest frame---perhaps the 4-velocity of its centroid. The Weyl tensor is then decomposed into its electric and magnetic components via $\mathcal{E}_{ab} \equiv C_{acbd} u^c u^d$ and $\mathcal{B}_{ab} \equiv C^*_{acbd} u^c u^d$, and one assumes that there are electric and magnetic deformabilities $\kappa_\mathrm{E}$ and $\kappa_\mathrm{B}$ such that
\begin{align}
	J_{abcd} = \kappa_\mathrm{E} u_{[a} \mathcal{E}_{b][c} u_{d]} - \kappa_\mathrm{B} ( u_{[a} \mathcal{B}_{b]e} \epsilon_{cd}{}^{ef} 
	\nonumber
	\\
	~  + u_{[c} \mathcal{B}_{d]e} \epsilon_{ab}{}^{ef} ) u_f.
	\label{JLove2}
\end{align}
See, e.g., Eq. (2.14) of \cite{Ramond2021} and references therein. While this model differs from \eqref{JLove} in general, there is overlap when $\kappa_- = 0$. Regardless, the model involving $\kappa_+$ and $\kappa_-$ results in considerably simpler force expressions than the one involving $\kappa_\mathrm{E}$ and $\kappa_\mathrm{B}$.

\bibliography{refs.bib}

\begin{thebibliography}{76}%
\makeatletter
\providecommand \@ifxundefined [1]{%
 \@ifx{#1\undefined}
}%
\providecommand \@ifnum [1]{%
 \ifnum #1\expandafter \@firstoftwo
 \else \expandafter \@secondoftwo
 \fi
}%
\providecommand \@ifx [1]{%
 \ifx #1\expandafter \@firstoftwo
 \else \expandafter \@secondoftwo
 \fi
}%
\providecommand \natexlab [1]{#1}%
\providecommand \enquote  [1]{``#1''}%
\providecommand \bibnamefont  [1]{#1}%
\providecommand \bibfnamefont [1]{#1}%
\providecommand \citenamefont [1]{#1}%
\providecommand \href@noop [0]{\@secondoftwo}%
\providecommand \href [0]{\begingroup \@sanitize@url \@href}%
\providecommand \@href[1]{\@@startlink{#1}\@@href}%
\providecommand \@@href[1]{\endgroup#1\@@endlink}%
\providecommand \@sanitize@url [0]{\catcode `\\12\catcode `\$12\catcode
  `\&12\catcode `\#12\catcode `\^12\catcode `\_12\catcode `\%12\relax}%
\providecommand \@@startlink[1]{}%
\providecommand \@@endlink[0]{}%
\providecommand \url  [0]{\begingroup\@sanitize@url \@url }%
\providecommand \@url [1]{\endgroup\@href {#1}{\urlprefix }}%
\providecommand \urlprefix  [0]{URL }%
\providecommand \Eprint [0]{\href }%
\providecommand \doibase [0]{http://dx.doi.org/}%
\providecommand \selectlanguage [0]{\@gobble}%
\providecommand \bibinfo  [0]{\@secondoftwo}%
\providecommand \bibfield  [0]{\@secondoftwo}%
\providecommand \translation [1]{[#1]}%
\providecommand \BibitemOpen [0]{}%
\providecommand \bibitemStop [0]{}%
\providecommand \bibitemNoStop [0]{.\EOS\space}%
\providecommand \EOS [0]{\spacefactor3000\relax}%
\providecommand \BibitemShut  [1]{\csname bibitem#1\endcsname}%
\let\auto@bib@innerbib\@empty
%</preamble>
\bibitem [{\citenamefont {Ehlers}\ and\ \citenamefont
  {Geroch}(2004)}]{Ehlers2004}%
  \BibitemOpen
  \bibfield  {author} {\bibinfo {author} {\bibfnamefont {J.}~\bibnamefont
  {Ehlers}}\ and\ \bibinfo {author} {\bibfnamefont {R.}~\bibnamefont
  {Geroch}},\ }\href {\doibase 10.1016/j.aop.2003.08.020} {\bibfield  {journal}
  {\bibinfo  {journal} {Ann. Phys.}\ }\textbf {\bibinfo {volume} {309}},\
  \bibinfo {pages} {232} (\bibinfo {year} {2004})}\BibitemShut {NoStop}%
\bibitem [{\citenamefont {Yang}(2013)}]{Yang2013}%
  \BibitemOpen
  \bibfield  {author} {\bibinfo {author} {\bibfnamefont {S.}~\bibnamefont
  {Yang}},\ }\href {\doibase 10.1007/s00220-013-1834-7} {\bibfield  {journal}
  {\bibinfo  {journal} {Comm. Math. Phys.}\ }\textbf {\bibinfo {volume}
  {325}},\ \bibinfo {pages} {997} (\bibinfo {year} {2013})}\BibitemShut
  {NoStop}%
\bibitem [{\citenamefont {Geroch}\ and\ \citenamefont
  {Weatherall}(2018)}]{Geroch2018}%
  \BibitemOpen
  \bibfield  {author} {\bibinfo {author} {\bibfnamefont {R.}~\bibnamefont
  {Geroch}}\ and\ \bibinfo {author} {\bibfnamefont {J.~O.}\ \bibnamefont
  {Weatherall}},\ }\href {https://doi.org/10.1007/s00220-018-3268-8} {\bibfield
   {journal} {\bibinfo  {journal} {Comm. Math. Phys.}\ }\textbf {\bibinfo
  {volume} {364}},\ \bibinfo {pages} {607} (\bibinfo {year}
  {2018})}\BibitemShut {NoStop}%
\bibitem [{\citenamefont {Gralla}\ and\ \citenamefont
  {Wald}(2008)}]{GrallaWald}%
  \BibitemOpen
  \bibfield  {author} {\bibinfo {author} {\bibfnamefont {S.~E.}\ \bibnamefont
  {Gralla}}\ and\ \bibinfo {author} {\bibfnamefont {R.~M.}\ \bibnamefont
  {Wald}},\ }\href {http://stacks.iop.org/0264-9381/25/i=20/a=205009}
  {\bibfield  {journal} {\bibinfo  {journal} {Class. Quantum Grav.}\ }\textbf
  {\bibinfo {volume} {25}},\ \bibinfo {pages} {205009} (\bibinfo {year}
  {2008})}\BibitemShut {NoStop}%
\bibitem [{\citenamefont {Poisson}\ \emph {et~al.}(2011)\citenamefont
  {Poisson}, \citenamefont {Pound},\ and\ \citenamefont {Vega}}]{PoissonRev}%
  \BibitemOpen
  \bibfield  {author} {\bibinfo {author} {\bibfnamefont {E.}~\bibnamefont
  {Poisson}}, \bibinfo {author} {\bibfnamefont {A.}~\bibnamefont {Pound}}, \
  and\ \bibinfo {author} {\bibfnamefont {I.}~\bibnamefont {Vega}},\ }\href
  {https://doi.org/10.12942/lrr-2011-7} {\bibfield  {journal} {\bibinfo
  {journal} {Living Rev. Relativ.}\ }\textbf {\bibinfo {volume} {14}} (\bibinfo
  {year} {2011})}\BibitemShut {NoStop}%
\bibitem [{\citenamefont {Harte}(2015)}]{HarteReview}%
  \BibitemOpen
  \bibfield  {author} {\bibinfo {author} {\bibfnamefont {A.~I.}\ \bibnamefont
  {Harte}},\ }in\ \href {https://doi.org/10.1007/978-3-319-18335-0_12} {\emph
  {\bibinfo {booktitle} {Equations of motion in relativistic gravity}}},\
  \bibinfo {series} {Fundamental Theories of Physics}, Vol.\ \bibinfo {volume}
  {179},\ \bibinfo {editor} {edited by\ \bibinfo {editor} {\bibfnamefont
  {D.}~\bibnamefont {Puetzfeld}}, \bibinfo {editor} {\bibfnamefont
  {C.}~\bibnamefont {L\"{a}mmerzahl}}, \ and\ \bibinfo {editor} {\bibfnamefont
  {B.}~\bibnamefont {Schutz}}}\ (\bibinfo  {publisher} {Springer},\ \bibinfo
  {year} {2015})\ p.\ \bibinfo {pages} {327}\BibitemShut {NoStop}%
\bibitem [{\citenamefont {Barack}\ and\ \citenamefont
  {Pound}(2018)}]{BarackPoundReview}%
  \BibitemOpen
  \bibfield  {author} {\bibinfo {author} {\bibfnamefont {L.}~\bibnamefont
  {Barack}}\ and\ \bibinfo {author} {\bibfnamefont {A.}~\bibnamefont {Pound}},\
  }\href {\doibase 10.1088/1361-6633/aae552} {\bibfield  {journal} {\bibinfo
  {journal} {Rept. Prog. Phys.}\ }\textbf {\bibinfo {volume} {82}},\ \bibinfo
  {pages} {016904} (\bibinfo {year} {2018})}\BibitemShut {NoStop}%
\bibitem [{\citenamefont {Dixon}(1979)}]{Dix79}%
  \BibitemOpen
  \bibfield  {author} {\bibinfo {author} {\bibfnamefont {W.~G.}\ \bibnamefont
  {Dixon}},\ }in\ \href@noop {} {\emph {\bibinfo {booktitle} {Isolated
  Gravitating Systems in General Relativity}}},\ \bibinfo {editor} {edited by\
  \bibinfo {editor} {\bibfnamefont {J.}~\bibnamefont {Ehlers}}}\ (\bibinfo
  {publisher} {North-Holland},\ \bibinfo {year} {1979})\BibitemShut {NoStop}%
\bibitem [{\citenamefont {Thorne}\ and\ \citenamefont
  {Hartle}(1985)}]{ThorneHartle}%
  \BibitemOpen
  \bibfield  {author} {\bibinfo {author} {\bibfnamefont {K.~S.}\ \bibnamefont
  {Thorne}}\ and\ \bibinfo {author} {\bibfnamefont {J.~B.}\ \bibnamefont
  {Hartle}},\ }\href {https://doi.org/10.1103/PhysRevD.31.1815} {\bibfield
  {journal} {\bibinfo  {journal} {Phys. Rev. D}\ }\textbf {\bibinfo {volume}
  {31}},\ \bibinfo {pages} {1815} (\bibinfo {year} {1985})}\BibitemShut
  {NoStop}%
\bibitem [{\citenamefont {{Hut}}(1981)}]{HutTideReview}%
  \BibitemOpen
  \bibfield  {author} {\bibinfo {author} {\bibfnamefont {P.}~\bibnamefont
  {{Hut}}},\ }\href {https://ui.adsabs.harvard.edu/abs/1981A&A....99..126H}
  {\bibfield  {journal} {\bibinfo  {journal} {Astron. and Astrophys.}\ }\textbf
  {\bibinfo {volume} {99}},\ \bibinfo {pages} {126} (\bibinfo {year}
  {1981})}\BibitemShut {NoStop}%
\bibitem [{\citenamefont {Tremaine}(2023)}]{Tremaine}%
  \BibitemOpen
  \bibfield  {author} {\bibinfo {author} {\bibfnamefont {S.}~\bibnamefont
  {Tremaine}},\ }\href@noop {} {\emph {\bibinfo {title} {Dynamics of Planetary
  Systems}}}\ (\bibinfo  {publisher} {Princeton University Press},\ \bibinfo
  {year} {2023})\BibitemShut {NoStop}%
\bibitem [{\citenamefont {Poisson}\ and\ \citenamefont
  {Will}(2014)}]{PoissonWill}%
  \BibitemOpen
  \bibfield  {author} {\bibinfo {author} {\bibfnamefont {E.}~\bibnamefont
  {Poisson}}\ and\ \bibinfo {author} {\bibfnamefont {C.~M.}\ \bibnamefont
  {Will}},\ }\href@noop {} {\emph {\bibinfo {title} {Gravity: Newtonian,
  Post-Newtonian, Relativistic}}}\ (\bibinfo  {publisher} {Cambridge University
  Press},\ \bibinfo {year} {2014})\BibitemShut {NoStop}%
\bibitem [{\citenamefont {Ferroglia}\ and\ \citenamefont
  {Fiolhais}(2020)}]{Ferroglia2020}%
  \BibitemOpen
  \bibfield  {author} {\bibinfo {author} {\bibfnamefont {A.}~\bibnamefont
  {Ferroglia}}\ and\ \bibinfo {author} {\bibfnamefont {M.~C.~N.}\ \bibnamefont
  {Fiolhais}},\ }\href {\doibase 10.1119/10.0001772} {\bibfield  {journal}
  {\bibinfo  {journal} {Am. J. Phys.}\ }\textbf {\bibinfo {volume} {88}},\
  \bibinfo {pages} {1059} (\bibinfo {year} {2020})}\BibitemShut {NoStop}%
\bibitem [{\citenamefont {Wisdom}\ \emph {et~al.}(1984)\citenamefont {Wisdom},
  \citenamefont {Peale},\ and\ \citenamefont {Mignard}}]{Wisdom1984}%
  \BibitemOpen
  \bibfield  {author} {\bibinfo {author} {\bibfnamefont {J.}~\bibnamefont
  {Wisdom}}, \bibinfo {author} {\bibfnamefont {S.~J.}\ \bibnamefont {Peale}}, \
  and\ \bibinfo {author} {\bibfnamefont {F.}~\bibnamefont {Mignard}},\ }\href
  {\doibase 10.1016/0019-1035(84)90032-0} {\bibfield  {journal} {\bibinfo
  {journal} {Icarus}\ }\textbf {\bibinfo {volume} {58}},\ \bibinfo {pages}
  {137} (\bibinfo {year} {1984})}\BibitemShut {NoStop}%
\bibitem [{\citenamefont {Wisdom}(1987)}]{WisdomChaosReview}%
  \BibitemOpen
  \bibfield  {author} {\bibinfo {author} {\bibfnamefont {J.}~\bibnamefont
  {Wisdom}},\ }\href {\doibase 10.1016/0019-1035(87)90175-8} {\bibfield
  {journal} {\bibinfo  {journal} {Icarus}\ }\textbf {\bibinfo {volume} {72}},\
  \bibinfo {pages} {241} (\bibinfo {year} {1987})}\BibitemShut {NoStop}%
\bibitem [{\citenamefont {Ogilvie}(2014)}]{Ogilvie2014}%
  \BibitemOpen
  \bibfield  {author} {\bibinfo {author} {\bibfnamefont {G.~I.}\ \bibnamefont
  {Ogilvie}},\ }\href {\doibase 10.1146/annurev-astro-081913-035941} {\bibfield
   {journal} {\bibinfo  {journal} {Ann. Rev. Astron. Astrophys.}\ }\textbf
  {\bibinfo {volume} {52}},\ \bibinfo {pages} {171} (\bibinfo {year}
  {2014})}\BibitemShut {NoStop}%
\bibitem [{\citenamefont {Flanagan}\ and\ \citenamefont
  {Hinderer}(2008)}]{Flanagan2008}%
  \BibitemOpen
  \bibfield  {author} {\bibinfo {author} {\bibfnamefont {{\'{E}}.~{\'{E}}.}\
  \bibnamefont {Flanagan}}\ and\ \bibinfo {author} {\bibfnamefont
  {T.}~\bibnamefont {Hinderer}},\ }\href
  {https://doi.org/10.1103/physrevd.77.021502} {\bibfield  {journal} {\bibinfo
  {journal} {Phys. Rev. D}\ }\textbf {\bibinfo {volume} {77}},\ \bibinfo
  {pages} {021502(R)} (\bibinfo {year} {2008})}\BibitemShut {NoStop}%
\bibitem [{\citenamefont {Chatziioannou}\ \emph {et~al.}(2018)\citenamefont
  {Chatziioannou}, \citenamefont {Haster},\ and\ \citenamefont
  {Zimmerman}}]{Chatziioannou2018}%
  \BibitemOpen
  \bibfield  {author} {\bibinfo {author} {\bibfnamefont {K.}~\bibnamefont
  {Chatziioannou}}, \bibinfo {author} {\bibfnamefont {C.-J.}\ \bibnamefont
  {Haster}}, \ and\ \bibinfo {author} {\bibfnamefont {A.}~\bibnamefont
  {Zimmerman}},\ }\href {https://doi.org/10.1103/physrevd.97.104036} {\bibfield
   {journal} {\bibinfo  {journal} {Phys. Rev. D}\ }\textbf {\bibinfo {volume}
  {97}},\ \bibinfo {pages} {104036} (\bibinfo {year} {2018})}\BibitemShut
  {NoStop}%
\bibitem [{\citenamefont {Baiotti}\ and\ \citenamefont
  {Rezzolla}(2017)}]{Baiotti2017}%
  \BibitemOpen
  \bibfield  {author} {\bibinfo {author} {\bibfnamefont {L.}~\bibnamefont
  {Baiotti}}\ and\ \bibinfo {author} {\bibfnamefont {L.}~\bibnamefont
  {Rezzolla}},\ }\href {\doibase 10.1088/1361-6633/aa67bb} {\bibfield
  {journal} {\bibinfo  {journal} {Rept. Prog. Phys.}\ }\textbf {\bibinfo
  {volume} {80}},\ \bibinfo {pages} {096901} (\bibinfo {year}
  {2017})}\BibitemShut {NoStop}%
\bibitem [{\citenamefont {Murakami}(1981)}]{Murakami1981}%
  \BibitemOpen
  \bibfield  {author} {\bibinfo {author} {\bibfnamefont {C.}~\bibnamefont
  {Murakami}},\ }\href {\doibase 10.1016/0094-5765(81)90014-x} {\bibfield
  {journal} {\bibinfo  {journal} {Acta Astron.}\ }\textbf {\bibinfo {volume}
  {8}},\ \bibinfo {pages} {733} (\bibinfo {year} {1981})}\BibitemShut {NoStop}%
\bibitem [{\citenamefont {Martinez-Sanchez}\ and\ \citenamefont
  {Gavit}(1987)}]{MartinezSanchez}%
  \BibitemOpen
  \bibfield  {author} {\bibinfo {author} {\bibfnamefont {M.}~\bibnamefont
  {Martinez-Sanchez}}\ and\ \bibinfo {author} {\bibfnamefont {S.~A.}\
  \bibnamefont {Gavit}},\ }\href {\doibase 10.2514/3.20208} {\bibfield
  {journal} {\bibinfo  {journal} {J. Guid. Cont. Dyn.}\ }\textbf {\bibinfo
  {volume} {10}},\ \bibinfo {pages} {233} (\bibinfo {year} {1987})}\BibitemShut
  {NoStop}%
\bibitem [{\citenamefont {Beletsky}(2001)}]{Beletsky}%
  \BibitemOpen
  \bibfield  {author} {\bibinfo {author} {\bibfnamefont {V.~V.}\ \bibnamefont
  {Beletsky}},\ }\href@noop {} {\emph {\bibinfo {title} {Essays on the Motion
  of Celestial Bodies}}}\ (\bibinfo  {publisher} {Birkhäuser Basel},\ \bibinfo
  {year} {2001})\BibitemShut {NoStop}%
\bibitem [{\citenamefont {Gratus}\ and\ \citenamefont
  {Tucker}(2003)}]{GratusTucker}%
  \BibitemOpen
  \bibfield  {author} {\bibinfo {author} {\bibfnamefont {J.}~\bibnamefont
  {Gratus}}\ and\ \bibinfo {author} {\bibfnamefont {R.}~\bibnamefont
  {Tucker}},\ }\href {\doibase 10.1016/s0094-5765(02)00202-3} {\bibfield
  {journal} {\bibinfo  {journal} {Acta Astron.}\ }\textbf {\bibinfo {volume}
  {53}},\ \bibinfo {pages} {161} (\bibinfo {year} {2003})}\BibitemShut
  {NoStop}%
\bibitem [{\citenamefont {Harte}\ and\ \citenamefont
  {Gaffney}(2021)}]{HarteNewtonian}%
  \BibitemOpen
  \bibfield  {author} {\bibinfo {author} {\bibfnamefont {A.~I.}\ \bibnamefont
  {Harte}}\ and\ \bibinfo {author} {\bibfnamefont {M.~T.}\ \bibnamefont
  {Gaffney}},\ }\href {\doibase
  https://doi.org/10.1016/j.actaastro.2020.09.038} {\bibfield  {journal}
  {\bibinfo  {journal} {Acta Astron.}\ }\textbf {\bibinfo {volume} {178}},\
  \bibinfo {pages} {625 } (\bibinfo {year} {2021})}\BibitemShut {NoStop}%
\bibitem [{\citenamefont {Wisdom}(2003)}]{Wisdom}%
  \BibitemOpen
  \bibfield  {author} {\bibinfo {author} {\bibfnamefont {J.}~\bibnamefont
  {Wisdom}},\ }\href {\doibase 10.1126/science.1081406} {\bibfield  {journal}
  {\bibinfo  {journal} {Science}\ }\textbf {\bibinfo {volume} {299}},\ \bibinfo
  {pages} {1865} (\bibinfo {year} {2003})}\BibitemShut {NoStop}%
\bibitem [{\citenamefont {Harte}(2007)}]{Harte2007}%
  \BibitemOpen
  \bibfield  {author} {\bibinfo {author} {\bibfnamefont {A.~I.}\ \bibnamefont
  {Harte}},\ }\href {\doibase 10.1088/0264-9381/24/20/015} {\bibfield
  {journal} {\bibinfo  {journal} {Class. Quantum Grav.}\ }\textbf {\bibinfo
  {volume} {24}},\ \bibinfo {pages} {5161} (\bibinfo {year}
  {2007})}\BibitemShut {NoStop}%
\bibitem [{\citenamefont {Avron}\ and\ \citenamefont {Kenneth}(2006)}]{Avron}%
  \BibitemOpen
  \bibfield  {author} {\bibinfo {author} {\bibfnamefont {J.~E.}\ \bibnamefont
  {Avron}}\ and\ \bibinfo {author} {\bibfnamefont {O.}~\bibnamefont
  {Kenneth}},\ }\href {\doibase 10.1088/1367-2630/8/5/068} {\bibfield
  {journal} {\bibinfo  {journal} {New J. Phys.}\ }\textbf {\bibinfo {volume}
  {8}},\ \bibinfo {pages} {68} (\bibinfo {year} {2006})}\BibitemShut {NoStop}%
\bibitem [{\citenamefont {Bergamin}\ \emph {et~al.}(2009)\citenamefont
  {Bergamin}, \citenamefont {Delva},\ and\ \citenamefont {Hees}}]{Bergamin}%
  \BibitemOpen
  \bibfield  {author} {\bibinfo {author} {\bibfnamefont {L.}~\bibnamefont
  {Bergamin}}, \bibinfo {author} {\bibfnamefont {P.}~\bibnamefont {Delva}}, \
  and\ \bibinfo {author} {\bibfnamefont {A.}~\bibnamefont {Hees}},\ }\href
  {\doibase 10.1088/0264-9381/26/18/185006} {\bibfield  {journal} {\bibinfo
  {journal} {Class. Quantum Grav.}\ }\textbf {\bibinfo {volume} {26}},\
  \bibinfo {pages} {185006} (\bibinfo {year} {2009})}\BibitemShut {NoStop}%
\bibitem [{\citenamefont {Silva}\ \emph {et~al.}(2016)\citenamefont {Silva},
  \citenamefont {Matsas},\ and\ \citenamefont {Vanzella}}]{Silva2016}%
  \BibitemOpen
  \bibfield  {author} {\bibinfo {author} {\bibfnamefont {R.~A.}\ \bibnamefont
  {Silva}}, \bibinfo {author} {\bibfnamefont {G.~E.~A.}\ \bibnamefont
  {Matsas}}, \ and\ \bibinfo {author} {\bibfnamefont {D.~A.~T.}\ \bibnamefont
  {Vanzella}},\ }\href {\doibase 10.1103/physrevd.94.121502} {\bibfield
  {journal} {\bibinfo  {journal} {Phys. Rev. D}\ }\textbf {\bibinfo {volume}
  {94}},\ \bibinfo {pages} {121502(R)} (\bibinfo {year} {2016})}\BibitemShut
  {NoStop}%
\bibitem [{\citenamefont {Mendes}\ and\ \citenamefont
  {Poisson}(2017)}]{PoissonSwim}%
  \BibitemOpen
  \bibfield  {author} {\bibinfo {author} {\bibfnamefont {R.~F.~P.}\
  \bibnamefont {Mendes}}\ and\ \bibinfo {author} {\bibfnamefont
  {E.}~\bibnamefont {Poisson}},\ }\href@noop {} {\  (\bibinfo {year} {2017})},\
  \Eprint {http://arxiv.org/abs/arXiv:1707.08870} {arXiv:1707.08870}
  \BibitemShut {NoStop}%
\bibitem [{\citenamefont {Vesel{\'{y}}}\ and\ \citenamefont
  {{\v{Z}}ofka}(2019)}]{Vesely2019}%
  \BibitemOpen
  \bibfield  {author} {\bibinfo {author} {\bibfnamefont {V.}~\bibnamefont
  {Vesel{\'{y}}}}\ and\ \bibinfo {author} {\bibfnamefont {M.}~\bibnamefont
  {{\v{Z}}ofka}},\ }\href {\doibase 10.1088/1361-6382/ab0976} {\bibfield
  {journal} {\bibinfo  {journal} {Class. Quantum Grav.}\ }\textbf {\bibinfo
  {volume} {36}},\ \bibinfo {pages} {075011} (\bibinfo {year}
  {2019})}\BibitemShut {NoStop}%
\bibitem [{\citenamefont {Harte}(2020)}]{HarteExtendedtypeD}%
  \BibitemOpen
  \bibfield  {author} {\bibinfo {author} {\bibfnamefont {A.~I.}\ \bibnamefont
  {Harte}},\ }\href {\doibase 10.1103/PhysRevD.102.124075} {\bibfield
  {journal} {\bibinfo  {journal} {Phys. Rev. D}\ }\textbf {\bibinfo {volume}
  {102}},\ \bibinfo {pages} {124075} (\bibinfo {year} {2020})}\BibitemShut
  {NoStop}%
\bibitem [{\citenamefont {Silva}(2022)}]{Silva2022}%
  \BibitemOpen
  \bibfield  {author} {\bibinfo {author} {\bibfnamefont {R.~A.~e.}\
  \bibnamefont {Silva}},\ }\href {https://arxiv.org/abs/2211.04654} {\
  (\bibinfo {year} {2022})},\ \Eprint {http://arxiv.org/abs/arXiv:2211.04654}
  {arXiv:2211.04654} \BibitemShut {NoStop}%
\bibitem [{\citenamefont {{Dixon}}(1970)}]{Dixon70a}%
  \BibitemOpen
  \bibfield  {author} {\bibinfo {author} {\bibfnamefont {W.~G.}\ \bibnamefont
  {{Dixon}}},\ }\href {\doibase 10.1098/rspa.1970.0020} {\bibfield  {journal}
  {\bibinfo  {journal} {Proc. Roy. Soc. A}\ }\textbf {\bibinfo {volume}
  {314}},\ \bibinfo {pages} {499} (\bibinfo {year} {1970})}\BibitemShut
  {NoStop}%
\bibitem [{\citenamefont {{Dixon}}(1974)}]{Dixon74}%
  \BibitemOpen
  \bibfield  {author} {\bibinfo {author} {\bibfnamefont {W.~G.}\ \bibnamefont
  {{Dixon}}},\ }\href {\doibase 10.1098/rsta.1974.0046} {\bibfield  {journal}
  {\bibinfo  {journal} {Phil. Trans. Roy. Soc. A}\ }\textbf {\bibinfo {volume}
  {277}},\ \bibinfo {pages} {59} (\bibinfo {year} {1974})}\BibitemShut
  {NoStop}%
\bibitem [{\citenamefont {Ehlers}\ and\ \citenamefont
  {Rudolph}(1977)}]{EhlersRudolph}%
  \BibitemOpen
  \bibfield  {author} {\bibinfo {author} {\bibfnamefont {J.}~\bibnamefont
  {Ehlers}}\ and\ \bibinfo {author} {\bibfnamefont {E.}~\bibnamefont
  {Rudolph}},\ }\href {\doibase 10.1007/bf00763547} {\bibfield  {journal}
  {\bibinfo  {journal} {Gen. Rel. Grav.}\ }\textbf {\bibinfo {volume} {8}},\
  \bibinfo {pages} {197} (\bibinfo {year} {1977})}\BibitemShut {NoStop}%
\bibitem [{\citenamefont {Harte}(2008{\natexlab{a}})}]{HarteScalar}%
  \BibitemOpen
  \bibfield  {author} {\bibinfo {author} {\bibfnamefont {A.~I.}\ \bibnamefont
  {Harte}},\ }\href {http://stacks.iop.org/0264-9381/25/i=23/a=235020}
  {\bibfield  {journal} {\bibinfo  {journal} {Class. Quantum Grav.}\ }\textbf
  {\bibinfo {volume} {25}},\ \bibinfo {pages} {235020} (\bibinfo {year}
  {2008}{\natexlab{a}})}\BibitemShut {NoStop}%
\bibitem [{\citenamefont {Dixon}(2015)}]{DixonReview}%
  \BibitemOpen
  \bibfield  {author} {\bibinfo {author} {\bibfnamefont {W.~G.}\ \bibnamefont
  {Dixon}},\ }in\ \href {\doibase 10.1007/978-3-319-18335-0_1} {\emph {\bibinfo
  {booktitle} {Equations of Motion in Relativistic Gravity}}},\ \bibinfo
  {series} {Fundamental Theories of Physics}, Vol.\ \bibinfo {volume} {179},\
  \bibinfo {editor} {edited by\ \bibinfo {editor} {\bibfnamefont
  {D.}~\bibnamefont {Puetzfeld}}, \bibinfo {editor} {\bibfnamefont
  {C.}~\bibnamefont {L\"{a}mmerzahl}}, \ and\ \bibinfo {editor} {\bibfnamefont
  {B.}~\bibnamefont {Schutz}}}\ (\bibinfo  {publisher} {Springer},\ \bibinfo
  {year} {2015})\ p.~\bibinfo {pages} {1}\BibitemShut {NoStop}%
\bibitem [{\citenamefont {Harte}(2008{\natexlab{b}})}]{HarteSyms}%
  \BibitemOpen
  \bibfield  {author} {\bibinfo {author} {\bibfnamefont {A.~I.}\ \bibnamefont
  {Harte}},\ }\href {https://doi.org/10.1088/0264-9381/25/20/205008} {\bibfield
   {journal} {\bibinfo  {journal} {Class. Quantum Grav.}\ }\textbf {\bibinfo
  {volume} {25}},\ \bibinfo {pages} {205008} (\bibinfo {year}
  {2008}{\natexlab{b}})}\BibitemShut {NoStop}%
\bibitem [{\citenamefont {Harte}(2012)}]{HarteGrav}%
  \BibitemOpen
  \bibfield  {author} {\bibinfo {author} {\bibfnamefont {A.~I.}\ \bibnamefont
  {Harte}},\ }\href {https://doi.org/10.1088/0264-9381/29/5/055012} {\bibfield
  {journal} {\bibinfo  {journal} {Class. Quantum Grav.}\ }\textbf {\bibinfo
  {volume} {29}},\ \bibinfo {pages} {055012} (\bibinfo {year}
  {2012})}\BibitemShut {NoStop}%
\bibitem [{\citenamefont {Schattner}(1979{\natexlab{a}})}]{CM1}%
  \BibitemOpen
  \bibfield  {author} {\bibinfo {author} {\bibfnamefont {R.}~\bibnamefont
  {Schattner}},\ }\href {\doibase 10.1007/bf00760221} {\bibfield  {journal}
  {\bibinfo  {journal} {Gen. Rel. Grav.}\ }\textbf {\bibinfo {volume} {10}},\
  \bibinfo {pages} {377} (\bibinfo {year} {1979}{\natexlab{a}})}\BibitemShut
  {NoStop}%
\bibitem [{\citenamefont {Schattner}(1979{\natexlab{b}})}]{CM2}%
  \BibitemOpen
  \bibfield  {author} {\bibinfo {author} {\bibfnamefont {R.}~\bibnamefont
  {Schattner}},\ }\href {\doibase 10.1007/bf00760222} {\bibfield  {journal}
  {\bibinfo  {journal} {Gen. Rel. Grav.}\ }\textbf {\bibinfo {volume} {10}},\
  \bibinfo {pages} {395} (\bibinfo {year} {1979}{\natexlab{b}})}\BibitemShut
  {NoStop}%
\bibitem [{\citenamefont {Harte}(2010)}]{HarteTrenorm}%
  \BibitemOpen
  \bibfield  {author} {\bibinfo {author} {\bibfnamefont {A.~I.}\ \bibnamefont
  {Harte}},\ }\href {https://doi.org/10.1088/0264-9381/27/13/135002} {\bibfield
   {journal} {\bibinfo  {journal} {Class. Quantum Grav.}\ }\textbf {\bibinfo
  {volume} {27}},\ \bibinfo {pages} {135002} (\bibinfo {year}
  {2010})}\BibitemShut {NoStop}%
\bibitem [{\citenamefont {Bini}\ and\ \citenamefont
  {Geralico}(2014)}]{Bini2014}%
  \BibitemOpen
  \bibfield  {author} {\bibinfo {author} {\bibfnamefont {D.}~\bibnamefont
  {Bini}}\ and\ \bibinfo {author} {\bibfnamefont {A.}~\bibnamefont
  {Geralico}},\ }\href {\doibase 10.1103/physrevd.89.044013} {\bibfield
  {journal} {\bibinfo  {journal} {Phys. Rev. D}\ }\textbf {\bibinfo {volume}
  {89}},\ \bibinfo {pages} {044013} (\bibinfo {year} {2014})}\BibitemShut
  {NoStop}%
\bibitem [{\citenamefont {Jezierski}\ and\ \citenamefont
  {{\L}ukasik}(2006)}]{jezierski2006}%
  \BibitemOpen
  \bibfield  {author} {\bibinfo {author} {\bibfnamefont {J.}~\bibnamefont
  {Jezierski}}\ and\ \bibinfo {author} {\bibfnamefont {M.}~\bibnamefont
  {{\L}ukasik}},\ }\href {https://doi.org/10.1088/0264-9381/23/9/008}
  {\bibfield  {journal} {\bibinfo  {journal} {Class. Quantum Grav.}\ }\textbf
  {\bibinfo {volume} {23}},\ \bibinfo {pages} {2895} (\bibinfo {year}
  {2006})}\BibitemShut {NoStop}%
\bibitem [{\citenamefont {Carter}(1977)}]{Carter1977}%
  \BibitemOpen
  \bibfield  {author} {\bibinfo {author} {\bibfnamefont {B.}~\bibnamefont
  {Carter}},\ }\href {\doibase 10.1103/physrevd.16.3395} {\bibfield  {journal}
  {\bibinfo  {journal} {Phys. Rev. D}\ }\textbf {\bibinfo {volume} {16}},\
  \bibinfo {pages} {3395} (\bibinfo {year} {1977})}\BibitemShut {NoStop}%
\bibitem [{\citenamefont {Carter}\ and\ \citenamefont
  {McLenaghan}(1979)}]{Carter1979}%
  \BibitemOpen
  \bibfield  {author} {\bibinfo {author} {\bibfnamefont {B.}~\bibnamefont
  {Carter}}\ and\ \bibinfo {author} {\bibfnamefont {R.~G.}\ \bibnamefont
  {McLenaghan}},\ }\href {\doibase 10.1103/physrevd.19.1093} {\bibfield
  {journal} {\bibinfo  {journal} {Phys. Rev. D}\ }\textbf {\bibinfo {volume}
  {19}},\ \bibinfo {pages} {1093} (\bibinfo {year} {1979})}\BibitemShut
  {NoStop}%
\bibitem [{\citenamefont {Kastor}\ and\ \citenamefont
  {Traschen}(2004)}]{Kastor2004}%
  \BibitemOpen
  \bibfield  {author} {\bibinfo {author} {\bibfnamefont {D.}~\bibnamefont
  {Kastor}}\ and\ \bibinfo {author} {\bibfnamefont {J.}~\bibnamefont
  {Traschen}},\ }\href {\doibase 10.1088/1126-6708/2004/08/045} {\bibfield
  {journal} {\bibinfo  {journal} {J. High En. Phys.}\ }\textbf {\bibinfo
  {volume} {2004}},\ \bibinfo {pages} {045} (\bibinfo {year}
  {2004})}\BibitemShut {NoStop}%
\bibitem [{\citenamefont {Andersson}\ \emph {et~al.}(2015)\citenamefont
  {Andersson}, \citenamefont {B\"{a}ckdahl},\ and\ \citenamefont
  {Blue}}]{LarsSpinReview}%
  \BibitemOpen
  \bibfield  {author} {\bibinfo {author} {\bibfnamefont {L.}~\bibnamefont
  {Andersson}}, \bibinfo {author} {\bibfnamefont {T.}~\bibnamefont
  {B\"{a}ckdahl}}, \ and\ \bibinfo {author} {\bibfnamefont {P.}~\bibnamefont
  {Blue}},\ }\href {\doibase 10.4310/sdg.2015.v20.n1.a8} {\bibfield  {journal}
  {\bibinfo  {journal} {Surv. Diff. Geom.}\ }\textbf {\bibinfo {volume} {20}},\
  \bibinfo {pages} {183} (\bibinfo {year} {2015})}\BibitemShut {NoStop}%
\bibitem [{\citenamefont {Grant}\ and\ \citenamefont
  {Flanagan}(2020{\natexlab{a}})}]{Grant2020EM}%
  \BibitemOpen
  \bibfield  {author} {\bibinfo {author} {\bibfnamefont {A.~M.}\ \bibnamefont
  {Grant}}\ and\ \bibinfo {author} {\bibfnamefont {{\'{E}}.~{\'{E}}.}\
  \bibnamefont {Flanagan}},\ }\href {\doibase 10.1088/1361-6382/ab995a}
  {\bibfield  {journal} {\bibinfo  {journal} {Class. Quantum Grav.}\ }\textbf
  {\bibinfo {volume} {37}},\ \bibinfo {pages} {185021} (\bibinfo {year}
  {2020}{\natexlab{a}})}\BibitemShut {NoStop}%
\bibitem [{\citenamefont {Grant}\ and\ \citenamefont
  {Flanagan}(2020{\natexlab{b}})}]{Grant2020}%
  \BibitemOpen
  \bibfield  {author} {\bibinfo {author} {\bibfnamefont {A.~M.}\ \bibnamefont
  {Grant}}\ and\ \bibinfo {author} {\bibfnamefont {{\'{E}}.~{\'{E}}.}\
  \bibnamefont {Flanagan}},\ }\href {\doibase 10.1088/1361-6382/abc3f7}
  {\bibfield  {journal} {\bibinfo  {journal} {Class. Quantum Grav.}\ }\textbf
  {\bibinfo {volume} {38}},\ \bibinfo {pages} {055004} (\bibinfo {year}
  {2020}{\natexlab{b}})}\BibitemShut {NoStop}%
\bibitem [{\citenamefont {Jezierski}(2002)}]{Jezierski2002}%
  \BibitemOpen
  \bibfield  {author} {\bibinfo {author} {\bibfnamefont {J.}~\bibnamefont
  {Jezierski}},\ }\href {\doibase 10.1088/0264-9381/19/16/313} {\bibfield
  {journal} {\bibinfo  {journal} {Class. Quantum Grav.}\ }\textbf {\bibinfo
  {volume} {19}},\ \bibinfo {pages} {4405} (\bibinfo {year}
  {2002})}\BibitemShut {NoStop}%
\bibitem [{\citenamefont {Grant}\ and\ \citenamefont
  {Flanagan}(2015)}]{Grant2015}%
  \BibitemOpen
  \bibfield  {author} {\bibinfo {author} {\bibfnamefont {A.~M.}\ \bibnamefont
  {Grant}}\ and\ \bibinfo {author} {\bibfnamefont {{\'{E}}.~{\'{E}}.}\
  \bibnamefont {Flanagan}},\ }\href {\doibase 10.1088/0264-9381/32/15/157001}
  {\bibfield  {journal} {\bibinfo  {journal} {Class. Quantum Grav.}\ }\textbf
  {\bibinfo {volume} {32}},\ \bibinfo {pages} {157001} (\bibinfo {year}
  {2015})}\BibitemShut {NoStop}%
\bibitem [{\citenamefont {R\"{u}diger}(1981)}]{Ruediger1}%
  \BibitemOpen
  \bibfield  {author} {\bibinfo {author} {\bibfnamefont {R.}~\bibnamefont
  {R\"{u}diger}},\ }\href {\doibase 10.1098/rspa.1981.0046} {\bibfield
  {journal} {\bibinfo  {journal} {Proc. Roy. Soc. London A}\ }\textbf {\bibinfo
  {volume} {375}},\ \bibinfo {pages} {185} (\bibinfo {year}
  {1981})}\BibitemShut {NoStop}%
\bibitem [{\citenamefont {R\"{u}diger}(1983)}]{Ruediger2}%
  \BibitemOpen
  \bibfield  {author} {\bibinfo {author} {\bibfnamefont {R.}~\bibnamefont
  {R\"{u}diger}},\ }\href {\doibase 10.1098/rspa.1983.0012} {\bibfield
  {journal} {\bibinfo  {journal} {Proc. Roy. Soc. London A}\ }\textbf {\bibinfo
  {volume} {385}},\ \bibinfo {pages} {229} (\bibinfo {year}
  {1983})}\BibitemShut {NoStop}%
\bibitem [{\citenamefont {Compère}\ and\ \citenamefont
  {Druart}(2022)}]{CompereKT1}%
  \BibitemOpen
  \bibfield  {author} {\bibinfo {author} {\bibfnamefont {G.}~\bibnamefont
  {Compère}}\ and\ \bibinfo {author} {\bibfnamefont {A.}~\bibnamefont
  {Druart}},\ }\href {\doibase 10.21468/SciPostPhys.12.1.012} {\bibfield
  {journal} {\bibinfo  {journal} {SciPost Phys.}\ }\textbf {\bibinfo {volume}
  {12}},\ \bibinfo {pages} {012} (\bibinfo {year} {2022})}\BibitemShut
  {NoStop}%
\bibitem [{\citenamefont {Santos}\ and\ \citenamefont
  {Batista}(2020)}]{Santos2020}%
  \BibitemOpen
  \bibfield  {author} {\bibinfo {author} {\bibfnamefont {E.~B.}\ \bibnamefont
  {Santos}}\ and\ \bibinfo {author} {\bibfnamefont {C.}~\bibnamefont
  {Batista}},\ }\href {https://doi.org/10.1103/physrevd.101.104049} {\bibfield
  {journal} {\bibinfo  {journal} {Phys. Rev. D}\ }\textbf {\bibinfo {volume}
  {101}},\ \bibinfo {pages} {104049} (\bibinfo {year} {2020})}\BibitemShut
  {NoStop}%
\bibitem [{\citenamefont {Compère}\ \emph {et~al.}(2023)\citenamefont
  {Compère}, \citenamefont {Druart},\ and\ \citenamefont
  {Vines}}]{CompereKT2}%
  \BibitemOpen
  \bibfield  {author} {\bibinfo {author} {\bibfnamefont {G.}~\bibnamefont
  {Compère}}, \bibinfo {author} {\bibfnamefont {A.}~\bibnamefont {Druart}}, \
  and\ \bibinfo {author} {\bibfnamefont {J.}~\bibnamefont {Vines}},\ }\href
  {https://arxiv.org/abs/2302.14549} {\  (\bibinfo {year} {2023})},\ \Eprint
  {http://arxiv.org/abs/arXiv:2302.14549} {arXiv:2302.14549} \BibitemShut
  {NoStop}%
\bibitem [{\citenamefont {Barnes}(2015)}]{Barnes2015}%
  \BibitemOpen
  \bibfield  {author} {\bibinfo {author} {\bibfnamefont {A.}~\bibnamefont
  {Barnes}},\ }\href {\doibase 10.1088/1742-6596/600/1/012067} {\bibfield
  {journal} {\bibinfo  {journal} {J. Phys.: Conf. Ser.}\ }\textbf {\bibinfo
  {volume} {600}},\ \bibinfo {pages} {012067} (\bibinfo {year}
  {2015})}\BibitemShut {NoStop}%
\bibitem [{\citenamefont {Griffiths}\ and\ \citenamefont
  {Podolsk\'{y}}(2010)}]{GriffithsExact}%
  \BibitemOpen
  \bibfield  {author} {\bibinfo {author} {\bibfnamefont {J.~B.}\ \bibnamefont
  {Griffiths}}\ and\ \bibinfo {author} {\bibfnamefont {J.}~\bibnamefont
  {Podolsk\'{y}}},\ }\href@noop {} {\emph {\bibinfo {title} {Exact Space-Times
  in Einstein's General Relativity}}}\ (\bibinfo  {publisher} {Cambridge
  University Press},\ \bibinfo {year} {2010})\BibitemShut {NoStop}%
\bibitem [{\citenamefont {Gralla}\ \emph {et~al.}(2010)\citenamefont {Gralla},
  \citenamefont {Harte},\ and\ \citenamefont {Wald}}]{Bobbing}%
  \BibitemOpen
  \bibfield  {author} {\bibinfo {author} {\bibfnamefont {S.~E.}\ \bibnamefont
  {Gralla}}, \bibinfo {author} {\bibfnamefont {A.~I.}\ \bibnamefont {Harte}}, \
  and\ \bibinfo {author} {\bibfnamefont {R.~M.}\ \bibnamefont {Wald}},\ }\href
  {\doibase 10.1103/PhysRevD.81.104012} {\bibfield  {journal} {\bibinfo
  {journal} {Phys. Rev. D}\ }\textbf {\bibinfo {volume} {81}},\ \bibinfo
  {pages} {104012} (\bibinfo {year} {2010})}\BibitemShut {NoStop}%
\bibitem [{\citenamefont {Stephani}\ \emph {et~al.}(2009)\citenamefont
  {Stephani}, \citenamefont {Kramer}, \citenamefont {MacCallum}, \citenamefont
  {Hoenselaers},\ and\ \citenamefont {Herlt}}]{ExactSolns}%
  \BibitemOpen
  \bibfield  {author} {\bibinfo {author} {\bibfnamefont {H.}~\bibnamefont
  {Stephani}}, \bibinfo {author} {\bibfnamefont {D.}~\bibnamefont {Kramer}},
  \bibinfo {author} {\bibfnamefont {M.}~\bibnamefont {MacCallum}}, \bibinfo
  {author} {\bibfnamefont {C.~L.~U.}\ \bibnamefont {Hoenselaers}}, \ and\
  \bibinfo {author} {\bibfnamefont {E.}~\bibnamefont {Herlt}},\ }\href@noop {}
  {\emph {\bibinfo {title} {Exact Solutions of Einstein's Field Equations}}}\
  (\bibinfo  {publisher} {Cambridge University Press},\ \bibinfo {year}
  {2009})\BibitemShut {NoStop}%
\bibitem [{\citenamefont {Jordan}\ \emph {et~al.}(2009)\citenamefont {Jordan},
  \citenamefont {Ehlers},\ and\ \citenamefont {Kundt}}]{EhlersKundt}%
  \BibitemOpen
  \bibfield  {author} {\bibinfo {author} {\bibfnamefont {P.}~\bibnamefont
  {Jordan}}, \bibinfo {author} {\bibfnamefont {J.}~\bibnamefont {Ehlers}}, \
  and\ \bibinfo {author} {\bibfnamefont {W.}~\bibnamefont {Kundt}},\ }\href
  {https://dx.doi.org/10.1007/s10714-009-0869-8} {\bibfield  {journal}
  {\bibinfo  {journal} {Gen. Relativ. Grav.}\ }\textbf {\bibinfo {volume}
  {41}},\ \bibinfo {pages} {2191} (\bibinfo {year} {2009})}\BibitemShut
  {NoStop}%
\bibitem [{\citenamefont {Aichelburg}\ and\ \citenamefont
  {Sexl}(1971)}]{AichelburgSexl}%
  \BibitemOpen
  \bibfield  {author} {\bibinfo {author} {\bibfnamefont {P.~C.}\ \bibnamefont
  {Aichelburg}}\ and\ \bibinfo {author} {\bibfnamefont {R.~U.}\ \bibnamefont
  {Sexl}},\ }\href {\doibase 10.1007/bf00758149} {\bibfield  {journal}
  {\bibinfo  {journal} {Gen. Rel. Grav.}\ }\textbf {\bibinfo {volume} {2}},\
  \bibinfo {pages} {303} (\bibinfo {year} {1971})}\BibitemShut {NoStop}%
\bibitem [{\citenamefont {Podolsk{\'{y}}}\ and\ \citenamefont
  {Griffiths}(1998)}]{Podolsk1998}%
  \BibitemOpen
  \bibfield  {author} {\bibinfo {author} {\bibfnamefont {J.}~\bibnamefont
  {Podolsk{\'{y}}}}\ and\ \bibinfo {author} {\bibfnamefont {J.~B.}\
  \bibnamefont {Griffiths}},\ }\href
  {https://doi.org/10.1103/physrevd.58.124024} {\bibfield  {journal} {\bibinfo
  {journal} {Phys. Rev. D}\ }\textbf {\bibinfo {volume} {58}},\ \bibinfo
  {pages} {124024} (\bibinfo {year} {1998})}\BibitemShut {NoStop}%
\bibitem [{\citenamefont {Penrose}(1976)}]{PenroseLimit}%
  \BibitemOpen
  \bibfield  {author} {\bibinfo {author} {\bibfnamefont {R.}~\bibnamefont
  {Penrose}},\ }in\ \href {\doibase 10.1007/978-94-010-1508-0_23} {\emph
  {\bibinfo {booktitle} {Differential Geometry and Relativity}}}\ (\bibinfo
  {publisher} {Springer},\ \bibinfo {year} {1976})\ p.\ \bibinfo {pages}
  {271}\BibitemShut {NoStop}%
\bibitem [{\citenamefont {Blau}\ \emph {et~al.}(2006)\citenamefont {Blau},
  \citenamefont {Frank},\ and\ \citenamefont {Weiss}}]{Blau}%
  \BibitemOpen
  \bibfield  {author} {\bibinfo {author} {\bibfnamefont {M.}~\bibnamefont
  {Blau}}, \bibinfo {author} {\bibfnamefont {D.}~\bibnamefont {Frank}}, \ and\
  \bibinfo {author} {\bibfnamefont {S.}~\bibnamefont {Weiss}},\ }\href
  {https://doi.org/10.1088/0264-9381/23/11/020} {\bibfield  {journal} {\bibinfo
   {journal} {Class. Quantum Grav.}\ }\textbf {\bibinfo {volume} {23}},\
  \bibinfo {pages} {3993} (\bibinfo {year} {2006})}\BibitemShut {NoStop}%
\bibitem [{\citenamefont {Xanthopoulos}(1978)}]{Xanthopoulos1978}%
  \BibitemOpen
  \bibfield  {author} {\bibinfo {author} {\bibfnamefont {B.~C.}\ \bibnamefont
  {Xanthopoulos}},\ }\href {\doibase 10.1063/1.523851} {\bibfield  {journal}
  {\bibinfo  {journal} {J. Math. Phys.}\ }\textbf {\bibinfo {volume} {19}},\
  \bibinfo {pages} {1607} (\bibinfo {year} {1978})}\BibitemShut {NoStop}%
\bibitem [{\citenamefont {Harte}\ and\ \citenamefont
  {Vines}(2016)}]{HarteVines}%
  \BibitemOpen
  \bibfield  {author} {\bibinfo {author} {\bibfnamefont {A.~I.}\ \bibnamefont
  {Harte}}\ and\ \bibinfo {author} {\bibfnamefont {J.}~\bibnamefont {Vines}},\
  }\href {\doibase 10.1103/physrevd.94.084009} {\bibfield  {journal} {\bibinfo
  {journal} {Phys. Rev. D}\ }\textbf {\bibinfo {volume} {94}},\ \bibinfo
  {pages} {084009} (\bibinfo {year} {2016})}\BibitemShut {NoStop}%
\bibitem [{\citenamefont {Sippel}\ and\ \citenamefont
  {Goenner}(1986)}]{SippelSyms}%
  \BibitemOpen
  \bibfield  {author} {\bibinfo {author} {\bibfnamefont {R.}~\bibnamefont
  {Sippel}}\ and\ \bibinfo {author} {\bibfnamefont {H.}~\bibnamefont
  {Goenner}},\ }\href {\doibase 10.1007/bf00763448} {\bibfield  {journal}
  {\bibinfo  {journal} {Gen. Rel. Grav.}\ }\textbf {\bibinfo {volume} {18}},\
  \bibinfo {pages} {1229} (\bibinfo {year} {1986})}\BibitemShut {NoStop}%
\bibitem [{\citenamefont {Harvey}(1990)}]{HarveyKasner}%
  \BibitemOpen
  \bibfield  {author} {\bibinfo {author} {\bibfnamefont {A.}~\bibnamefont
  {Harvey}},\ }\href {\doibase 10.1007/bf00756841} {\bibfield  {journal}
  {\bibinfo  {journal} {Gen. Rel. Grav.}\ }\textbf {\bibinfo {volume} {22}},\
  \bibinfo {pages} {1433} (\bibinfo {year} {1990})}\BibitemShut {NoStop}%
\bibitem [{\citenamefont {Wald}(1984)}]{Wald}%
  \BibitemOpen
  \bibfield  {author} {\bibinfo {author} {\bibfnamefont {R.~M.}\ \bibnamefont
  {Wald}},\ }\href@noop {} {\emph {\bibinfo {title} {General Relativity}}}\
  (\bibinfo  {publisher} {University of Chicago Press},\ \bibinfo {year}
  {1984})\BibitemShut {NoStop}%
\bibitem [{\citenamefont {Chandrasekhar}(1998)}]{Chandra}%
  \BibitemOpen
  \bibfield  {author} {\bibinfo {author} {\bibfnamefont {S.}~\bibnamefont
  {Chandrasekhar}},\ }\href@noop {} {\emph {\bibinfo {title} {The Mathematical
  Theory of Black Holes}}}\ (\bibinfo  {publisher} {Oxford University Press},\
  \bibinfo {year} {1998})\BibitemShut {NoStop}%
\bibitem [{\citenamefont {Hall}(2004)}]{Hall}%
  \BibitemOpen
  \bibfield  {author} {\bibinfo {author} {\bibfnamefont {G.~S.}\ \bibnamefont
  {Hall}},\ }\href@noop {} {\emph {\bibinfo {title} {Symmetries and Curvature
  Structure in General Relativity}}}\ (\bibinfo  {publisher} {World
  Scientific},\ \bibinfo {year} {2004})\BibitemShut {NoStop}%
\bibitem [{\citenamefont {Pravda}\ \emph {et~al.}(2002)\citenamefont {Pravda},
  \citenamefont {Pravdov\'{a}}, \citenamefont {Coley},\ and\ \citenamefont
  {Milson}}]{Pravda2002}%
  \BibitemOpen
  \bibfield  {author} {\bibinfo {author} {\bibfnamefont {V.}~\bibnamefont
  {Pravda}}, \bibinfo {author} {\bibfnamefont {A.}~\bibnamefont
  {Pravdov\'{a}}}, \bibinfo {author} {\bibfnamefont {A.}~\bibnamefont {Coley}},
  \ and\ \bibinfo {author} {\bibfnamefont {R.}~\bibnamefont {Milson}},\ }\href
  {\doibase 10.1088/0264-9381/19/23/318} {\bibfield  {journal} {\bibinfo
  {journal} {Class. Quantum Grav.}\ }\textbf {\bibinfo {volume} {19}},\
  \bibinfo {pages} {6213} (\bibinfo {year} {2002})}\BibitemShut {NoStop}%
\bibitem [{\citenamefont {Ramond}\ and\ \citenamefont {{Le
  Tiec}}(2021)}]{Ramond2021}%
  \BibitemOpen
  \bibfield  {author} {\bibinfo {author} {\bibfnamefont {P.}~\bibnamefont
  {Ramond}}\ and\ \bibinfo {author} {\bibfnamefont {A.}~\bibnamefont {{Le
  Tiec}}},\ }\href {\doibase 10.1088/1361-6382/abebef} {\bibfield  {journal}
  {\bibinfo  {journal} {Class. Quantum Grav.}\ }\textbf {\bibinfo {volume}
  {38}},\ \bibinfo {pages} {135022} (\bibinfo {year} {2021})}\BibitemShut
  {NoStop}%
\end{thebibliography}%

\end{document}